\documentclass[12pt]{article}
\usepackage{axodraw}
\usepackage{pstricks}
\usepackage{color}
\usepackage{cite}
\usepackage{array}
\usepackage{epsfig}
\usepackage{amssymb}
\usepackage{graphics,graphpap}
\usepackage{amssymb}
\usepackage{amsmath}
\usepackage{slashed}
\usepackage{dsfont}
\def\no{\nonumber}

\newcommand{\model}{\overline{331}}

\def\eps{\varepsilon}
\def\epe{\varepsilon'/\varepsilon}

\newcommand{\tev}{\, {\rm TeV}}
\newcommand{\gev}{\, {\rm GeV}}
\newcommand{\mev}{\, {\rm MeV}}

\newcommand{\Heff}{{\cal H}_\text{ eff}}

\newcommand{\be}{\begin{equation}}
\newcommand{\ee}{\end{equation}}
\newcommand{\bea}{\begin{eqnarray}}
\newcommand{\eea}{\end{eqnarray}}

\newcommand{\bi}{\begin{itemize}}
\newcommand{\ei}{\end{itemize}}
\newcommand{\ord}{{\cal O}}

\newcommand{\vcb}{|V_{cb}|}
\newcommand{\vtd}{|V_{td}|}
\newcommand{\vub}{|V_{ub}|}
\newcommand{\vts}{|V_{ts}|}
\newcommand{\vus}{|V_{us}|}
\newcommand{\Br}{{\rm Br}}

\def\kpn{K^+\rightarrow\pi^+\nu\bar\nu}

\def\klpn{K_{L}\rightarrow\pi^0\nu\bar\nu}

\usepackage{graphicx}

\medskipamount % besser als explizite Angabe in pt; etwas Platz nach neuem Absatz

\usepackage{fancyhdr}
\advance \headheight by 3.0truept       % for 12pt mandatory...
\newlength{\textlength}
\newlength{\overlinelength}

 \def\s#1{\setbox0=\hbox{$#1$}%
   \rlap{\ifdim\wd0>.7em\kern.22\wd0\else\kern.1\wd0\fi /}#1}


\begin{document}

%%%%%%%%%% Title page
\begin{titlepage}
\begin{flushright}
%\begin{tabular}{l}
{FLAVOUR(267104)-ERC-26}\\
 {BARI-TH/12-665}
%\end{tabular}
\end{flushright}
\vskip0.7cm
\begin{center}
{\Large \bf \boldmath The Anatomy of
$Z'$ and $Z$ with Flavour Changing Neutral Currents in the Flavour Precision Era}
\vskip1.0cm
{\bf Andrzej~J.~Buras$^{a,b}$, Fulvia~De~Fazio$^{c}$ and Jennifer Girrbach$^{a,b}$
 \\[0.4 cm]}
{\small
$^a$TUM Institute for Advanced Study, Lichtenbergstr. 2a, D-85747 Garching, Germany\\
$^b$Physik Department, Technische Universit\"at M\"unchen,
James-Franck-Stra{\ss}e, \\D-85747 Garching, Germany\\
$^c$Istituto Nazionale di Fisica Nucleare, Sezione di Bari, Via Orabona 4,
I-70126 Bari, Italy}
\vskip0.71cm

%{\em Version of \today}

%\vskip0.1cm

{\large\bf Abstract\\[10pt]} \parbox[t]{\textwidth}{\small
The simplest extension of the Standard Model (SM)
that generally introduces
new sources of flavour violation and CP violation
as well as right-handed (RH) currents is the
addition of a $U(1)$ gauge symmetry to the SM gauge group. If the
corresponding heavy gauge boson $(Z')$ mediates FCNC processes in the
quark sector at tree-level, these new physics (NP) contributions imply
a pattern of deviations from SM expectations for FCNC processes that depends
only on the couplings of $Z'$ to fermions  and on its mass. This implies
stringent correlations between $\Delta F=2$ and $\Delta F=1$ observables
which  govern the landscape of the allowed parameter space
for $Z'$-models. Anticipating the Flavour Precision Era (FPE) ahead
of us we illustrate this by searching for allowed oases in this
landscape assuming significantly smaller uncertainties in CKM and hadronic
parameters than presently available. To this end we  analyze
$\Delta F=2$ observables in $K^0-\bar K^0$ and
$B^0_{s,d}-\bar B^0_{s,d}$ systems and rare $K$ and $B$ decays including
both left-handed and right-handed $Z'$-couplings to quarks in various
combinations.  We identify a number of correlations
between various flavour observables that could test and distinguish these different $Z'$ scenarios. The important
role of $b\to s \ell^+\ell^-$  and $b\to s \nu\bar\nu$ transitions in these
studies is emphasized. Imposing the existing flavour constraints, a rich
pattern of deviations from the SM expectations in $B_{s,d}$ and $K$ meson
systems emerges provided $M_{Z'}\le 3\tev$.
While for  $M_{Z'}\ge 5\tev$ $Z'$ effects in rare $B_{s,d}$ decays are found
typically below $10\%$ and hard to measure even in the FPE, $K\to\pi\nu\bar\nu$
and $K_L\to\pi^0\ell^+\ell^-$ decays  provide an important portal to scales  beyond those explored
by the LHC.
We apply our
formalism to NP scenarios with induced flavour changing neutral $Z$-couplings
to quarks. We find that in the case of $B_d$ and $K$ decays such 
$Z$-couplings still allow for sizable departures from the SM.  On the other 
hand in the $B_s$ system, constraints on $b\to s \ell^+\ell^-$ transitions 
 basically eliminate NP effects from such couplings.
}

\vfill
\end{center}
\end{titlepage}

\setcounter{footnote}{0}

\newpage

\section{Introduction}
\label{sec:1}
Elementary particle physicists are  eagerly waiting for clear signals of
New Physics (NP) from the LHC. While the recent discovery of a scalar
particle with a mass
of $125\gev$ and the unexpectedly high direct CP violation in the charm
decays could already be such signals, presently in both cases the SM
explanations of these events are possible. In the first case it should be
possible with increased statistic to answer the question whether the  new
particle observed at the LHC is the SM Higgs
boson or another one belonging to a particular  NP scenario. In the second case the situation
is less optimistic in view of hadronic uncertainties but the measurements
of other flavour observables in charm decays may tell us in due time whether
the events seen by the LHCb is NP or not.

After numerous proposals for the physics beyond the SM in the last 35 years
it is really time that we know which one if any of these proposal
is realized in nature. In particular, an exciting question is whether beyond
the SM Higgs, the first new particle to be discovered will be a new heavy
gauge boson, a new heavy fermion or a new heavy scalar. If this discovery
is to be made directly in high energy collisions then the only collider in
this decade that could achieve this goal is the LHC. But what if {\it nature}
is not nice to us and the lightest new particle has a mass of $5-10\tev$ and
will just escape a convincing detection at the LHC. While this is fortunately
only a nightmare at present
and many new particles could still be discovered by
the LHC in the coming months and years, we cannot presently exclude
the possibility that we will have to search for new particles first
indirectly. In such a case the high precision flavour dedicated experiments
will be of paramount importance. However, this will require the measurements
of very many observables and a significant reduction of hadronic uncertainties
in several of them through improved treatment of QCD effects, in particular
improved lattice calculations.

Now over the last decades significant efforts have been made by theorists
to suppress flavour changing neutral current (FCNC) processes so that they are absent at  tree-level.
In addition to the GIM mechanism \cite{Glashow:1970gm} that governs
the flavour physics in the SM, the frameworks of
constrained Minimal Flavour Violation (CMFV) \cite{Buras:2000dm,Buras:2003jf,Blanke:2006ig} and Minimal Flavour Violation at large (MFV) \cite{D'Ambrosio:2002ex} were very instrumental in suppressing new flavour and CP-violating
phenomena below the present experimental bounds even in the presence of new
particles with masses of a few hundreds GeV. Selected reviews with
comprehensive list of references can be found in
\cite{Isidori:2010kg,Buras:2012ts}.

However, if the scale of NP is shifted to $5-10\tev$ or even higher energy
scales this kind of suppression is less important as FCNC processes are
then naturally suppressed by the large scales of heavy particles mediating
these phenomena. In fact
 while loop diagrams, like
penguin diagrams of various sorts and box diagrams dominated the physics of
FCNC processes
in the last thirty years both within the SM and several of its extensions,
we should hope  at first sight that
in the case of new particles
with masses above $10\tev$ this role will be taken over by tree-level diagrams.
The reason is simple. Internal particles with such large masses, if hidden
in loop diagrams, will quite generally imply very small effects in FCNC processes that will
be very difficult to measure. On the other hand tree diagrams could still
in principle
provide a large window to these very short distance scales.

 We will demonstrate in the present paper that
in the simplest extensions
of the SM which contain just a new heavy neutral gauge boson ($Z'$) with flavour-changing quark couplings, the correlations between $\Delta F=2$ and $\Delta F=1$ observables in the
quark sector, in the absence of new heavy fermions and scalars have a
significant impact on this optimistic expectations.
In
fact we find that
these correlations preclude NP effects
above $10\%$ in rare $B$ decay branching ratios and CP-asymmetries if
$M_{Z'}\ge 5\tev$. Much larger effects are still possible in rare $K$ decays.

The reason is simple. A tree-level $Z'$ contribution to $\Delta F=2$
observables depends quadratically on $\Delta_{L,R}^{i,j}(Z')/M_{Z'}$, where
   $\Delta_{L,R}^{i,j}(Z')$ are flavour-violating couplings with $i,j$ denoting
quark flavours. For any high value of $M_{Z'}$, even beyond the reach of the
LHC, it is possible to find couplings  $\Delta_{L,R}^{i,j}(Z')$  which are not
only  consistent with the existing data but    can even remove certain
tensions found within the SM. The larger $M_{Z'}$, the larger
$\Delta_{L,R}^{i,j}(Z')$  are allowed: $\Delta_{L,R}^{i,j}(Z')\approx a_{ij}M_{Z'}$
with $a_{ij}$ sufficiently small to agree with $\Delta F=2$ data. Once
$\Delta_{L,R}^{i,j}(Z')$ are fixed in this manner, they can be used to
predict $Z'$ effects in $\Delta F=1$ observables. However here
NP contributions to the amplitudes are proportional to
$\Delta_{L,R}^{i,j}(Z')/M^2_{Z'}$ and with the couplings proportional to $M_{Z'}$, $Z'$ contributions to $\Delta F=1$ observables decrease with increasing
$M_{Z'}$.

Our analysis demonstrates that for $1\tev \le M_{Z'}\le 3\tev$, still in the
reach of the LHC, indirect $Z'$ effects can be well
tested by means of
rare $K$ and $B$ decays. For such values of $ M_{Z'}$ effects up to $50\%$
at the level of the branching ratios and measurable effects in CP-asymmetries
are possible for $B_{s,d}$ meson systems. However for  $M_{Z'}\ge 5\tev$, this begins to be very difficult
even in the FPE as NP effects in rare and CP-violating $B$ decays
turn out to be typically below $10\%$. Significantly larger effects
are still allowed in rare $K$ decays.

On the other hand it is evident from this discussion that flavour-violating
$Z$-couplings, that arise in various extensions of the SM, could in the
presence of much lower value of $M_Z$ provide
clear NP effects in rare $K$ and $B$ decays even if NP generating these
couplings is outside the reach of the LHC. In this manner
flavour-violating $Z$ couplings, { similarly to $Z'$ couplings in rare $K$ 
decays,}
could turn out to be an important portal to short distance
scales which cannot be explored
by the LHC. We will demonstrate that this is still the case for rare $B_d$
and $K$ decays but not any longer for $B_s$ decays and related CP asymmetries.

In this spirit, we will first ask in the present paper  the following question:
\begin{itemize}
\item What can be learned about NP from
precise measurements of flavour observables to be performed in this decade
if the lightest messenger of NP is a heavy $Z'$ gauge boson with arbitrary
couplings to quarks and arbitrary mass? In particular we will ask the
question whether it is possible to determine all these couplings entirely with the help of quark-flavour violating observables
for $M_{Z'}$ in the reach
of the LHC.
To this end we will assume that the flavour diagonal couplings of $Z'$ to
leptons have
been determined in pure leptonic processes.
\end{itemize}

While there is a very rich literature on FCNC processes mediated by a
$Z'(Z)$ gauge boson and several extensive analyses  have been
presented on various occasions \footnote{It is not possible to refer to all these papers. Selected analyses can be found in
\cite{Langacker:2000ju,Buchalla:2000sk,delAguila:2000rc,Barger:2003hg,Barger:2004qc,Chiang:2006we,Baek:2006bv,Cheung:2006tm,He:2006bk,Barger:2009qs,delAguila:2011yd,Li:2012xc,Chang:2010zy,Chang:2009tx}. See
also the review in \cite{Langacker:2008yv}.},
to our knowledge this specific question
has not been addressed so far. After having positively answered this question
we will ask the second question:
\begin{itemize}
\item
What are the correlations between various flavour observables in this
simple framework and how do they compare with the stringent correlations
implied by the simplest BSM frameworks on the market,
the class of models with
 constrained Minimal Flavour Violation (CMFV) \cite{Buras:2000dm,Buras:2003jf,Blanke:2006ig} and the
models with $U(2)^3$ flavour symmetry
\cite{Barbieri:2011ci,Barbieri:2011fc,Barbieri:2012uh,Barbieri:2012bh,Crivellin:2011fb,Crivellin:2011sj,Crivellin:2008mq,Buras:2012sd}?
\end{itemize}

The simple model analyzed  here can be considered as a part of a bigger
theory as already analyzed in numerous papers in the literature. Moreover
its simplicity provides an analytic insight into the departure
from SM expectations in flavour physics and the role of right-handed (RH) currents.
In fact
certain lessons gained from the involved studies of the LHT model 
\cite{Blanke:2006eb,Blanke:2009am} and
RS scenario with custodial protection (RSc) \cite{Blanke:2008yr} as
summarized in \cite{Blanke:2009pq}
will be seen here in a much
simpler setting. In particular our analysis in the $K$ system, where we
investigate the correlation between $\varepsilon_K$ and the $K\to\pi\nu\bar\nu$
decays, can be considered as an explicit dynamical example for the findings of
\cite{Blanke:2009pq}.

While the violation of stringent relations of CMFV is evident in this
framework due to new sources of flavour and CP violation, it is of interest
to impose  $U(2)^3$ symmetry on $Z'$ couplings and study its phenomenological
implications. Such a study is more transparent than in more
complicated
models in which loop diagrams with heavy fermions and scalars
accompanied by many free parameters dominate the NP contributions to
FCNC processes.

The analysis of $Z'$ flavour physics presented here can be considered
as a generalization of our recent paper \cite{Buras:2012xx} in which we have analyzed
in detail the pattern of flavour violating $Z'$ tree-level contributions in
a specific $331$ model (the $\model$ model). The generalization in question
is three-fold:
\begin{itemize}
\item
First of all we consider general structure of $Z'$ couplings to SM quarks so
that at the fundamental level there are no correlations between flavour
violation from NP in $K$, $B_d$ and $B_s$ meson systems. While certain
correlations between them could be generated once the constraints from experimental
data are imposed, significant NP effects in $\varepsilon_K$ and in particular 
in rare $K$
decays are now possible, while this was not the case in the $\model$ model.
\item
Also a new important feature of NP contributions in the present paper
is the presence of
flavour-violating right-handed (RH) $Z'$ couplings to SM quarks, which has profound
implications for correlations between $\Delta F=2$ and $\Delta F=1$
observables
identified in scenarios with only left-handed (LH) couplings.
Also correlations  between rare decays with $\nu\bar\nu$ and $\mu^+\mu^-$
in the final state can be modified in a profound manner.
\item
While in the $\model$ model the flavour diagonal couplings of $Z'$ to neutrinos and muons were fixed and smaller than the corresponding ordinary $Z$ couplings,
in a general case considered here they could be larger than the latter,
enhancing thereby the branching ratios for rare leptonic and semi-leptonic
decays for fixed quark couplings.
\end{itemize}

 As advertised above, our anatomy of $Z'$ scenarios will lead us to the
conclusion that the correlations between various flavour observables
will test this type of NP in the FPE provided $M_{Z'}\le 3\tev$. For
$M_{Z'}\ge 5\tev$ this will be very difficult, except for rare $K$
decays and in the second part
of our paper we will apply our formalism to the case of flavour-violating $Z$
couplings. Here the effects in rare $B_d$ decays and in particular
$K$ decays can be much larger than
those allowed in the case of $Z'$ for $M_{Z'}\ge 1\tev$, but in the
$B_s$ system significant NP effects from flavour violating
 $Z$ coupling are already ruled out by present constraints from $b\to s\mu^+\mu^-$ transitions.  Similar conclusions have been
reached in 
\cite{Altmannshofer:2012ir,Beaujean:2012uj} in a more general context.

{For readers interested mainly in our results and less in the formalism 
     presented in subsequent sections we have made  
     an overview of all correlations and anticorrelations
      found by us and of the related figures 
      in Tables~\ref{tab:corrB} and~\ref{tab:corrK}.
The comments in the last column of this table indicate 
        the relevance of a given correlation or anticorrelation.}

 Our paper is organized as follows. In Section~\ref{sec:2} we describe
 our strategy by listing processes to be considered. Our analysis
 will only involve processes which are theoretically clean and have
 simple structure. Here we will also introduce a number of
 different scenarios for the $Z'$ couplings to quarks thereby
reducing the number
 of free parameters.
In Section~\ref{sec:3} we will first present a compendium of formulae for
master functions that govern FCNC processes with $Z'$ contributions
taken into account. Subsequently we present formulae
for flavour observables in $\Delta F=2$ transitions including for
the first time NLO QCD corrections to tree-level $Z'$ contributions.
Finally formulae for rare $K$
and $B$ decays
considered by us are collected. In
Section~\ref{sec:4B} we calculate $Z'$ contributions to the
$B\to X_s\gamma$ decay improving on the calculation of QCD corrections present
in the $Z'$-literature by
using the general formulae of \cite{Buras:2011zb}. In Section~\ref{sec:3a}
we present a general qualitative view on NP contributions to flavour
observables in four scenarios for the $Z'$ couplings.
In Section~\ref{sec:4} we present our strategy for the numerical analysis
and in Section~\ref{sec:Excursion} we execute our strategy for the determination of $Z'$
couplings and discuss several scenarios of its couplings
in question, identifying stringent correlations between various observables.
In Section~\ref{sec:U(2)} we investigate what the imposition of the $U(2)^3$ flavour
symmetry on $\Delta_L^{i,j}(Z')$ couplings would imply.
In Section~\ref{sec:ZSM} we apply our formalism to the SM $Z$ boson for
which the mass $M_Z$ and the diagonal lepton couplings are known.
 A summary of our main results and a brief outlook for the future
 are given in  Section~\ref{sec:5}.

\section{Strategy}
\label{sec:2}
\begin{figure}[!tb]
\includegraphics[width = 0.6\textwidth]{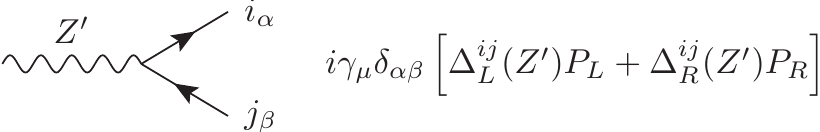}
\caption{\it Feynman rule for the coupling of a
colourless neutral gauge boson $Z'$ to quarks, where $i,\,j$ denote different
quark flavours and $\alpha,\,\beta$ the colours. $P_{L,R}=(1\mp\gamma_5)/2$.}\label{neutralZ}
\end{figure}

Our paper is dominated
by tree-level contributions to FCNC processes mediated
by a heavy neutral gauge boson $Z'$.
These contributions are governed by the couplings $\Delta_{L,R}^{ij}(Z')$
to quarks and the corresponding Feynman rule has
been shown in Fig~\ref{neutralZ}.
Here $(i,j)$ denote quark flavours. As we
will see in addition to a general form of these couplings
it will be instructive to consider the following four scenarios for them
keeping the pair $(i,j)$ fixed:
\begin{enumerate}
\item
Left-handed Scenario (LHS) with complex $\Delta_L^{bq}\not=0$  and $\Delta_R^{bq}=0$,
\item
Right-handed Scenario (RHS) with complex $\Delta_R^{bq}\not=0$  and $\Delta_L^{bq}=0$,
\item
Left-Right symmetric Scenario (LRS) with complex
$\Delta_L^{bq}=\Delta_R^{bq}\not=0$,
\item
Left-Right asymmetric Scenario (ALRS) with complex
$\Delta_L^{bq}=-\Delta_R^{bq}\not=0$,
\end{enumerate}
with analogous scenarios for the pair $(s,d)$.
We will see that these simple scenarios will give us a profound insight
into the flavour structure of models in which NP is dominated by left-handed
currents or right-handed currents or left-handed and right-handed currents
of the same size. In particular the last two scenarios will exhibit a very
clear distinction between $K\to\pi \nu\bar\nu$ decays and
$B_{s,d}\to \mu^+\mu^-$ which are governed by $V$ and $A$ couplings,
respectively.
Moreover we will consider a scenario with underlying
flavour $U(2)^3$ symmetry
which will imply relations between $\Delta_L^{bd}$  and $\Delta_L^{bs}$
couplings and interesting phenomenological consequences.

The idea of looking at the first  three NP scenarios is not new and has been in
particular motivated by a detailed study of supersymmetric flavour models
with NP dominated by LH currents, RH currents or equal amount of LH and RH
currents \cite{Altmannshofer:2009ne}\footnote{Similar scenarios have been
considered subsequently in \cite{Altmannshofer:2011gn,Altmannshofer:2012ir}}. Moreover, it has been found in several studies of non-supersymmetric
 frameworks
like LHT model \cite{Blanke:2006eb} or Randall-Sundrum scenario with custodial protection (RSc)
\cite{Blanke:2008yr}
that models with the dominance of LH or RH currents exhibit quite different
patterns of flavour violation. Our analysis will demonstrate it in
a transparent manner.

Let us then outline our strategy for the determination of $Z^\prime$
couplings to quarks and for finding correlations between flavour observables in
the context of the simple scenarios listed above.
Our strategy will only be fully effective in the second half of this
decade, when hadronic uncertainties will be reduced and the data on
various observables significantly improved. It involves ten steps including
a number of working assumptions:

{\bf Step 1:}

Determination of CKM parameters by means of tree-level decays
and of the necessary non-perturbative parameters by means of lattice calculations. This step will provide the results for all observables considered below
within the SM as well as all non-perturbative parameters entering the NP
contributions. As $\vub$ is presently poorly known, it will be interesting in
the spirit of our recent papers \cite{Buras:2012sd,Blanke:2011ry}
to investigate how the outcome of this step
depends on the value of $\vub$ with direct implications for the necessary
size of NP contributions which will be different in different observables.

{\bf Step 2:}

We will assume that the ratios
\be\label{leptonic}
\frac{\Delta_A^{\mu\bar\mu}(Z')}{ M_{Z'}}, \qquad \frac{\Delta_L^{\nu\bar\nu}(Z')}{ M_{Z'}}
\ee
have been determined in pure leptonic processes. We will further
assume that these
ratios are real but could have both signs. In principle these ratios
can be determined up to the sign from quark flavour violating processes
considered below but
their knowledge increases predictive power of our analysis. In particular
the knowledge of their signs allows to remove certain discrete ambiguities
and is crucial for the distinction between LHS and RHS scenarios in
$B_{s,d}\to\mu^+\mu^-$ decays.

{\bf Step 3:}

Here we will consider the $B_s^0$ system and the observables
 \be\label{Step3}
\Delta M_s, \quad S_{\psi \phi}, \quad \mathcal{B}(B_s\to\mu^+\mu^-),\quad
S^s_{\mu^+\mu^-},
\ee
where $S^s_{\mu^+\mu^-}$ measures CP violation in $B_s\to\mu^+\mu^-$ decay
 \cite{deBruyn:2012wj,deBruyn:2012wk}. Explicit expressions for these observables in terms of the
relevant couplings can be found in  Section~\ref{sec:3}.

Concentrating in this step on the LHS scenario, NP contributions to these
three observables are fully described by
\be\label{Step3a}
\frac{\Delta_L^{bs}(Z')}{M_{Z'}}=-\frac{\tilde s_{23}}{M_{Z'}}
e^{-i\delta_{23}},\quad \frac{\Delta_A^{\mu\bar\mu}(Z')}{ M_{Z'}},
\ee
with the second ratio known from Step 2. Here $\tilde s_{23}\ge 0$
and it is found to be below unity but it does not represent any mixing
parameter as in \cite{Buras:2012xx}. The minus sign is introduced
to cancel the minus sign in $V_{ts}$ in the phenomenological formulae listed
in the
next section.

Thus we have four observables
to our disposal and two parameters in the quark sector to determine. This
allows to remove certain discrete ambiguities, determine all parameters
uniquely and predict correlations between these four observables  that
are characteristic for this scenario.

{\bf Step 4:}

Repeating this exercise in the $B_d^0$ system we have to
our disposal
 \be\label{Step4}
\Delta M_d, \quad S_{\psi K_S}, \quad \mathcal{B}(B_d\to\mu^+\mu^-), \quad
S^d_{\mu^+\mu^-}
\ee
 Explicit expressions for these observables in terms of the
relevant couplings can be found in  Section~\ref{sec:3}.

Now NP contributions  to these
three observables are fully described by
\be\label{Step4a}
\frac{\Delta_L^{bd}(Z')}{M_{Z'}}=\frac{\tilde s_{13}}{M_{Z'}}
e^{-i\delta_{13}},\quad \frac{\Delta_A^{\mu\bar\mu}(Z')}{ M_{Z'}},
\ee
with the last one known from Step 2. Again we can determine all the couplings uniquely to be used in the steps
below. Our notations and sign conventions are as in Step 3 with
 $\tilde s_{13}\ge 0$  but no minus sign as $V_{td}$ has no such sign.

{\bf Step 5:}

Moving to the $K$ system we have to our disposal
\be\label{Step5}
\varepsilon_K, \quad \kpn, \quad \klpn,\quad K_L\to\pi^0\ell^+\ell^-,\quad K_L\to\mu^+\mu^-,
\ee
where in view of hadronic uncertainties the last decay on this list will
only be used to make sure that the existing rough bound on its branching ratio
is satisfied.  In the present paper we do not study the ratio $\epe$, which
is rather accurately measured but subject to much larger hadronic uncertainties
than observables listed in (\ref{Step5}).
Explicit expressions for the  observables in the $K$ system
in terms of the
relevant couplings can be found in  Section~\ref{sec:3}.

Now NP contributions  to these
four observables are fully described by
\be\label{Step5a}
\frac{\Delta_L^{sd}(Z')}{M_{Z'}}=-\frac{\tilde s_{12}}{M_{Z'}}
e^{-i\delta_{12}},\quad \frac{\Delta_L^{\nu\bar\nu}(Z')}{ M_{Z'}}, \quad
\frac{\Delta_A^{\mu\bar\mu}(Z')}{ M_{Z'}}
\ee
where we assumed the couplings to neutrinos to be left-handed and real. The
ratios involving leptonic couplings are known already from Step 2.
Consequently,
we can determine
all couplings involved by using the data on the observables in (\ref{Step5}).
Moreover we identify certain correlations characteristic for LHS scenario.
$\tilde s_{12}\ge 0$ and the minus sign is chosen to cancel the one of
$V_{ts}$.

{\bf Step 6:}
As all parameters of LHS scenario has been fixed in the first five steps we
are in the position to make predictions for the following processes
\be\label{Step6b}
B\to X_s\ell^+\ell^-, \quad B\to K\ell^+\ell^-, \quad B\to K^*\ell^+\ell^-
\ee
\be\label{Step6a}
B\to K \nu \bar \nu, \quad {B}\to K^* \nu \bar \nu,
\quad {B}\to X_s \nu \bar \nu,
\ee
\be\label{Step5b}
B\to X_s\gamma, \quad B\to K^*\gamma
\ee
and test whether they provide additional constraints on the couplings.

{\bf Step 7:}

We repeat Steps 3-6 for the case of RHS. We will see that in
               view of the change of the sign of NP contribution to
               $B_{s,d}\to\mu^+\mu^-$ and $ K_L\to\mu^+\mu^-$ decays
                the structure of the correlations between various observables
                will distinguish this scenario from the LHS one. Yet,
as we will find out, by going from LHS to RHS scenario we can keep
results of Steps 3-5 unchanged by interchanging
simultaneously two {\it big oases} in the parameter space  that we
encountered already in our study of the $\model$ model.
 This LH-RH invariance present in Steps 3-5 can be broken by the
$b\to s \ell^+\ell^-$ and
$b\to s\nu\bar\nu$ transitions listed in  (\ref{Step6b}) and (\ref{Step6a}),
respectively. They
 will allow us very clearly to
distinguish the physics of RH currents from LH ones.
As only
RH couplings are present in the NP contributions in this scenario we can use the
parametrization of these couplings as in (\ref{Step3a}), (\ref{Step4a}) and
(\ref{Step5a}) keeping in mind that now RH couplings are involved.

{\bf Step 8:}

We repeat Steps 3-6 for the case of LRS. Here the new feature
               is the vanishing of NP contributions
               to $B_{s,d}\to\mu^+\mu^-$ and $ K_L\to\mu^+\mu^-$ decays and
                rather sizable NP contributions to $\Delta F=2$ observables
               due to the presence of LR operators. As the LH and RH couplings
               are equal we can again  use the
parametrization of these couplings as in (\ref{Step3a}), (\ref{Step4a}) and
(\ref{Step5a}) but their values will change due to different constraints
from $\Delta F=2$ transitions. Also in this step the
$b\to s \ell^+\ell^-$ and $b\to s\nu\bar\nu$
transitions will play very important role.

{\bf Step 9:}

We repeat Steps 3-6 for the case of ALRS. Here the new feature
               is the vanishing of NP contributions
               to $\kpn$ and $\klpn$ decays, while
               $B_{s,d}\to\mu^+\mu^-$, including $S^{d,s}_{\mu^+\mu^-}$
               CP asymmetries  and again the $b\to s \ell^+\ell^-$ and
$ b\to s\nu\bar\nu$ transitions
                will exhibit their strength in testing the theory in
                 a different environment:
                rather sizable NP contributions to $\Delta F=2$ observables
               due to the presence of LR operators. As the LH and RH couplings
                differ only by a sign we can again  use the
parametrization of these couplings as in (\ref{Step3a}), (\ref{Step4a}) and
(\ref{Step5a}) but their values will change due to different constraints
from $\Delta F=2$ transitions.

{\bf Step 10:}

One can consider next the case of simultaneous LH and RH
               couplings that are unrelated to each other. This step is more challenging as one has more free parameters and in order to reach clear cut 
conclusions one would need a concrete model for $Z'$ couplings or a very 
involved numerical analysis \cite{Altmannshofer:2011gn,Altmannshofer:2012ir}.
 We will therefore leave out this step from our paper.

Once this analysis of $Z'$ contributions is completed it will be straightforward to apply it to the case of the SM $Z$ boson with flavour
violating couplings.

We should remark that we have left from this analysis $\epe$. This ratio 
is important for the tests of $Z'(Z)$ FCNC
scenarios as it is very sensitive to any NP contribution \cite{Buras:1998ed,Buras:1999da,Blanke:2007wr}.
However, due to significant  hadronic uncertainties
it is  less suitable for
the determination of the $Z'(Z)$  FCNC couplings than the decays used by us when the
latter will be precisely measured. On the other hand having these couplings
one can make predictions for $\epe$ and study correlations.

\boldmath
 \section{Compendium for the $Z'$ Contributions}
\unboldmath
 \label{sec:3}
\subsection{Parametrization}
First it will be useful to introduce a useful parametrization of $Z'$ contributions by generalizing the master functions known from CMFV models and the LHT
model to include in addition to left-handed currents also right-handed currents.
 In the case of the RSc model this has already been done in \cite{Blanke:2008yr} 
but our parametrization below is a bit different than the one in the latter 
paper.
For our purposes it will be sufficient to consider the following functions:
\begin{itemize}
\item
For $\Delta F=2$ processes
\be\label{FS}
S(K), \qquad  S(B_d), \qquad S(B_s),
\ee
where we will include in the definitions of these functions the contributions
of operators with $LL$, $RR$ and $LR$ Dirac structures.
\item
For decays with $\nu\bar\nu$ in the final state
\be\label{FX}
X_{\rm L,R}(K), \qquad X_{\rm L,R}(B_d), \qquad   X_{\rm L,R}(B_s).
\ee
\item
For decays with $\mu\bar\mu$ in the final state
\be\label{FY}
Y_{\rm A}(K), \qquad Y_{\rm A}(B_d), \qquad   Y_{\rm A}(B_s).
\ee
\end{itemize}
All these functions in contrast to the SM and more generally CMFV models
depend on the meson considered and moreover are complex valued.

In the SM the corresponding {\it flavour universal real valued} functions are
given as follows
($x_t=m_t^2/M_W^2$):
\begin{align}
S_0(x_t)  = \frac{4x_t - 11 x_t^2 + x_t^3}{4(1-x_t)^2}-\frac{3 x_t^2\log x_t}{2
(1-x_t)^3}~,
\end{align}

\begin{equation}\label{X0}
X_0(x_t)={\frac{x_t}{8}}\;\left[{\frac{x_t+2}{x_t-1}}
+ {\frac{3 x_t-6}{(x_t -1)^2}}\; \ln x_t\right],
\end{equation}

\be\label{YSM}
Y_0(x_t)=\frac{x_t}{8}\left(\frac{x_t-4}{x_t-1} + \frac{3 x_t \log x_t}{(x_t-1)^2}\right).
\ee
In other CMFV models they take different values still keeping flavour universality
and
being real valued. This implies very stringent relations between various
observables in three meson system in question
which have been reviewed in \cite{Buras:2003jf}. It is evident
that in the presence of $Z'$ tree-level contributions the
breakdown of flavour universality and also the presence of new complex phases
implies the violation of these relations. Generalizing the three SM functions
to twelve functions listed in (\ref{FS})--(\ref{FY}), allows to describe
these new effects in a transparent manner. In what follows
we will list explicit expressions for these functions. Subsequently we will show
how they enter the branching ratios for various decays.

The derivation of the formulae listed below
is so simple that we will not present it here.
From the normalization of the corrections from $Z'$ to the master functions in
question
and the formulae for observables given subsequently, it will be clear how
these functions have been defined in the corresponding effective Hamiltonians.
In any case, the compendium given below is self-contained as far as numerical
analysis is concerned.

\boldmath
\subsection{Master Functions Including $Z'$ Contributions}
\unboldmath
 Calculating the contributions of $Z'$ to various decays it is straightforward
 to write down the expressions for the master functions in terms of the
 couplings defined in Fig.~\ref{neutralZ}.
\boldmath
\subsubsection{$\Delta F=2$ Master Functions}
\unboldmath
We define the relevant CKM factors
\be
\lambda_i^{(K)} =V_{is}^*V_{id},\qquad
\lambda_t^{(d)} =V_{tb}^*V_{td}, \qquad \lambda_t^{(s)} =V_{tb}^*V_{ts},
\ee
and introduce
\be\label{gsm}
g_{\text{SM}}^2=4\frac{G_F}{\sqrt 2}\frac{\alpha}{2\pi\sin^2\theta_W}=1.78137\times 10^{-7} \gev^{-2}\,.
\ee

The $\Delta F=2$ master functions for $M=K,B_q$ are given as follows
\begin{equation}\label{Seff}
S(M)=S_0(x_t)+\Delta S(M)\equiv|S(M)|e^{i\theta_S^M}
\end{equation}
with $\Delta S(M)$ receiving contributions from various operators so that
it is useful to write
\begin{equation}\label{sum}
\Delta S(M)=[\Delta S(M)]_{\rm VLL}+[\Delta S(M)]_{\rm VRR}+[\Delta S(M)]_{\rm LR}.
\end{equation}

The contributing operators
are defined for the $K$ system as follows
\cite{Buras:2001ra,Buras:2012fs}
\begin{subequations}\label{equ:operatorsZ}
\bea
{Q}_1^\text{VLL}&=&\left(\bar s\gamma_\mu P_L d\right)\left(\bar s\gamma^\mu P_L d\right)\,,\\
{Q}_1^\text{VRR}&=&\left(\bar s\gamma_\mu P_R d\right)\left(\bar s\gamma^\mu P_R d\right)\,,\\
{Q}_1^\text{LR}&=&\left(\bar s\gamma_\mu P_L d\right)\left(\bar s\gamma^\mu P_R d\right)\,,\\
{Q}_2^\text{LR}&=&\left(\bar s P_L d\right)\left(\bar s P_R d\right)\,.
\eea
\end{subequations}
with analogous expressions for $B_{s,d}$ systems. For instance in the 
$B_s$ system ${Q}_1^\text{VLL}=\left(\bar b\gamma_\mu P_L s\right)\left(\bar b\gamma^\mu P_L s\right)$.
Here we suppressed colour indices as they are summed up in each factor. For instance $\bar s\gamma_\mu P_L d$ stands for $\bar s_\alpha\gamma_\mu P_L d_\alpha$ and similarly for other factors.

$[\Delta S(M)]_{\rm VLL}$ and $[\Delta S(M)]_{\rm VRR}$ can be obtained
directly from our previous paper \cite{Buras:2012xx}:
\be\label{Zprime1}
[\Delta S(B_q)]_{\rm VLL}=
\left[\frac{\Delta_L^{bq}(Z^\prime)}{\lambda_t^{(q)}}\right]^2
\frac{4\tilde r}{M^2_{Z^\prime}g_{\text{SM}}^2}, \qquad
[\Delta S(K)]_{\rm VLL}=
\left[\frac{\Delta_L^{sd}(Z^\prime)}{\lambda_t^{(K)}}\right]^2
\frac{4\tilde r}{M^2_{Z^\prime}g_{\text{SM}}^2},
\ee
where $\tilde r=0.985$ for $M_{Z'}=1\tev$.  $[\Delta S(M)]_{\rm VRR}$  is then
found from the formula above by simply replacing L by R. For the
case of tree-level $Z$ exchanges $\tilde r=1.068$.

In order to calculate the LR contributions we introduce
quantities familiar from SM expressions for mixing amplitudes
\be
T(B_q)=\frac{G_F^2}{12\pi^2}F_{B_q}^2\hat B_{B_q}m_{B_q}M_{W}^2
\left(\lambda_t^{(q)}\right)^2\eta_B,
\label{eq:3.6}
\ee
\be
T(K)=\frac{G_F^2}{12\pi^2}F_{K}^2\hat B_{K}m_{K}M_{W}^2
\left(\lambda_t^{(K)}\right)^2\eta_2
\label{eq:3.7},
\ee
where $\eta_i$ are QCD corrections and $\hat B_i$ known SM non-perturbative
factors.

Then
\be\label{DSK}
T(K)[\Delta S(K)]_{\rm LR}=
 \frac{\Delta_L^{sd}(Z')\Delta_R^{sd}(Z')}{
 M_{Z'}^2} \left [ C_1^\text{LR}(\mu_{Z'}) \langle Q_1^\text{LR}(\mu_{Z'},K)\rangle +
 C_2^\text{LR}(\mu_{Z'}) \langle Q_2^\text{LR}(\mu_{Z'},K)\rangle \right]\,.
\ee

Including NLO QCD corrections \cite{Buras:2012fs} the Wilson coefficients of
LR operators are given by
\begin{align}\label{equ:WilsonZ}
\begin{split}
 C_1^\text{LR}(\mu_{Z'})
& =1+\frac{\alpha_s}{4\pi}
\left(-\log\frac{M_{Z'}^2}{\mu_{Z'}^2}-\frac{1}{6}\right)\,,\end{split}\\
C_2^\text{LR}(\mu_{Z'}) &=\frac{\alpha_s}{4\pi}\left(-6\log\frac{M_{Z'}^2}{\mu_{Z'}^2}-1\right)\,.
\end{align}
Next
\be
\langle Q^a_i(\mu_{Z'},K)\rangle \equiv \frac{m_K F_K^2}{3} P^a_i(\mu_{Z'},K)
\ee
are the matrix elements of operators evaluated at the matching scale
$\mu_{Z'}=\ord(M_{Z'})$
and
 $P^a_i$ are  the coefficients introduced in \cite{Buras:2001ra}.
The $\mu_{Z'}$ dependence
of $C^a_i(\mu_{Z'})$ cancels the one of $P^a_i(\mu_{Z'})$ so that $S(K)$ does not depend  on $\mu_{Z'}$.

Similarly for $B_q$ systems we have
\be\label{DSBq}
T(B_q)[\Delta S(B_q)]_{\rm LR}=
 \frac{\Delta_L^{bq}(Z')\Delta_R^{bq}(Z')}{
 M_{Z'}^2} \left [ C_1^\text{LR}(\mu_{Z'}) \langle Q_1^\text{LR}(\mu_{Z'},B_q)\rangle +
 C_2^\text{LR}(\mu_{Z'}) \langle Q_2^\text{LR}(\mu_{Z'},B_q)\rangle \right]\,,
\ee
where the Wilson coefficients   $C^a_i(\mu_{Z'})$  are as in the $K$ system
and the matrix elements are given by
\be
\langle Q^a_i(\mu_{Z'},B_q)\rangle \equiv \frac{m_{B_q} F_{B_q}^2}{3} P^a_i(\mu_{Z'},B_q).
\ee

Finally, we collect in Table~\ref{tab:Qi}
central values of  $\langle Q^a_i(\mu_{Z'})\rangle$. They are given in the
$\overline{\text{MS}}$-NDR scheme and are based on lattice calculations
in \cite{Boyle:2012qb,Bertone:2012cu} for $K^0-\bar K^0$ system and  in
\cite{Bouchard:2011xj} for
$B_{d,s}^0-\bar B^0_{d,s}$ systems. For the $K^0-\bar K^0$ system we have just
used the average of the results in \cite{Boyle:2012qb,Bertone:2012cu} that
are consistent with each other.
As the values of the relevant $B_i$ parameters in these papers have been
evaluated at $\mu=3\gev$ and $4.2\gev$, respectively, we have used the
formulae in  \cite{Buras:2001ra} to obtain the values of the matrix
elements in question at $\mu_{Z'}$. For simplicity we choose this scale to
be $M_{Z'}$ but any scale of this order would give the same results for
the physical quantities up to NNLO QCD corrections that are negligible
at these high scales.  The renormalization scheme dependence of the 
matrix elements is canceled by the one of the Wilson coefficients.

In the case of tree-level $Z$-exchanges we evaluate the matrix elements
at $m_t(m_t)$ as the inclusion of NLO QCD corrections allows us to choose
any scale of $\ord(M_Z)$ without changing physical results. Then
in the formulae above
one should replace $M_{Z'}$ by $M_Z$ and $\mu_{Z'}$ by $m_t(m_t)$. The values
of hadronic matrix elements at $m_t(m_t)$ in the
$\overline{\text{MS}}$-NDR scheme
are given in Table~\ref{tab:Qi}.\footnote{We thank Robert Ziegler for checking 
the results in this table.}

\begin{table}[!ht]
{\renewcommand{\arraystretch}{1.3}
\begin{center}
\begin{tabular}{|c||c|c|c|c|}
\hline
&
$\langle Q_1^\text{LR}(M_{Z'})\rangle$& $\langle Q_2^\text{LR}(M_{Z'})\rangle$&
$\langle Q_1^\text{LR}(m_t)\rangle$& $\langle Q_2^\text{LR}(m_t)\rangle$
\\
\hline
\hline
$K^0$-$\bar K^0$ &  $-0.14$ &  $0.22$ & -0.11 & 0.18  \\
\hline
$B_d^0$-$\bar B_d^0$&  $-0.25$ & $0.34$ & -0.21 & 0.27 \\
\hline
$B_s^0$-$\bar B_s^0$&   $-0.37$ & $0.51$ & -0.30 & 0.40\\
\hline
\end{tabular}
\end{center}}
\caption{\it Hadronic matrix elements $\langle Q_i^a\rangle$  in units of GeV$^3$ at $M_{Z'}=1\tev$  and at $m_t(m_t)$
\label{tab:Qi}}
\end{table}

\boldmath
\subsubsection{$\Delta F=1$ Master Functions}
\unboldmath
We find
\be\label{XLK}
X_{\rm L}(K)=\eta_X X_0(x_t)+\frac{\Delta_L^{\nu\bar\nu}(Z')}{g^2_{\rm SM}M_{Z'}^2}
                                       \frac{\Delta_L^{sd}(Z')}{V_{ts}^* V_{td}},
\ee
\be\label{XRK}
X_{\rm R}(K)=\frac{\Delta_L^{\nu\bar\nu}(Z')}{g^2_{\rm SM}M_{Z'}^2}
                                       \frac{\Delta_R^{sd}(Z')}{V_{ts}^* V_{td}},
\ee

\be\label{XLB}
 X_{\rm L}(B_q)=\eta_X X_0(x_t)+\left[\frac{\Delta_{L}^{\nu\nu}(Z')}{M_{Z'}^2g^2_{\rm SM}}\right]
\frac{\Delta_{L}^{qb}(Z')}{ V_{tq}^\ast V_{tb}},
\ee
\be\label{XRB}
 X_{\rm R}(B_q)=\left[\frac{\Delta_{L}^{\nu\nu}(Z')}{M_{Z'}^2g^2_{\rm SM}}\right]
\frac{\Delta_{R}^{qb}(Z')}{ V_{tq}^\ast V_{tb}},
\ee

\be\label{YAK}
Y_{\rm A}(K)= \eta _Y Y_0(x_t)
+\frac{\left[\Delta_A^{\mu\bar\mu}(Z')\right]}{M_{Z'}^2g_\text{SM}^2}
\left[\frac{\Delta_L^{sd}(Z')-\Delta_R^{sd}(Z')}{V_{ts}^\star V_{td}}\right]\,
\equiv |Y_A(K)|e^{i\theta_Y^K},
\ee

\be\label{YAB}
Y_{\rm A}(B_q)= \eta_Y Y_0(x_t)
+\frac{\left[\Delta_A^{\mu\bar\mu}(Z')\right]}{M_{Z'}^2g_\text{SM}^2}
\left[\frac{\Delta_L^{qb}(Z')-\Delta_R^{qb}(Z')}{V_{tq}^\star V_{tb}}\right]\,
\equiv |Y_A(B_q)|e^{i\theta_Y^{B_q}}.
\ee

Here $\eta_{X,Y}$ are QCD factors which for $m_t=m_t(m_t)$ are close to unity
\cite{Buchalla:1998ba,Misiak:1999yg}.
\be
\eta_X=0.994, \qquad \eta_Y=1.012~.
\ee

\boldmath
\subsubsection{Effective Hamiltonian for $b\to s\ell^+\ell^-$}\label{sec:bqll}
\unboldmath
For our discussion of constraints from
$b\to s\ell^+\ell^-$ transitions we will need the corresponding effective
Hamiltonian which is
a generalization of the SM one:
\be\label{eq:Heffqll}
 \Heff(b\to s \ell\bar\ell)
= \Heff(b\to s\gamma)
-  \frac{4 G_{\rm F}}{\sqrt{2}} \frac{\alpha}{4\pi}V_{ts}^* V_{tb} \sum_{i = 9,10} [C_i(\mu)Q_i(\mu)+C^\prime_i(\mu)Q^\prime_i(\mu)]
\end{equation}
where
\be\label{QAQVL}
Q_9  = (\bar s\gamma_\mu P_L b)(\bar \ell\gamma^\mu\ell),\qquad
Q_{10}  = (\bar s\gamma_\mu P_L b)(\bar \ell\gamma^\mu\gamma_5\ell)
\ee
\be\label{QAQVR}
Q^\prime_9  = (\bar s\gamma_\mu P_R b)(\bar \ell\gamma^\mu\ell), \qquad
Q^\prime_{10}  = (\bar s\gamma_\mu P_R b)(\bar \ell\gamma^\mu\gamma_5\ell)\,.
\ee
Here  $\Heff(b\to s\gamma)$ stands for the effective Hamiltonian for the
$b\to s\gamma$ transition that involves
the dipole operators.
An explicit formula
for the latter Hamiltonian  will be presented in the next section.
For  the  Wilson coefficients we find
\begin{align}
 \sin^2\theta_W C_9 &=[\eta_Y Y_0(x_t)-4\sin^2\theta_W Z_0(x_t)]
-\frac{1}{g_{\text{SM}}^2}\frac{1}{M_{Z'}^2}
\frac{\Delta_L^{sb}(Z')\Delta_V^{\mu\bar\mu}(Z')} {V_{ts}^* V_{tb}} ,\\
   \sin^2\theta_W C_{10} &= -\eta_Y Y_0(x_t) -\frac{1}{g_{\text{SM}}^2}\frac{1}{M_{Z'}^2}
\frac{\Delta_L^{sb}(Z')\Delta_A^{\mu\bar\mu}(Z')}{V_{ts}^* V_{tb}},\\
  \sin^2\theta_W C^\prime_9         &=-\frac{1}{g_{\text{SM}}^2}\frac{1}{M_{Z'}^2}
\frac{\Delta_R^{sb}(Z')\Delta_V^{\mu\bar\mu}(Z')}{V_{ts}^* V_{tb}},\\
  \sin^2\theta_W C_{10}^\prime   &= -\frac{1}{g_{\text{SM}}^2}\frac{1}{M_{Z'}^2}
\frac{\Delta_R^{sb}(Z')\Delta_A^{\mu\bar\mu}(Z')}{V_{ts}^* V_{tb}},
 \end{align}
where we have defined
\begin{align}\label{DeltasVA}
\begin{split}
 &\Delta_V^{\mu\bar\mu}(Z')= \Delta_R^{\mu\bar\mu}(Z')+\Delta_L^{\mu\bar\mu}(Z'),\\
&\Delta_A^{\mu\bar\mu}(Z')= \Delta_R^{\mu\bar\mu}(Z')-\Delta_L^{\mu\bar\mu}(Z').
\end{split}
\end{align}
Here $Z_0(x_t)$ is the  SM one-loop function, analogous to $X_0(x_t)$
and $Y_0(x_t)$, that represents gauge invariant combination of $Z-$ and photon 
penguin diagrams:
\begin{align}
  Z_0 (x) & = -\frac{1}{9} \log x + \frac{18 x^4 - 163 x^3 + 259 x^2 - 108 x}{144 (x-1)^3} + \frac{32 x^4 - 38 x^3 - 15 x^2 + 18 x}{72
(x-1)^4}\log x \,.
\end{align}
 The presence of additional
coupling $\Delta_R^{\mu\bar\mu}(Z')$ or $\Delta_L^{\mu\bar\mu}(Z')$, in
addition to $\Delta_A^{\mu\bar\mu}(Z')$, introduces two new parameters and
allows thereby to avoid present constraints on the coefficients $C_9$ and $C_9^\prime$.
 Therefore only the constraints on $C_{10}$ and $C_{10}^\prime$ from
$B\to K^*\ell^+\ell^-$, $B\to K\ell^+\ell^-$ and $B\to X_s\ell^+\ell^-$ will be relevant in
the case of $Z'$. In the case of FCNC processes mediated by $Z$, which
will be discussed in Section~\ref{sec:ZSM}, all leptonic couplings are
known and  also the constraints on the coefficients $C_9$
and $C_9^\prime$ have to be taken into account.

The formulae above do not include QCD renormalization group effects which
influence only  $C_9$ and $C_9^\prime$. They will be taken into account
in the model independent bounds on these coefficients in  Section~\ref{sec:ZSM}.

\subsection{Basic Formulae for Observables}

\boldmath
 \subsubsection{$\Delta F=2$ Observables}
\unboldmath

The $\Delta B=2$ mass differences are given as follows:
\be\label{DMd}
\Delta M_d=\frac{G_F^2}{6 \pi^2}M_W^2 m_{B_d}|\lambda_t^{(d)}|^2   F_{B_d}^2\hat B_{B_d} \eta_B |S(B_d)|\,,
\ee
\be\label{DMs}
\Delta M_s =\frac{G_F^2}{6 \pi^2}M_W^2 m_{B_s}|\lambda_t^{(s)}|^2   F_{B_s}^2\hat B_{B_s} \eta_B |S(B_s)|\,.
\ee
The corresponding mixing induced CP-asymmetries are then given
by
\begin{equation}
S_{\psi K_S} = \sin(2\beta+2\varphi_{B_d})\,, \qquad
S_{\psi\phi} =  \sin(2|\beta_s|-2\varphi_{B_s})\,,
\label{eq:3.44}
\end{equation}
where the phases $\beta$ and $\beta_s$ are defined by
\be\label{vtdvts}
V_{td}=\vtd e^{-i\beta}, \qquad V_{ts}=-\vts e^{-i\beta_s}.
\ee
$\beta_s\simeq -1^\circ\,$.
 The new phases $\varphi_{B_q}$  are directly related to the phases of the functions
$S(B_q)$:
\be
2\varphi_{B_q}=-\theta_S^{B_q}.
\ee
Our phase conventions are as in \cite{Buras:2012xx} and our previous papers 
quoted in this work.

For the CP-violating parameter $\varepsilon_K$  and $\Delta M_K$ we have
respectively
\be
\varepsilon_K=\frac{\kappa_\eps e^{i\varphi_\eps}}{\sqrt{2}(\Delta M_K)_\text{exp}}\left[\Im\left(M_{12}^K\right)\right]\,,\qquad \Delta M_K=2\Re\left(M_{12}^K\right),
\label{eq:3.35}
\ee
where
\be\label{eq:3.4}
\left(M_{12}^K\right)^*=\frac{G_F^2}{12\pi^2}F_K^2\hat
B_K m_K M_{W}^2\left[
\lambda_c^{2}\eta_1x_c +\lambda_t^{2}\eta_2S(K) +
2\lambda_c\lambda_t\eta_3S_0(x_c,x_t)
\right]\,.
\ee
Here, $S_0(x_c,x_t)$
is a {\it real valued} one-loop box function for which explicit expression is given e.\,g.~in \cite{Blanke:2006sb}. The
factors $\eta_i$ are QCD {corrections} evaluated at the NLO level in
\cite{Herrlich:1993yv,Herrlich:1995hh,Herrlich:1996vf,Buras:1990fn,Urban:1997gw}. For $\eta_1$ and $\eta_3$ also NNLO corrections
have been recently
calculated \cite{Brod:2010mj,Brod:2011ty}.
Next $\varphi_\eps = (43.51\pm0.05)^\circ$ and $\kappa_\eps=0.94\pm0.02$ \cite{Buras:2008nn,Buras:2010pza} takes into account
that $\varphi_\eps\ne \tfrac{\pi}{4}$ and includes long distance  effects in $\Im( \Gamma_{12})$ and $\Im (M_{12})$.

In the rest of the paper, unless otherwise stated, we will assume that all four parameters in the CKM
matrix have been determined through tree-level decays without any NP pollution
and pollution from QCD-penguin diagrams so that their values can be used
universally  in
all NP models considered by us.

\boldmath
\subsubsection{$B_{d,s} \to \mu^+ \mu^-$}
\unboldmath
With the assumption that the CKM parameters
have been determined independently
of NP and are universal we find
\be\label{GB/SM}
\frac{\mathcal{B}(B_q\to\mu^+\mu^-)}{\mathcal{B}(B_q\to\mu^+\mu^-)^{\rm SM}}
=\left|\frac{Y_A(B_q)}{\eta_Y Y_0(x_t)}\right|^2,
\ee
where $Y_A(B_q)$ is given in (\ref{YAB}).

As stressed in 
\cite{DescotesGenon:2011pb,deBruyn:2012wj,deBruyn:2012wk} \footnote{We follow here presentation and notations of \cite{deBruyn:2012wj,deBruyn:2012wk}.},
when comparing
the theoretical branching 
ratio $\mathcal{B}(B_s\to\mu^+\mu^-)$ with experimental data quoted by LHCb, ATLAS and CMS,
a correction factor has to be included which takes care of $\Delta\Gamma_s$
effects
that influence the extraction of this branching ratio from the data:
\be
\label{Fleischer1}
\mathcal{B}(B_{s}\to\mu^+\mu^-)_{\rm th} =
r(y_s)~\mathcal{B}(B_{s}\to\mu^+\mu^-)_{\rm exp}, \quad r(0)=1.
\ee
Here
\be
r(y_s)\equiv\frac{1-y_s^2}{1+\mathcal{A}^\lambda_{\Delta\Gamma} y_s}
\approx 1 - \mathcal{A}^\lambda_{\Delta\Gamma} y_s
\ee
with
\be
y_s\equiv\tau_{B_s}\frac{\Delta\Gamma_s}{2}=0.088\pm0.014.
\ee
The quantity $\mathcal{A}^\lambda_{\Delta\Gamma}$ is discussed below.

It is a matter of choice whether the factor $r(y_s)$
is included in the experimental branching ratio or
in the theoretical calculation,
provided $r(y_s)$ is not significantly affected by NP.
Once it is measured, its inclusion in the experimental value, as advocated in
\cite{deBruyn:2012wj}, should be favoured as it would have no impact on the theoretical
calculations of branching ratios that do not depend on $\Delta\Gamma_s$. As in the SM
and CMFV  $\mathcal{A}^\lambda_{\Delta\Gamma}=1$ \cite{deBruyn:2012wk} and
the factor $r(y_s)$ is universal, it is also a good idea to include this factor
in experimental branching ratio. In this manner various CMFV relations remain intact.

 If a given model predicts  $\mathcal{A}^\lambda_{\Delta\Gamma}$ significantly different from unity and
the dependence of $r(y_s)$ on
model parameters is large  one may include this factor in
the theoretical branching ratio:
\be
\label{Fleischer2}
\mathcal{B}(B_{s}\to\mu^+\mu^-)_{\rm corr}=\frac{\mathcal{B}(B_{s}\to\mu^+\mu^-)_{\rm th}}{r(y_s)}.
\ee

The branching ratios $\mathcal{B}(B_q\to\mu^+\mu^-)$ are only sensitive to
the absolute value of $Y_A(B_q)$. However,
as pointed out in \cite{deBruyn:2012wj,deBruyn:2012wk}  in
the flavour precision era these decays could allow to get also some information
on the phase of $Y_A(B_q)$ and we want to investigate whether in
the models considered this effect is significant.
The authors of \cite{deBruyn:2012wk,Fleischer:2012fy} provide general expressions for
$\mathcal{A}^\lambda_{\Delta\Gamma}$ and
$S_{\mu^+\mu^-}^s$ as functions of Wilson coefficients involved. Using
these formulae we find in $Z'$ models very simple formulae that
reflect the fact that $Z'$ and not scalar operators dominate NP 
contributions:
\be\label{Smumus}
\mathcal{A}^\lambda_{\Delta\Gamma}=\cos (2\theta^{B_s}_Y-2\varphi_{B_s}), \quad
S_{\mu^+\mu^-}^s=\sin (2\theta^{B_s}_Y-2\varphi_{B_s})
\ee
Both $\mathcal{A}^\lambda_{\Delta\Gamma}$ and
$S_{\mu^+\mu^-}^s$ are theoretically clean observables.

In the formulae (\ref{Smumus}) and (\ref{Smumud}) 
we took into account new phases in the $B_q-\bar B_q$ mixings as we deal 
here with the mixing induced CP violation. While 
smaller than the phases of $Y_A(B_q)$ their inclusion could be 
relevant one day. The SM phases cancel in this asymmetry 
\cite{deBruyn:2012wk,Fleischer:2012fy}\footnote{We thank Robert Knegjens and Robert Fleischer  
for discussion of this point.}.

In the SM and CMFV models
\be\label{ADG}
\mathcal{A}^\lambda_{\Delta\Gamma}=1, \quad S_{\mu^+\mu^-}^s=0,
\quad r(y_s)=0.912\pm0.014
\ee
independently of NP parameters.

While $\Delta\Gamma_d$ is very small and $y_d$ can be set to zero,
in the case of $B_d\to\mu^+\mu^-$ one can still consider the CP asymmetry
$S_{\mu^+\mu^-}^d$ \cite{Fleischer:2012fy}, for which we simply find
\be\label{Smumud}
S_{\mu^+\mu^-}^d=\sin(2\theta^{B_d}_Y-2\varphi_{B_d}).
\ee

The most recent  results from LHCb read \cite{Aaij:2012ac,LHCbBsmumu}
\be\label{LHCb2}
\mathcal{B}(B_{s}\to\mu^+\mu^-) = (3.2^{+1.5}_{-1.2}) \times 10^{-9}, \quad
\mathcal{B}(B_{s}\to\mu^+\mu^-)^{\rm SM}=(3.23\pm0.27)\times 10^{-9},
\ee
\be\label{LHCb3}
\mathcal{B}(B_{d}\to\mu^+\mu^-) \le  9.4\times 10^{-10}, \quad
\mathcal{B}(B_{d}\to\mu^+\mu^-)^{\rm SM}=(1.07\pm0.10)\times 10^{-10}.
\ee
We have shown also SM predictions for these
observables \cite{Buras:2012ru} that do not include the correction
${r(y_s)}$. If this factor is included one finds \cite{deBruyn:2012wj,deBruyn:2012wk}
\be
\label{FleischerSM}
\mathcal{B}(B_{s}\to\mu^+\mu^-)^{\rm SM}_{\rm corr}= (3.5\pm0.3)\cdot 10^{-9}.
\ee
It is this branching that should be compared in such a case
with the results of LHCb given above. 
For the latest discussions of these issues see 
\cite{deBruyn:2012wj,deBruyn:2012wk,Buras:2012ru,Fleischer:2012fy}.

As we will see below in
the $Z'$ models considered by us  $0.5\le\mathcal{A}^\lambda_{\Delta\Gamma}\le 1$  with the smallest values corresponding to the largest allowed 
values of $|S_{\psi\phi}|$. Thus the $\Delta\Gamma_s$ effect in question 
varies from $5\%$ to $9\%$. In view of still large experimental error 
we will approximately include this effect in the
experimental branching ratio using the values in (\ref{ADG}).
If this is done the experimental results in (\ref{LHCb2}) is reduced by $9\%$
 and 
we find 
\be\label{LHCb2corr}
\mathcal{B}(B_{s}\to\mu^+\mu^-)_{\rm corr} =(2.9^{+1.4}_{-1.1}) \times 10^{-9}, 
\ee
that should be compared with the SM result in (\ref{LHCb2}). While the central 
theoretical value agrees very well with experiment, the large experimental 
error still allows for  NP contributions. In our plots we will show 
the result in (\ref{LHCb2corr}).

\boldmath
\subsubsection{$K_L\to\mu^+\mu^-$}\label{sec:KLmumu}
\unboldmath
Only the so-called short distance (SD)
part to a dispersive contribution
to $K_L\to\mu^+\mu^-$ can be reliably calculated. Therefore in what follows
this decay will be treated only as an additional constraint to be sure
that the rough upper bound given below is not violated.
We have then
following \cite{Buras:2004ub}
($\lambda=0.226$)
\be
\mathcal{B}(K_L\to\mu^+\mu^-)_{\rm SD} =
 2.08\cdot 10^{-9} \left[\bar P_c\left(Y_K\right)+
A^2 R_t\left|Y_A(K)\right|\cos\bar\beta_{Y}^K\right]^2\,,
\ee
where $R_t $ is given in (\ref{eq:Rt_beta}), $\vcb\equiv=A\lambda^2$ and
\be
\bar\beta_{Y}^K \equiv \beta-\beta_s-\theta^K_Y\,,
\qquad
\bar P_c\left(Y_K\right) \equiv \left(1-\frac{\lambda^2}{2}\right)P_c\left(Y_K\right)\,,
\ee
with $P_c\left(Y_K\right)=0.113\pm 0.017$
\cite{Gorbahn:2006bm}. Here
$\beta$ and $\beta_s$ are the phases of $V_{td}$ and $V_{ts}$ defined in
(\ref{vtdvts}).

The extraction of the short distance
part from the data is subject to considerable uncertainties. The most recent
estimate gives \cite{Isidori:2003ts}
\be\label{eq:KLmm-bound}
\mathcal{B}(K_L\to\mu^+\mu^-)_{\rm SD} \le 2.5 \cdot 10^{-9}\,,
\ee
to be compared with $(0.8\pm0.1)\cdot 10^{-9}$ in the SM
\cite{Gorbahn:2006bm}.

\boldmath
\subsubsection{$K^+ \rightarrow \pi^+\nu\bar\nu$ and $K_L \rightarrow \pi^0\nu\bar\nu$}
\unboldmath
These are the two theoretically cleanest rare decays in quark flavour physics.
 Reviews of these two decays can be found in
\cite{Buras:2004uu,Isidori:2006yx,Smith:2006qg}.
The
branching ratios for these two
 modes
 can be written generally as
\begin{gather} \label{eq:BRSMKp}
  \Br (K^+\to \pi^+ \nu\bar\nu) = \kappa_+ \left [ \left ( \frac{{\rm Im} X_{\rm eff} }{\lambda^5}
  \right )^2 + \left ( \frac{{\rm Re} X_{\rm eff} }{\lambda^5}
  - P_c(X)  \right )^2 \right ] \, , \\
\label{eq:BRSMKL} \Br( K_L \to \pi^0 \nu\bar\nu) = \kappa_L \left ( \frac{{\rm Im}
    X_{\rm eff} }{\lambda^5} \right )^2 \, ,
\end{gather}
where \cite{Mescia:2007kn}
\begin{equation}\label{kapp}
\kappa_+=(5.36\pm0.026)\cdot 10^{-11}\,, \quad \kappa_{\rm L}=(2.31\pm0.01)\cdot 10^{-10}
\ee
and \cite{Buras:2005gr,Buras:2006gb,Brod:2008ss,Isidori:2005xm}.
\be
P_c(X)=0.42\pm0.03.
\end{equation}
The short distance contributions are described by
\be\label{Xeff}
X_{\rm eff} = V_{ts}^* V_{td} (X_{L}(K) + X_{R}(K))
\ee
where $X_{L,R}(K)$ are given in (\ref{XLK}) and (\ref{XRK}).

 Experimentally we have \cite{Artamonov:2008qb}
\be\label{EXP1}
\mathcal{B}(\kpn)_\text{exp}=(17.3^{+11.5}_{-10.5})\cdot 10^{-11}\,,
\ee
and the $90\%$ C.L. upper bound \cite{Ahn:2009gb}
\be\label{EXP2}
\mathcal{B}(\klpn)_\text{exp}\le 2.6\cdot 10^{-8}\,.
\ee

In the SM one finds
\cite{Brod:2008ss,Brod:2010hi}
\be
\mathcal{B}(\kpn)_\text{SM} =(8.5\pm 0.7)\cdot 10^{-11}\,,
\ee
\be
\mathcal{B}(\klpn)_\text{SM} =(2.6\pm 0.4)\cdot 10^{-11}\,,
\ee
where the errors are dominated by CKM uncertainties. This should be compared
with the experimental values given in (\ref{EXP1}) and (\ref{EXP2}). Clearly we have to wait for improved data.

\subsubsection{\boldmath $B \to \{X_s,K, K^*\} \nu\bar \nu$}\label{sec:Bnunu}
Following the analysis of \cite{Altmannshofer:2009ma},
the branching ratios of the $B \to \{X_s,K, K^*\}\nu\bar \nu$
modes in the presence of RH currents can be written as follows
 \bea
 \mathcal{B}(B\to K \nu \bar \nu) &=&
 \mathcal{B}(B\to K \nu \bar \nu)_{\rm SM} \times\left[1 -2\eta \right] \epsilon^2~, \label{eq:BKnn}\\
 \mathcal{B}(B\to K^* \nu \bar \nu) &=&
 \mathcal{B}(B\to K^* \nu \bar \nu)_{\rm SM}\times\left[1 +1.31\eta \right] \epsilon^2~, \\
 \mathcal{B}(B\to X_s \nu \bar \nu) &=&
 \mathcal{B}(B\to X_s \nu \bar \nu)_{\rm SM} \times\left[1 + 0.09\eta \right] \epsilon^2~,\label{eq:Xsnn}
 \eea
 where we have introduced the variables
 \be\label{etaepsilon}
 \epsilon^2 = \frac{ |X_{\rm L}(B_s)|^2 + |X_{\rm R}(B_s)|^2 }{
 |\eta_X X_0(x_t)|^2 }~,  \qquad
 \eta = \frac{ - {\rm Re} \left( X_{\rm L}(B_s) X_{\rm R}^*(B_s)\right) }
{ |X_{\rm L}(B_s)|^2 + |X_{\rm R}(B_s)|^2 }~,
 \ee
with $X_{L,R}(B_s)$ defined in (\ref{XLB}) and (\ref{XRB}).

Moreover the average of the $K^*$ longitudinal polarization fraction $F_L$
also used in the studies of $B\to K^*\ell^+\ell^-$ is a useful variable as
it depends only on  $\eta$:
\be
\label{eq:epseta-FL}
 \langle F_L \rangle = 0.54 \, \frac{(1 + 2 \,\eta)}{(1 + 1.31 \,\eta)}~.
\ee

We should remark that the expressions in Eqs.~(\ref{eq:BKnn})--(\ref{eq:Xsnn}),
as well as the SM results in (\ref{eq:BKnnSM}), refer only to the short-distance contributions
to these decays. The latter are obtained from the corresponding total rates
subtracting the reducible long-distance effects pointed out in~\cite{Kamenik:2009kc}.

The predictions for the SM branching  ratios
are~\cite{Bartsch:2009qp,Kamenik:2009kc,Altmannshofer:2009ma}
\bea
\mathcal{B}(B\to K \nu \bar \nu)_{\rm SM}   &=& (3.64 \pm 0.47)\times 10^{-6}~, \no \\
\mathcal{B}(B\to K^* \nu \bar \nu)_{\rm SM} &=& (7.2 \pm 1.1)\times 10^{-6}~, \no \\
\mathcal{B}(B\to X_s \nu \bar \nu)_{\rm SM} &=& (2.7 \pm 0.2)\times 10^{-5}~,
\label{eq:BKnnSM}
\eea
to be compared with the experimental bounds~\cite{Barate:2000rc,:2007zk,:2008fr}
\bea
\mathcal{B}(B\to K \nu \bar \nu)   &<&  1.4 \times 10^{-5}~, \no \\
\mathcal{B}(B\to K^* \nu \bar \nu) &<&  8.0 \times 10^{-5}~, \no \\
\mathcal{B}(B\to X_s \nu \bar \nu)  &<&  6.4 \times 10^{-4}~.
\label{eq:BKnn_exp}
\eea

As $\epsilon$ and $\eta$ can be calculated in any model by means of
(\ref{etaepsilon}) the expressions given above can be considered as
fundamental formulae for any phenomenological analysis of these decays
and  a given model can be represented by a point in
the $\epsilon-\eta$ plane. Measuring the three branching ratios allows
uniquely to determine experimentally the point  $(\epsilon,\eta)$ and
to compare with any model result.  We will illustrate this for 
$Z'$ scenarios.

\boldmath
\subsubsection{$K_L\to\pi^0\ell^+\ell^-$}
\unboldmath

The rare decays $K_L\to\pi^0e^+e^-$ and $K_L\to\pi^0\mu^+\mu^-$ are dominated
by CP-violating contributions. The indirect CP-violating
contributions are determined by the measured decays 
$K_S\to\pi^0 \ell^+\ell^-$ and the parameter $\varepsilon_K$ in 
a model independent manner. It is the dominant contribution within the SM 
where one finds
\cite{Mescia:2006jd}
\begin{gather}
\mathcal{B}(K_L\to\pi^0e^+e^-)_\text{SM}=
3.54^{+0.98}_{-0.85}\left(1.56^{+0.62}_{-0.49}\right)\cdot 10^{-11}\,,\label{eq:KLpee}\\
\mathcal{B}(K_L\to\pi^0\mu^+\mu^-)_\text{SM}= 1.41^{+0.28}_{-0.26}\left(0.95^{+0.22}_{-0.21}\right)\cdot
10^{-11}\label{eq:KLpmm}\,,
\end{gather}
with the values in parentheses corresponding to the destructive interference
between directly and indirectly CP-violating contributions. 
The last discussion  of the theoretical status of this interference
sign can be found in \cite{Prades:2007ud} where the results of \cite{Isidori:2004rb,Friot:2004yr,Bruno:1992za} are
critically analysed. From this discussion, constructive interference
seems to be  favoured though more work is necessary. In view of significant
uncertainties in the SM prediction we will mostly use these decays 
to test whether the correlations of them with $\klpn$ and $\kpn$ decays 
can have an impact on the latter. To this end we will confine our analysis 
to the case of the constructive interference between the directly and 
indirectly CP-violating contributions.

The present experimental bounds
\be
\mathcal{B}(K_L\to\pi^0e^+e^-)_\text{exp} <28\cdot10^{-11}\quad\text{\cite{AlaviHarati:2003mr}}\,,\qquad
\mathcal{B}(K_L\to\pi^0\mu^+\mu^-)_\text{exp} <38\cdot10^{-11}\quad\text{\cite{AlaviHarati:2000hs}}\,,
\ee
are still by one order of magnitude larger than the SM predictions, leaving 
thereby large room for NP contributions. In fact as our numerical analysis in 
Sections~\ref{sec:Excursion} and \ref{sec:ZSM} demonstrates, these bounds 
have no impact on $\kpn$ and $\klpn$ decays but the present data on 
$\kpn$ do not allow to reach the above bounds in the $Z'(Z)$ scenarios considered.

In the LHT model the branching
ratios for both decays can be enhanced at most 
by a factor of 1.5 \cite{Blanke:2006eb,Blanke:2008ac}. Slightly larger 
effects are still allowed in RSc \cite{Blanke:2008yr}. 

In the LHT model, where only SM operators are present
the effects 
of NP can be compactly summarised by generalisation of the 
real SM functions $Y_0(x_t)$ and $Z_0(x_t)$ to two complex functions $Y_K$ and 
$Z_K$, respectively. As demonstrated in the context of the corresponding 
analysis within RSc  \cite{Blanke:2008yr}, also in the presence of RH 
currents two complex functions $Y_K$ and $Z_K$
are sufficient to describe jointly the SM and NP contributions.
Consequently the LHT formulae (8.1)--(8.8) of \cite{Blanke:2006eb} with 
$Y_K$ and $Z_K$ given below can be used to study these decays
in the context of tree-level $Z'$ and $Z$ exchanges. 
The original papers behind these formulae can 
be found in 
\cite{Buchalla:2003sj,Isidori:2004rb,Friot:2004yr,Mescia:2006jd,Buras:1994qa}.

Using the formulae of \cite{Blanke:2008yr} we find
\be\label{YK}
Y_K=\eta_Y Y_0(x_t)+ \left[\frac{\Delta_{A}^{\mu\bar\mu}(Z')}{M_{Z'}^2g^2_{\rm SM}}\right]
\frac{\Delta_{V}^{sd}(Z')}{ V_{ts}^\ast V_{td}},
\ee
\be
Z_K=Z_0(x_t)+\frac{1}{4\sin^2\theta_W}\left[\frac{2\Delta_{R}^{\mu\bar\mu}(Z')}{M_{Z'}^2g^2_{\rm SM}}\right]
\frac{\Delta_{V}^{sd}(Z')}{ V_{ts}^\ast V_{td}},
\ee
where $\Delta_V^{sd}$ is defined as in (\ref{DeltasVA}). These formulae with obvious changes can
also be used for tree-level 
$Z$ exchanges considered in Section~\ref{sec:ZSM}.

 The presence of additional
coupling $\Delta_R^{\mu\bar\mu}(Z')$ in
addition to $\Delta_A^{\mu\bar\mu}(Z')$, introduces as in $B\to K^*\ell^+\ell^-$, $B\to K\ell^+\ell^-$ and $B\to X_s\ell^+\ell^-$ 
two new parameters and
allows thereby to avoid present constraints if necessary. In our analysis 
we will set $\Delta_R^{\mu\bar\mu}(Z')$ to its SM value.
In the case of FCNC processes mediated by $Z$, which
will be discussed in Section~\ref{sec:ZSM}, all leptonic couplings are
known and the predictions for  $K_L\to\pi^0e^+e^-$ and $K_L\to\pi^0\mu^+\mu^-$ 
are more specific.
The numerical results are presented for $Z'$ and $Z$ contributions in 
 Sections~\ref{sec:Excursion} and \ref{sec:ZSM}, respectively.

\boldmath
\section{$B\to X_s\gamma$ Decay}\label{sec:4B}
\unboldmath

\subsection{Preliminaries}
The $B\to X_s\gamma$ decay being the first loop induced B-decay determined
experimentally has been extensively studied within the SM and its various
extensions. For our calculation of $Z'$ contributions to the relevant
Wilson coefficients very useful turned out to be recent study
of this decay within
gauged flavour models in \cite{Buras:2011zb}. Indeed several formulae of this
paper could be easily adapted to our analysis.

Let us recall that
in the SM the LH structure of
the $W$ couplings to quarks requires the chirality flip, necessary for $b\to s\gamma$
transition to occur, only through the mass of the initial or the final state
quark. Consequently the amplitude is proportional to $m_b$ or $m_s$.
In contrast in models like LR models RH couplings of $W_R^\pm$ to quarks allow the chirality flip on the
internal top quark line resulting in an enhancement factor $m_t/m_b$
of NP contribution relative to the SM one at the level of the amplitude.
However, in the present analysis $Z'$ contributions to $B\to X_s\gamma$
involve only SM quarks with electric charge $-1/3$ and such an enhancement is
absent. Therefore we do not expect large corrections to $B\to X_s\gamma$ from
$Z'$ exchanges, which is good as the SM agrees well with the data. Still it
is of interest to check the size of these contributions. In doing this
we include QCD corrections to $Z'$ contributions at the LO using
the general formulae of  \cite{Buras:2011zb}, while the SM contributions
are included at the NNLO level.

Adopting the overall normalization of the SM effective Hamiltonian
we have
{\begin{equation} \label{Heff_at_mu}
{\cal H}_{\rm eff}(b\to s\gamma) = - \frac{4 G_{\rm F}}{\sqrt{2}} V_{ts}^* V_{tb}
\left[  C_{7\gamma}(\mu_b) Q_{7\gamma} +  C_{8G}(\mu_b) Q_{8G} \right]\,,
\end{equation}}
where $\mu_b=\ord(m_b)$.
The dipole operators are defined as
\begin{equation}\label{O6B}
Q_{7\gamma}  =  \frac{e}{16\pi^2} m_b \bar{s}_\alpha \sigma^{\mu\nu}
P_R b_\alpha F_{\mu\nu}\,,\qquad
Q_{8G}     =  \frac{g_s}{16\pi^2} m_b \bar{s}_\alpha \sigma^{\mu\nu}
P_R T^a_{\alpha\beta} b_\beta G^a_{\mu\nu}\,.
\end{equation}
In writing (\ref{Heff_at_mu}) we have dropped
the primed operators that are obtained from (\ref{O6B}) by  replacing $P_{R}$
by $P_L$. In the SM the primed operators (RL) are suppressed by $m_s/m_b$
relative to the ones in (\ref{Heff_at_mu}). This is not the case for
RL operators but as such contributions do not interfere with SM contributions
that is dominant in any case we will neglect these contributions in the
case of $Z'$ as well.
We have also suppressed
current-current operators which are important for the QCD analysis. We will
include these effects in the final formulae at the end of this section.

The coefficients $C_i(\mu_b)$ are calculated from their initial values at
high energy scales by means of renormalization group methods. We distinguish
between SM quark contributions with the matching scale  $\mu_t=\ord(m_t)$
and the $Z'$ quark contributions with the matching scale
$\mu_{Z'}=\ord(M_{Z'})$.
While in the LO approximation the results depend on the choice of the
matching scale, the experience shows that taking as the matching scale
the largest mass in the diagram appears to be a very good choice at LO.
The choices made above follow this strategy.

We  decompose next the Wilson coefficients at the scale $\mu_b=\ord(m_b)$
as the sum of the SM contribution and the $Z'$ contributions:
\be
C_i(\mu_b)=C_i^{\rm SM}(\mu_b)
+\Delta C^{Z'}_i(\mu_b).
\label{cstart}\,
\ee

We recall that for the SM coefficients at $\mu_t=\ord(m_t)$
we have ($x_t=m_t^2/M_W^2$) without QCD corrections
\begin{equation}\label{c7}
C^{\rm SM}_{7\gamma} (\mu_t) = \frac{3 x_t^3-2 x_t^2}{4(x_t-1)^4}\ln x_t +
\frac{-8 x_t^3 - 5 x_t^2 + 7 x_t}{24(x_t-1)^3}\equiv C^{\rm SM}_{7\gamma}(x_t)\,,
\end{equation}
\begin{equation}\label{c8}
C^{\rm SM}_{8G}(\mu_t) = \frac{-3 x_t^2}{4(x_t-1)^4}\ln x_t +
\frac{-x_t^3 + 5 x_t^2 + 2 x_t}{8(x_t-1)^3}\,\equiv C^{\rm SM}_{8G}(x_t).
\end{equation}

In the next subsection, we summarize the results for $Z'$ contributions
to the Wilson coefficients of the dipole operators
at the relevant matching scale $\mu_{Z'}=\ord(M_{Z'})$.
Subsequently we will
present renormalization group QCD corrections to
these coefficients.
The final formula for the branching ratio for the $B\to X_s\gamma$ decay that
includes SM and $Z'$ contributions will be presented at the end of this section.

\boldmath
\subsection{$Z'$ contribution without QCD Corrections}\label{eq:bsgamma-neutralBoson}
\unboldmath
A general analysis of neutral gauge boson contributions to $B\to X_s\gamma$
decay has been presented in \cite{Buras:2011zb}.
In addition to
SM-like LL contribution from $Z'$ we have a new LR one, where $L$ ($R$) stands for the $P_L$ ($P_R$) projector in the basic penguin diagram
involving the $s$($b$)-quark.

In what follows
we present the results for a contribution of a fermion $f$ carrying electric
charge $-1/3$ and having the mass $m_f$. This will allow us to compute
the contribution from SM down-quarks and in the future 
if necessary contributions involving
new heavy quarks.

We first decompose the Wilson coefficients $\Delta C^{Z^\prime}_i$ at the
$\mu_{Z^\prime}$
scale as the sum of the SM-like LL contribution and a new LR one:
\be
\begin{aligned}
\Delta C^{Z^\prime}_{7\gamma}(\mu_{Z^\prime})&=\Delta^{LL} C^{Z^\prime}_{7\gamma}(\mu_{Z^\prime}) +\Delta^{LR}C^{Z^\prime}_{7\gamma}(\mu_{Z^\prime})\,,\\[2mm]
\Delta C^{Z^\prime}_{8G}(\mu_{Z^\prime})     &=\Delta^{LL} C^{Z^\prime}_{8G}(\mu_{Z^\prime}) +\Delta^{LR}C^{Z^\prime}_{8G}(\mu_{Z^\prime})\,.
\end{aligned}
\label{eq:wilsonatmh}
\ee

Adapting the formulae of \cite{Buras:2011zb}
to our notation and denoting by $f$ the down-quark exchanged in the diagram
we find
\begin{equation}
\begin{aligned}
\Delta^{LL}C^{Z^\prime}_{7\gamma}(\mu_{Z^\prime}) &=-\frac{2}{3}\,\frac{1}{g_2^2}\,\frac{M_W^2}{M_{Z^\prime}^2}\,\sum_f\frac{\Delta_L^{fs*}(Z^\prime)\,\Delta_L^{fb}(Z^\prime)}{V_{ts}^*\,V_{tb}}\,\left(C_{8G}^{SM}(x_f)+\frac{1}{3}\right),\\[2mm]
\Delta^{LL}C^{Z^\prime}_{8 G}(\mu_{Z^\prime}) &= - 3 \Delta^{LL}C^{Z^\prime}_{7\gamma}(\mu_{Z^\prime})\,,
\end{aligned}
\label{LLnew}
\end{equation}
with $x_f=m_f^2/M_{Z^\prime}^2$ and summation is over the SM down-quarks.\\

For LR Wilson coefficients we find:
\begin{equation}
\begin{aligned}
\Delta^{LR}C^{Z'}_{7\gamma}(\mu_{Z'})&=-\frac{2}{3}\,\frac{1}{g_2^2}\,\frac{M_W^2}{ M_{Z'}^2}\,\sum_f \frac{m_f}{m_b}\,
\frac{\Delta_L^{fs*}(Z')\,\Delta_R^{fb}(Z')}{V_{ts}^*\,V_{tb}}
\,C^{LR}_{8G}(x_f)\,,\\[2mm]
\Delta^{LR}C^{Z'}_{8G}(\mu_{Z'})&= -3\Delta^{LR}C^{Z'}_{7\gamma}(\mu_{Z'})\,,
\end{aligned}
\label{LRnew}
\end{equation}
with
\be
C^{LR}_{8G}(x)=\frac{-3x}{2(1-x)^3}\ln{x}+ \frac{3 x (x-3)}{4(x-1)^2} -1\,.
\label{CLR8Gcoeff}
\ee
The summation is over SM down-quarks.

The following properties should be noted:
\begin{description}
\item 1)\quad As opposed to the case of $W^\pm$ contributions
the factor $m_f/m_b$ is either
$\ord(1)$ or smaller and LR contributions are not dominant.
\item 2)\quad $C^{LR}_{8G}(x)$ is a non-vanishing monotonic function of $x$ and takes values in the range $[-1,\,-1/4]$ for $x$ from $0$ to $\infty$.
\end{description}

\subsection{Final Results including QCD corrections}\label{FinalWC}
In order to complete the analysis of $B\to X_s\gamma$ we have to include
QCD corrections which play a very important role in this decay.
In the SM these corrections are known at the NNLO level \cite{Misiak:2006zs}.
In the LR model
a complete LO analysis has been done by Cho and Misiak \cite{Cho:1993zb}
and after proper modification we can use their results in our model.
In this context the recent analyses \cite{Buras:2011zb,Blanke:2011ry}
turned out to be very useful.

We find then
\begin{equation}
\Delta C^{Z'}_{7\gamma}(\mu_b)=
\kappa_7(\mu_{Z'})~\Delta C^{Z^\prime}_{7\gamma}(\mu_{Z^\prime}) +
\kappa_8(\mu_{Z'})~\Delta C^{Z'}_{8G}(\mu_{Z^\prime})+
\Delta^{\rm current}_{Z^\prime}(\mu_b)\,.
\label{eq:DeltaC7effA3}
\end{equation}
The last contribution in (\ref{eq:DeltaC7effA3}) results from the mixing of
new neutral current-current operators generated from the $Z'$ exchange
that mix with the dipole operators. The renormalization group analysis
of this contribution is very involved but fortunately the LO result is known
from  \cite{Buras:2011zb}. Therefore adapting the formulae (4.16), (4.17) and
(5.6) of this paper to our notation we find
\be\label{Zcurrent}
\Delta^{\rm current}_{Z^\prime}(\mu_b)=
\sum_{\substack{A=L,R\\f=u,c,t,d,s,b}}\!\!\!\!\! \kappa^{f}_{LA}~\Delta ^{LA} C_2^f(\mu_{Z'}) +\!\!\!\sum_{A=L,R}\!\!\!\!\hat{\kappa}^{d}_{LA}~\Delta ^{LA}
\hat{C}_2^d(\mu_{Z^\prime}),
\ee
where
\begin{equation}
\Delta^{AB}C^f_2(\mu_{Z'})=-\frac{2}{g_2^2}\frac{M_W^2}{M_{Z'}^2}
\frac{\Delta_A^{sb*}(Z')\,\tilde\Delta_B^{ff}(Z')}{V_{ts}^*\,V_{tb}}\,,
\label{InitialConditionQnn}
\end{equation}
and
\begin{equation}
\Delta^{AB}\hat{C}_2^d(\mu_{Z'})=-\frac{2}{g_2^2}\frac{M_W^2}{M_{Z'}^2}
\frac{\Delta_A^{sd*}(Z')\,\Delta_B^{bd}(Z')}{V_{ts}^*\,V_{tb}}\,.
\label{InitialConditionQnnHat}
\end{equation}

The diagonal couplings $\tilde\Delta_B^{ff}(Z')$
introduce additional parameters. For our numerical estimate we use their SM 
values.

Finally, $\kappa$'s are the NP magic numbers listed in Tab.~\ref{tab:Magic}
that is based on
\cite{Buras:2011zb} which used
$\alpha_s(M_Z=91.1876\,\text{GeV})=0.118$. They have been obtained for
$\mu_b=2.5\gev$ as used in the SM calculations. We add that for
$\mu_{Z'}=2.5\tev$ we have $\kappa_7=0.427$ and $\kappa_8=0.128$.

\begin{table}[h!]
\begin{center}
\begin{tabular}{|c||r|r|r|r||r|}
  \hline
  &&&&&\\[-4mm]
  $\mu_H$	 	 & 200 GeV 	& 1 TeV		& 5 TeV		& 10 TeV & $M_Z$\\[1mm]
  \hline\hline
  $\kappa_7$		 & 0.524	& 0.457		& 0.408		& 0.390	 &	 0.566 \\			
  $\kappa_8$		 & 0.118	& 0.125		& 0.129		& 0.130	 & 0.111	\\[1mm]		
  \hline
  &&&&&\\[-4mm]
  $\kappa_{LL}^{u,c}$	 & 0.039	& 0.057		& 0.076		& 0.084	 & 0.030	\\			
  $\kappa_{LL}^{t}$	 &-0.002	&-0.003		&-0.002		&-0.001	 & --	\\			
  $\kappa_{LL}^{d}$	 &-0.040	&-0.057		&-0.072		&-0.079	 & -0.032	\\			
  $\kappa_{LL}^{s,b}$	 & 0.087	& 0.090		& 0.090		& 0.090	 & 0.084	\\			
  $\hat{\kappa}_{LL}^{d}$& 0.128	& 0.147		& 0.163		& 0.168	 & 0.116	\\[1mm]
  \hline
  &&&&&\\[-4mm]                                                                            	
  $\kappa_{LR}^{u,c}$	 & 0.085	& 0.128	 	& 0.173		& 0.193	 & 0.065	\\			
  $\kappa_{LR}^{t}$	 & 0.004	& 0.012	 	& 0.023		& 0.028	 & --	\\			
  $\kappa_{LR}^{d}$	 &-0.015	&-0.025		&-0.036		&-0.041	 & -0.011	\\			
  $\kappa_{LR}^{s,b}$	 &-0.078	&-0.092	 	&-0.106		&-0.111	 & -0.070	\\			
  $\hat{\kappa}_{LR}^{d}$& 0.473	& 0.665		& 0.865		& 0.953	 & 0.383	\\[1mm]	
  \hline                                                                                         	
\end{tabular}
\caption{The NP magic numbers relevant for QCD calculations \cite{Buras:2011zb}.
 For completeness, in the last column the case of a flavour-violating $Z$ is
included. }
\label{tab:Magic}
\end{center}
\end{table}

 Using these formulae we find for $M_{Z'}=1\tev$
\begin{equation}
\Delta C^{Z'}_{7\gamma}(\mu_b)= \mathcal{O}(10^{-4}).
\end{equation}
While due to the presence of RH couplings, this contribution is by one order of magnitude larger than found in the $\model$ model \cite{Buras:2012xx}, it
is still negligible when compared with the SM value of $-0.353$. Therefore
we will not consider $B\to X_s\gamma$ decay further.

\section{General Structure of New Physics Contributions}\label{sec:3a}
\subsection{Preliminaries}
We have seen in Section~\ref{sec:2} that the small number of free parameters in each of LHS, RHS, LRS and ALRS scenarios allows to expect definite correlations between  flavour observables in each step of the strategy outlined there. These
expectations will be confirmed through the numerical analysis below but it is instructive to develop first  a qualitative general view on NP
contributions in different scenarios before entering the details.

First, it should be realized that the confrontation of correlations in question
with future precise data will not only depend on the size of theoretical, parametric and experimental uncertainties, but also in an important manner on the
size of allowed deviations from SM expectations. The latter deviations are
presently constrained dominantly by $\Delta F=2$ observables and $B\to X_s\gamma$ decay. But as already demonstrated in \cite{Altmannshofer:2011gn,Altmannshofer:2012ir,Buras:2012xx} after the new data from the LHCb, ATLAS and CMS
 also the decays $B_{s,d}\to\mu^+\mu^-$ and $b\to s\ell^+ \ell^-$ begin
to play important roles in this context. We will see their impact on our
analysis as well.

Now, in general NP scenarios in which there are many free parameters, it is
possible with the help of some amount of fine-tuning to satisfy constraints
from $\Delta F=2$ processes without a large impact on the size of NP contributions to $\Delta F=1$ processes. However, in the $Z'$ scenarios considered here,
in which NP in both $\Delta F=2$ and $\Delta F=1$ processes is governed by tree-diagrams, the situation is different.
Indeed, due to the property of {\it factorization} of decay
amplitudes into vertices and the propagator at the tree-level, the same quark flavour violating couplings and the same mass $M_{Z'}$ enter  $\Delta F=2$ and $\Delta F=1$ processes undisturbed by the presence of fermions entering the
usual box and penguin diagrams. Let us exhibit these correlations in
explicit terms.
\boldmath
\subsection{$\Delta F=1$ vs. $\Delta F=2$ Correlations}\label{CORR}
\unboldmath
In order to obtain transparent expressions we introduce
\be
r^\text{VLL}=r^\text{VRR}=\frac{8 \tilde r}{g^2_{\rm SM}}
\ee
which is the same for $K$ and $B_q$ systems. $\tilde r\approx 1$ is defined 
in (\ref{Zprime1}). In the case of LR contributions
to $\Delta S$ let
us rewrite (\ref{DSK}) and
(\ref{DSBq}) as follows
\be\label{DSKa}
[\Delta S(K)]_{\rm LR}=\frac{ r^\text{LR}(K)}{M_{Z'}^2}
  \frac{\Delta_L^{sd}(Z')\Delta_R^{sd}(Z')}{[\lambda_t^{(K)}]^2}
 \ee
\be\label{DSBqa}
[\Delta S(B_q)]_{\rm LR}=\frac{ r^\text{LR}(B_q)}{M_{Z'}^2}
 \frac{\Delta_L^{bq}(Z')\Delta_R^{bq}(Z')}{[\lambda_t^{(q)}]^2}
 \ee
where the quantities $r^\text{LR}$ can be found by comparing these expressions with
(\ref{DSK}) and (\ref{DSBq}), respectively. They depend on low energy
parameters, in particular on the meson system and logarithmically on $M_{Z'}$. The latter dependence can be
neglected for all practical purposes as long as  $M_{Z'}$ is in the range
of a few TeV.

We can then derive the following relations between shifts in the basic
functions in $\Delta F=1$ and $\Delta F=2$ processes which are independent
of any parameters like $\tilde s_{ij}$ but depend sensitively on $M_{Z'}$ \footnote{Similar
relations have been derived in \cite{Buras:2012xx} in the context of
$\model$ model but they involved only LHS scenario.}. In particular
they do not depend explicitly on whether S1 or S2 scenarios for $\vub$
are considered. This dependence is hidden in the allowed shifts in
$\Delta S(K)$ and $\Delta S(B_d)$ both in magnitudes and phases. We have then \footnote{The numerical values on the r.h.s of these equations correspond to $M_{Z'}=1\tev$.}

{\bf LHS Scenario}

\be\label{REL1}
\frac{\Delta X_L(K)}{\sqrt{\Delta S(K)}}=
\frac{\sqrt{2}\Delta_L^{\nu\bar\nu}(Z')}{M_{Z'}g^2_{\rm SM}\sqrt{r^\text{VLL}}}=0.597,
\ee

\be\label{REL2}
\frac{\Delta X_L(B_q)}{\sqrt{\Delta S(B_q)^*}}=
\frac{\sqrt{2}\Delta_L^{\nu\bar\nu}(Z')}{M_{Z'}g^2_{\rm SM}\sqrt{r^\text{VLL}}}=0.597
\ee
and
\be\label{REL3}
\Delta Y_A(K)=\Delta X_L(K) \frac{\Delta_A^{\mu\bar\mu}(Z')}{\Delta_L^{\nu\bar\nu}(Z')}, \qquad  \Delta Y_A(B_q)=\Delta X_L(B_q) \frac{\Delta_A^{\mu\bar\mu}(Z')}{\Delta_L^{\nu\bar\nu}(Z')}.
\ee

{\bf RHS Scenario}

\be\label{REL4}
\Delta X_R(K)=\Delta X_L(K)=-\Delta Y_A(K) \frac{\Delta_L^{\nu\bar\nu}(Z')}{\Delta_A^{\mu\bar\mu}(Z')},
\ee
\be\label{REL4a}
\Delta X_R(B_q)=\Delta X_L(B_q)=-\Delta Y_A(B_q) \frac{\Delta_L^{\nu\bar\nu}(Z')}{\Delta_A^{\mu\bar\mu}(Z')}.
\ee

{\bf LRS Scenario}

\be\label{REL5}
\frac{\Delta X_L(K)}{\sqrt{-\Delta S(K)}}=\frac{\Delta X_R(K)}{\sqrt{\Delta S(K)}}=
\frac{\Delta_L^{\nu\bar\nu}(Z')}{M_{Z'}g^2_{\rm SM}
\sqrt{-r^\text{VLL}-r^\text{LR}(K)}}=0.048,
\ee

\be\label{REL6}
\frac{\Delta X_L(B_q)}{\sqrt{-\Delta S(B_q)^*}}=\frac{\Delta X_R(B_q)}{\sqrt{-\Delta S(B_q)^*}}=
\frac{\Delta_L^{\nu\bar\nu}(Z')}{M_{Z'}g^2_{\rm SM}
\sqrt{-r^\text{VLL}-r^\text{LR}(B_q)}}
\ee
with $\displaystyle{\frac{\Delta X_L(B_d)}{\sqrt{-\Delta S(B_d)^*}}}= 0.204$ and  $\displaystyle{\frac{\Delta X_L(B_s)}{\sqrt{-\Delta S(B_s)^*}}}= 0.212 $.

There are no NP contributions to $Y_A$ functions in this scenario.

{\bf ALRS Scenario}

\be\label{REL7}
\frac{\Delta Y_A(K)}{\sqrt{\Delta S(K)}}=
2\frac{\Delta_A^{\mu\bar\mu}(Z')}{M_{Z'}g^2_{\rm SM}
\sqrt{r^\text{VLL}-r^\text{LR}(K)}}=0.094,
\ee

\be\label{REL8}
\frac{\Delta Y_A(B_q)}{\sqrt{\Delta S(B_q)^*}}=
2\frac{\Delta_A^{\mu\bar\mu}(Z')}{M_{Z'}g^2_{\rm SM}
\sqrt{r^\text{VLL}-r^\text{LR}(B_q)}}
\ee
with $\displaystyle{\frac{\Delta Y_A(B_d)}{\sqrt{\Delta S(B_d)^*}}}=0.337$
and $\displaystyle{\frac{\Delta Y_A(B_s)}{\sqrt{\Delta S(B_s)^*}}}=0.346$.

There are no NP contributions to $X_{L,R}$ functions in this scenario.

{\bf General Scenario}

Finally we give for completeness general formulae for the correlations in question that do
not assume any particular relation between LH and RH couplings. To this
end we write

\be\label{generalrel}
\Delta_R^{ij}=a_{ij} \Delta_L^{ij}, \qquad  a_{ji}=a^*_{ij},
\ee
where $a_{ij}$ are complex numbers.

We find then in the $K$ system

\be\label{REL9}
\frac{\Delta X_{L}(K)}{\sqrt{\Delta S(K)}}=
\frac{\sqrt{2}\Delta_L^{\nu\bar\nu}(Z')}{M_{Z'}g^2_{\rm SM}}
   \frac{1}{\sqrt{r^\text{VLL}(1+a_{sd}^2)+2 a_{sd}r^\text{LR}(K)}},
\ee
\be\label{REL9a}
\Delta X_{R}(K)=  a_{sd}\Delta X_{L}(K),
\ee

\be\label{REL10}
\frac{\Delta Y_{A}(K)}{\sqrt{\Delta S(K)}}=
\frac{\sqrt{2}\Delta_A^{\mu\bar\mu}(Z')}{M_{Z'}g^2_{\rm SM}}
   \frac{1-a_{sd}}{\sqrt{r^\text{VLL}(1+a_{sd}^2)+2 a_{sd}r^\text{LR}(K)}}.
\ee

Similarly in the $B_{s,d}$ systems we have
\be\label{REL11}
\frac{\Delta X_{L}(B_q)}{\sqrt{\Delta S(B_q)^*}}=
\frac{\sqrt{2}\Delta_L^{\nu\bar\nu}(Z')}{M_{Z'}g^2_{\rm SM}}
   \frac{1}{\sqrt{r^\text{VLL}(1+a_{qb}^2)+2 a_{qb}r^\text{LR}(B_q)}},
\ee

\be\label{REL12}
\Delta X_{R}(B_q)=  a_{qb}\Delta X_{L}(B_q)
\ee

\be\label{REL13}
\frac{\Delta Y_{A}(B_q)}{\sqrt{\Delta S(B_q)^*}}=
\frac{\sqrt{2}\Delta_A^{\mu\bar\mu}(Z')}{M_{Z'}g^2_{\rm SM}}
   \frac{1-a_{qb}}{\sqrt{r^\text{VLL}(1+a_{qb}^2)+2 a_{qb}r^\text{LR}(B_q)}},
\ee

Evidently these general relations involve more free parameters than in 
the scenarios considered in our paper but they could turn out to be useful in 
concrete $Z'$ models and models with tree-level FCNC's mediated by $Z$ boson.

\subsection{Implications}
Inspecting these formulae we observe that
if the SM prediction for $\varepsilon_K$ is very close to its
experimental value $\Delta S(K)$ cannot be large
and consequently at first sight
the shifts $\Delta X_{L,R}(K)$ and $\Delta Y_{A}(K)$
cannot be large implying
suppressed NP contributions
to rare $K$ decays unless $Z'$ couplings to neutrinos and charged leptons in the final state are enhanced. The details depend on the value of $M_{Z'}$.
However, as we will find below, the present theoretical and parametric
uncertainties in $\varepsilon_K$ and $\Delta M_K$ still allow for large
effects in rare $K$ decays both in S1 and S2 scenarios.

Similarly in the $B_d$ and $B_s$ systems if the SM
predictions for $\Delta M_{s,d}$, $S_{\psi K_S}$ and $S_{\psi\phi}$ are very
close to the data, it is unlikely that large NP contributions to rare
$B_d$ and $B_s$ decays, in particular the asymmetries $S^{s,d}_{\mu^+\mu^-}$, will be found, unless again
$Z'$ couplings to neutrinos and charged leptons in
the final state are enhanced. Here the situation concerning theoretical
and parametric uncertainties is better than in the $K$ system and
the presence of several additional constraints from $b\to s$ transitions
allows to reach in the $B_s$ system clear cut conclusions.

In this context it is fortunate that within the SM there appears  to be a
tension between the values of $\varepsilon_K$ and
$S_{\psi K_S}$ so that some action from
NP is required. Moreover, parallel to this tension,
the values of $\vub$
extracted from inclusive and exclusive decays differ significantly
from each other. For a recent review see
\cite{Ricciardi:2012pf}.

If one does not average the inclusive and exclusive values of $\vub$ and
takes into account the tensions mentioned above, one is lead naturally to
two scenarios for NP:
 \begin{itemize}
\item
{\bf Exclusive (small) $\vub$ Scenario 1:}
$|\varepsilon_K|$ is smaller than its experimental determination,
while $S_{\psi K_S}$ is rather close to the central experimental value.
\item
{\bf Inclusive (large) $\vub$ Scenario 2:}
$|\varepsilon_K|$ is consistent with its experimental determination,
while $S_{\psi K_S}$ is significantly higher than its  experimental value.
\end{itemize}

Thus dependently which scenario is considered we need either
{\it constructive} NP contributions to $|\varepsilon_K|$
(Scenario 1)
or {\it destructive} NP contributions to  $S_{\psi K_S}$ (Scenario 2).
However this  NP should not spoil the agreement with the data
for $S_{\psi K_S}$ (Scenario 1) and for $|\varepsilon_K|$ (Scenario 2).

While introducing these two scenarios, one should emphasize the following difference between them.
In Scenario 1, the central value of $|\varepsilon_K|$ is visibly smaller than
the very precise data  but the still  significant parametric uncertainty
due to $\vcb^4$ dependence in $|\varepsilon_K|$ and a large uncertainty
in the charm contribution found at the NNLO level in \cite{Brod:2011ty}
does not make this problem as pronounced as this is the case of
Scenario 2, where large $\vub$ implies definitely a value of $S_{\psi K_S}$
that is by $3\sigma$ above the data.

Our previous discussion allows to expect larger NP effects in rare
$B_d$ decays in scenario S2 than in S1. This will be indeed confirmed
by our numerical analysis. In the $K$ system one would expect larger NP
effects in scenario S1 than S2 but the present uncertainties in
$\varepsilon_K$ and $\Delta M_K$ do not allow to see this clearly.
The $B_s$ system is not affected by the choice of these scenarios and in fact
our results in S1 and S2 are basically indistinguishable from each other
as long as there is no correlation with the $B_d$ system. However, we
will demonstrate that the imposition of $U(2)^3$ symmetry on $Z'$ couplings
will introduce such correlation with interesting implications for the $B_s$
system.

We do not include $B⁺\to\tau^+\nu_\tau$ in this discussion as NP related
to this decay has nothing to do with $Z'$. Moreover, the disagreement
of the data with the SM in this case softened significantly with the new
data from Belle Collaboration \cite{BelleICHEP}. The
new world average provided by the UTfit collaboration of
$\mathcal{B}(B^+ \to \tau^+ \nu)_{\rm exp} = (0.99 \pm 0.25) \times 10^{-4}~$
\cite{Tarantino:2012mq}
is in perfect agreement with the SM in scenario S2 and only by $1.5\sigma$
above the SM value in scenario S1.

Evidently $\vub$ could be some average between the inclusive and exclusive
values, in which significant NP effects will be in principle allowed
simultaneously in $K$ and $B_d$ decays. This discussion shows how important
is the determination of the value of $\vub$.

As already remarked above, the case of $B_s$ mesons is different as the
$B^0_s-\bar B_s^0$ system is not involved in the tensions discussed above.
Here the visible deviation of the $\Delta M_s$ in the SM from the data and the
asymmetry $S_{\psi\phi}$, still being not  accurately measured, govern the possible size of NP contributions in rare decays.

With this general picture in mind we can now proceed to numerical analysis.

\boldmath
\section{Strategy for Numerical Analysis}\label{sec:4}
\unboldmath
\subsection{Preliminaries}
Similarly to our analysis in \cite{Buras:2012xx} it is not the goal of the next  section to present a full-fledged numerical
analysis of all correlations including present theoretical, parametric and experimental
uncertainties as this would only wash out the effects we want to emphasize.
Yet, these uncertainties will be significantly
reduced in the coming years \cite{Antonelli:2009ws,Bediaga:2012py} and it is of interest to ask how
 the $Z'$ scenarios considered here
 would face precision flavour data and the reduction
of hadronic and CKM uncertainties. In this respect as emphasized above
correlations between
various observables are very important and we would like to exhibit these
correlations by assuming reduced uncertainties in question. This strategy
will also be used for the case of flavour violating $Z$-couplings.

Therefore, in our numerical analysis we will  choose
as nominal values for three out of four CKM parameters:\vspace{1ex}
\be\label{fixed}
\vus=0.2252, \qquad \vcb=0.0406, \qquad \gamma=68^\circ,
\ee
and instead of taking into account their uncertainties directly, we will
take them effectively at a reduced level by increasing the experimental
uncertainties in $\Delta M_{s,d}$ and $\varepsilon_K$.
Here the values for
 $|V_{us}|$ and  $|V_{cb}|$ have been measured
 in tree level
decays. The value for $\gamma$ is consistent with CKM fits and as the
ratio $\Delta M_d/\Delta M_s$ in the SM agrees well with the data, this
choice is a legitimate one.
Other inputs are collected in
Table~\ref{tab:input}. For $\vub$ we will use as two values
\be\label{Vubrange}
|V_{ub}|=3.1\cdot 10^{-3}\qquad |V_{ub}|=4.0\cdot 10^{-3}
\ee
corresponding to central values of exclusive and inclusive determinations of
this CKM element
and representing thereby S1 and S2 scenarios, respectively.

\begin{table}[!tb]
\center{\begin{tabular}{|l|l|}
\hline
$G_F = 1.16637(1)\times 10^{-5}\gev^{-2}$\hfill\cite{Nakamura:2010zzi} 	&  $m_{B_d}= 5279.5(3)\mev$\hfill\cite{Nakamura:2010zzi}\\
$M_W = 80.385(15) \gev$\hfill\cite{Nakamura:2010zzi}  								&	$m_{B_s} =
5366.3(6)\mev$\hfill\cite{Nakamura:2010zzi}\\
$\sin^2\theta_W = 0.23116(13)$\hfill\cite{Nakamura:2010zzi} 				& 	$F_{B_d} =
(190.6\pm4.6)\mev$\hfill\cite{Laiho:2009eu}\\
$\alpha(M_Z) = 1/127.9$\hfill\cite{Nakamura:2010zzi}									& 	$F_{B_s} =
(227.7\pm6.2)\mev$\hfill\cite{Laiho:2009eu}\\
$\alpha_s(M_Z)= 0.1184(7) $\hfill\cite{Nakamura:2010zzi}								&  $\hat B_{B_d} =
1.26(11)$\hfill\cite{Laiho:2009eu}\\\cline{1-1}
$m_u(2\gev)=(2.1\pm0.1)\mev $ 	\hfill\cite{Laiho:2009eu}						&  $\hat B_{B_s} =
1.33(6)$\hfill\cite{Laiho:2009eu}\\
$m_d(2\gev)=(4.73\pm0.12)\mev$	\hfill\cite{Laiho:2009eu}							& $\hat B_{B_s}/\hat B_{B_d}
= 1.05(7)$ \hfill \cite{Laiho:2009eu} \\
$m_s(2\gev)=(93.4\pm1.1) \mev$	\hfill\cite{Laiho:2009eu}				&
$F_{B_d} \sqrt{\hat
B_{B_d}} = 226(13)\mev$\hfill\cite{Laiho:2009eu} \\
$m_c(m_c) = (1.279\pm 0.013) \gev$ \hfill\cite{Chetyrkin:2009fv}					&
$F_{B_s} \sqrt{\hat B_{B_s}} =
279(13)\mev$\hfill\cite{Laiho:2009eu} \\
$m_b(m_b)=4.19^{+0.18}_{-0.06}\gev$\hfill\cite{Nakamura:2010zzi} 			& $\xi =
1.237(32)$\hfill\cite{Laiho:2009eu}
\\
$m_t(m_t) = 163(1)\gev$\hfill\cite{Laiho:2009eu,Allison:2008xk} &  $\eta_B=0.55(1)$\hfill\cite{Buras:1990fn,Urban:1997gw}  \\
$M_t=172.9\pm0.6\pm0.9 \gev$\hfill\cite{Nakamura:2010zzi} 						&  $\Delta M_d = 0.507(4)
\,\text{ps}^{-1}$\hfill\cite{Nakamura:2010zzi}\\\cline{1-1}
$m_K= 497.614(24)\mev$	\hfill\cite{Nakamura:2010zzi}								&  $\Delta M_s = 17.73(5)
\,\text{ps}^{-1}$\hfill\cite{Abulencia:2006ze,Aaij:2011qx}\\	
$F_K = 156.1(11)\mev$\hfill\cite{Laiho:2009eu}												&
$S_{\psi K_S}= 0.679(20)$\hfill\cite{Nakamura:2010zzi}\\
$\hat B_K= 0.767(10)$\hfill\cite{Laiho:2009eu}												&
$S_{\psi\phi}= 0.0002\pm 0.087$\hfill\cite{Clarke:1429149}\\\cline{2-2}
$\kappa_\epsilon=0.94(2)$\hfill\cite{Buras:2008nn,Buras:2010pza}										&
$\mathcal{B}(B^+\to\tau^+\nu)=(1.64\pm0.34)\times10^{-4}$\hfill\cite{Nakamura:2010zzi}\\	
$\eta_1=1.87(76)$\hfill\cite{Brod:2011ty}												
	& $\tau_{B^\pm}=(1641\pm8)\times10^{-3}\,\text{ps}$\hfill\cite{Nakamura:2010zzi} \\\cline{2-2}		
$\eta_2=0.5765(65)$\hfill\cite{Buras:1990fn}												
&$|V_{us}|=0.2252(9)$\hfill\cite{Nakamura:2010zzi}\\
$\eta_3= 0.496(47)$\hfill\cite{Brod:2010mj}												
& $|V_{cb}|=(40.6\pm1.3)\times
10^{-3}$\hfill\cite{Nakamura:2010zzi}\\
$\Delta M_K= 0.5292(9)\times 10^{-2} \,\text{ps}^{-1}$\hfill\cite{Nakamura:2010zzi}	&
$|V^\text{incl.}_{ub}|=(4.27\pm0.38)\times10^{-3}$\hfill\cite{Nakamura:2010zzi}\\
$|\eps_K|= 2.228(11)\times 10^{-3}$\hfill\cite{Nakamura:2010zzi}					&
$|V^\text{excl.}_{ub}|=(3.12\pm0.26)\times10^{-3}$\hfill\cite{Laiho:2009eu}	\\
\hline
\end{tabular}  }
\caption {\textit{Values of the experimental and theoretical
    quantities used as input parameters.}}
\label{tab:input}~\\[-2mm]\hrule
\end{table}

Having fixed the three parameters of the CKM matrix to the values in (\ref{fixed}), for a given $\vub$  the {``true''} values
of the angle $\beta$  and
of the element $\vtd$
are  obtained from the unitarity of the CKM matrix:
\begin{equation} \label{eq:Rt_beta}
\vtd=\vus \vcb R_t,\quad
R_t=\sqrt{1+R_b^2-2 R_b\cos\gamma} ~,\quad
\cot\beta=\frac{1-R_b\cos\gamma}{R_b\sin\gamma}~,
\end{equation}
where
\be\label{Rb}
 R_b=\left(1-\frac{\lambda^2}{2}\right)\frac{1}{\lambda}\frac{|V_{ub}|}{\vcb}.
\ee

\begin{table}[!tb]
\centering
\begin{tabular}{|c||c|c|c|}
\hline
 & Scenario 1: & Scenario 2:   & Experiment\\
\hline
\hline
  \parbox[0pt][1.6em][c]{0cm}{} $|\varepsilon_K|$ & $1.72(22)  \cdot 10^{-3}$  & $2.15(32)\cdot 10^{-3}$ &$ 2.228(11)\times 10^{-3}$ \\
 \parbox[0pt][1.6em][c]{0cm}{}$(\sin2\beta)_\text{true}$ & 0.623(25) &0.770(23)  & $0.679(20)$\\
 \parbox[0pt][1.6em][c]{0cm}{}$\Delta M_s\, [\text{ps}^{-1}]$ &19.0(21)&  19.0(21) &$17.73(5)$ \\
 \parbox[0pt][1.6em][c]{0cm}{} $\Delta M_d\, [\text{ps}^{-1}]$ &0.56(6) &0.56(6)   &  $0.507(4)$\\
\parbox[0pt][1.6em][c]{0cm}{}$\mathcal{B}(B^+\to \tau^+\nu_\tau)$&  $0.62(14) \cdot 10^{-4}$&$1.02(20)\cdot 10^{-4}$ & $0.99(25) \times
10^{-4}$\\
\hline
\end{tabular}
\caption{\it SM prediction for various observables for  $|V_{ub}|=3.1\cdot 10^{-3}$ and $|V_{ub}|=4.0\cdot 10^{-3}$ and $\gamma =
68^\circ$ compared to experiment.
}\label{tab:SMpred}~\\[-2mm]\hrule
\end{table}

In Table~\ref{tab:SMpred} we
summarize for completeness the SM results for $|\varepsilon_K|$,  $\Delta M_{s,d}$,
$\left(\sin 2\beta\right)_\text{true}$ and $\mathcal{B}(B^+\to \tau^+\nu_\tau)$, obtained from (\ref{eq:Rt_beta}),
setting
$\gamma = 68^\circ$ and  choosing the two values for $\vub$ in (\ref{Vubrange}).
We observe that for both choices of $\vub$ the data show significant deviations from the SM predictions but
the character of the NP which could cure these tensions depends on the
choice of $\vub$ as already discussed in detail in \cite{Buras:2012ts} and in
the previous section.

What is
striking in this table is that
the predicted central values of $\Delta M_s$  and $\Delta M_d$, although
slightly above the data,  are both in  good agreement with the latter
when hadronic uncertainties are taken into account. In particular
the central value of the ratio $\Delta M_s/\Delta M_d$ is
 very close to  the data:
\be\label{Ratio}
\left(\frac{\Delta M_s}{\Delta M_d}\right)_{\rm SM}= 34.5\pm 3.0\qquad {\rm exp:~~ 35.0\pm 0.3}\,.
\ee
These results depend on the lattice input and in the case
of $\Delta M_d$ on the value of $\gamma$. Therefore to get a better insight
both lattice input and the tree level determination of $\gamma$
have to improve.

In \cite{Buras:2012xx} we have analyzed a particular 331 model, the so-called
$\model$ model. Because of suppressed contributions to $\varepsilon_K$, this model favoured the inclusive value of $\vub$. Moreover only left-handed couplings
of $Z'$ to quarks were present. As already described in Section~\ref{sec:2}
the present analysis can be considered as the generalization of \cite{Buras:2012xx}
to include also exclusive values of $\vub$ and the right-handed couplings
of $Z'$ to quarks. Thus with two scenarios for $\vub$ and four scenarios LHS, RHS, LRS, ALRS
for flavour violating couplings of $Z'$ to quarks we are led
to eight scenarios of $Z'$-physics to be denoted by
\be\label{generalS}
{\rm LHS1,\quad LHS2,\quad RHS1,\quad RHS2,\quad LRS1, \quad
LRS2, \quad ALRS1, \quad ALRS2}
\ee
with S1 and S2 indicating the $\vub$ scenarios.

We should emphasize that in each case we have only two
free parameters describing
the $Z'$-quark couplings in each meson system except for the universal
$M_{Z'}$. Therefore, as in the case of the $\model$ model it is possible
to determine these couplings from flavour observables (see Section~\ref{sec:2})
provided flavour conserving $Z'$ couplings to neutrinos and muons are known.
This was the case of the $\model$ model. Here these couplings are not fixed
by the theory and have to be determined in purely leptonic processes. In principle one could also get some insight about them from semi-leptonic meson decays
but determining them in purely leptonic processes increases the predictive
power of the theory.

Following Step 2 of our general strategy of Section~\ref{sec:2}, in what follows we will assume that
$\Delta_A^{\mu\bar\mu}(Z')$ and $\Delta_L^{\nu\bar\nu}(Z')$ have been determined
in purely leptonic processes. For definiteness
we set the lepton couplings
at the following values
\be\label{leptonicset}
\Delta_L^{\nu\bar\nu}(Z')=0.5, \qquad
\Delta_A^{\mu\bar\mu}(Z')=0.5,
\ee
to be compared with $\Delta_L^{\nu\bar\nu}(Z')=0.14$ and
$\Delta_A^{\mu\bar\mu}(Z')=-0.26$ in the $\model$ model \cite{Buras:2012xx}.
In the SM both couplings of $Z$ are equal to $0.372$.

The  specification of signs in (\ref{leptonicset}) is crucial for the identification of various enhancements
and suppressions with respect to SM branching ratios and CP asymmetries and
is at the basis of our search for successful oases in the space of parameters.
If these signs will be identified in the future to be different from the ones
assumed here, it will be straightforward to find out by inspecting our 
results how the landscape of
oases changes for each of the four possibilities for the signs of leptonic
couplings.

\boldmath
\subsection{Dependence on $M_{Z'}$}
\unboldmath
The correlations between $\Delta F=1$ and $\Delta F=2$ derived in subsection \ref{CORR} imply that when free NP parameters have been bounded by $\Delta F=2$ 
constraints, the modifications of $\Delta X_i$ and $\Delta Y_i$ are {\it inversely } proportional to  $M_{Z'}$. This means that in the case of NP contributions 
significantly smaller than the SM contributions, the modifications of rare 
decay branching ratios due to NP will be governed by the interference of 
SM and NP contributions and consequently will also be inversely  proportional to  $M_{Z'}$. This is the case of all observables in $B_s$ and $B_d$ systems, 
but not in $K$ system where NP contributions could be much larger than the 
SM contribution for sufficiently low values of  $M_{Z'}$. In the latter case 
the NP modifications of branching ratios will decrease faster than $1/M_{Z'}$ 
($1/M^2_{Z'}$ in the limit of full NP dominance) until NP contributions 
are sufficiently small so that the  $1/M_{Z'}$ dependence is again valid.

Concerning the direct lower bound  on  $M_{Z'}$ from collider experiments, the 
most stringent bounds are provided by CMS experiment \cite{Chatrchyan:2012it}. 
The precise value depends 
on the model considered. While for the so-called sequential $Z'$ the lower 
bound for $M_{Z'}$ is in the ballpark of $2.5\tev$, in other models values 
as low as $1\tev$ are still possible. In order to cover large set of models, 
we will choose as our nominal value $M_{Z'}=1\tev$. With the help of 
the formulae in subsection \ref{CORR} it should be possible to 
estimate approximately, how our results would change for 
$1\tev\le M_{Z'}\le 3\tev$. For much larger values of $M_{Z'}$, considered 
mainly in $K$ physics, explicit results will be provided.

\subsection{Simplified Analysis}
As in \cite{Buras:2012xx} we will perform a simplified analysis of $\varepsilon_K$,
$\Delta M_{d,s}$, $S_{\psi K_S}$
and $S_{\psi\phi}$
in order to identify oases in the space of new parameters (see Section~\ref{sec:2})
for which these five observables are consistent with experiment.
To this end we set all other input parameters at their central values
but in order to take partially hadronic
and experimental uncertainties into account we require the theory in
each of the eight scenarios in (\ref{generalS})
to reproduce the data for $\varepsilon_K$ within $\pm 10\%$, $\Delta M_{s,d}$ within $\pm 5\%$ and the
data on $S_{\psi K_S}$ and $S_{\psi\phi}$ within experimental
$2\sigma$. We choose larger uncertainty for  $\varepsilon_K$  than  $\Delta M_{s,d}$ because of its strong $\vcb^4$ dependence. For $\Delta M_K$ we will
only require the agreement within $\pm 25\%$ because of potential long
distance uncertainties.

Specifically, our search is governed by the following allowed ranges:
\be\label{C1}
16.9/{\rm ps}\le \Delta M_s\le 18.7/{\rm ps},
\quad  -0.18\le S_{\psi\phi}\le 0.18,
\ee
\be\label{C2}
0.48/{\rm ps}\le \Delta M_d\le 0.53/{\rm ps},\quad
0.64\le S_{\psi K_S}\le 0.72 .
\ee

\be\label{C3}
0.75\le \frac{\Delta M_K}{(\Delta M_K)_{\rm SM}}\le 1.25,\qquad
2.0\times 10^{-3}\le |\varepsilon_K|\le 2.5 \times 10^{-3}.
\ee

The search for these oases in each of the scenarios in (\ref{generalS})
is simplified by the fact that for fixed $M_{Z'}$  each of the pairs
$(\Delta M_s,S_{\psi\phi})$,
$(\Delta M_d,S_{\psi K_S})$  and $(\Delta M_K,|\varepsilon_K|)$ depend only
on two variables. The fact that in the $K$ system we have only one
powerful constraint at present is rather unfortunate. The situation will
improve by much when the branching ratios for $\kpn$ and $\klpn$ will be
measured.

In what follows we will first for each scenario identify the allowed
oases. As in the case of the $\model$ model there will be several oases
allowed by the constraints in (\ref{C1})-(\ref{C3}) and we will have to
invoke other observables, which are experimentally only weakly bounded
at present in order to find the optimal oasis in each case. Yet, our plots
will show that once these observables will be measured precisely one day
not only unique oasis in the parameter space will be identified but the
specific correlations in this oasis will provide a powerful test of the
$Z'$ scenarios.

As in \cite{Buras:2012xx}, inspecting the expressions for various observables
in different oases, we
have identified the fastest route to the optimal oasis in
each scenario. We will describe this
route in each case below.
We will also see
how the correlations between various observables can give additional tests
once the analysis is confined to a particular oasis.

It turns out that considerable progress in the search for the optimal oasis in
each scenario can be made by identifying some special observables for whom
the sign of departure from SM expectations is sufficient to identify this
oasis uniquely. For instance in the case of LHS scenario, in the 
$B_d$ and $B_s$ meson systems these special observables turn out to be
\be
\mathcal{B}(B_d\to\mu^+\mu^-), \qquad S_{\mu^+\mu^-}^s,
\ee
respectively.
Already the sign of shifts of
them with respect to the SM values allows to make significant
progress towards the identification of  the optimal oasis in each scenario
considered. However, in contrast to LHS scenario considered in
\cite{Buras:2012xx}, in the presence of RH currents the two observables
will not be sufficient to identify optimal oasis. As already
advertised, in the case of the $B_s$ system, the rescue will come from
$B\to K^*\mu^+\mu^-$, $B\to K\mu^+\mu^-$ and
$b\to s\nu\bar\nu$ transitions.

\section{An Excursion through $Z'$ Scenarios}\label{sec:Excursion}
\subsection{The LHS1 and LHS2 Scenarios}
\boldmath
\subsubsection{The $B_s$ Meson System}
\unboldmath
We begin the search for the oases with the $B_s$ system as here the choice
of $\vub$ is immaterial and the results for LHS1 and LHS2 scenarios are almost identical. Basically only the asymmetry $S_{\psi\phi}$ within the
SM and $\vts$ are slightly modified because of the unitarity of the CKM
matrix. But this changes  $S_{\psi\phi}$ in the SM from $0.032$ to  $0.042$
and can be neglected.

\begin{figure}[!tb]
\begin{center}
\includegraphics[width=0.45\textwidth] {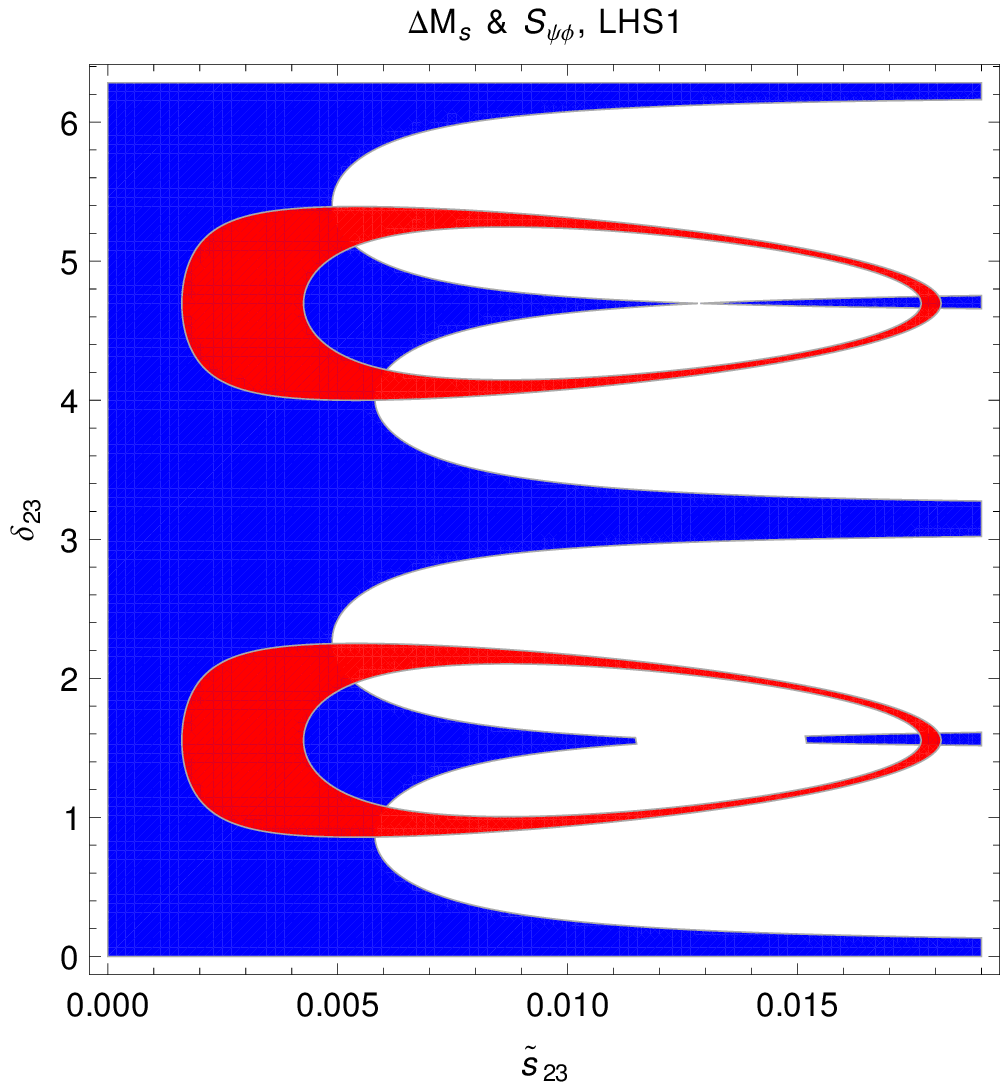}
\caption{\it  Ranges for $\Delta M_s$ (red region) and $S_{\psi \phi}$ (blue region) for $M_{Z^\prime}=1$~TeV in LHS1 satisfying the bounds
in Eq.~(\ref{C1}).
}\label{fig:oasesBsLHS1}~\\[-2mm]\hrule
\end{center}
\end{figure}

The result of this search for $M_{Z'}=1\tev$ is shown in Fig.~\ref{fig:oasesBsLHS1},
where we show the allowed ranges
for $(\tilde s_{23},\delta_{23})$.
The {\it red} regions correspond to the allowed ranges for $\Delta M_{s}$,
while the {\it blue} ones to the corresponding ranges for  $S_{\psi\phi}$. The overlap between red and blue regions identifies the
oases we were looking for. We observe that the requirement of suppression
of $\Delta M_s$ implies $\tilde s_{23}\not=0$.

From these plots we extract several oases that are collected in
Table~\ref{s23oases}. We denote by
$A_i(S1)$ and $A_i(S2)$ the oases found for the two values of $\vub$ but
as in the $B_s$-case there is no change in these oases when moving from S1 to S2  we will
show the results only for LHS1
scenario.
We observe the following pattern:
\begin{itemize}
\item
For each oasis with a given $\delta_{23}$ there is another oasis with $\delta_{23}$ shifted
by $180^\circ$ but the range for  $\tilde s_{23}$ is unchanged. This discrete
ambiguity results from the fact that $\Delta M_s$ and $S_{\psi\phi}$ are
governed by  $2\delta_{23}$. However, as already seen in Table~\ref{s23oases}
and discussed below this ambiguity can be resolved by other observables.
In this context we just investigate whether in a given oasis various
branching ratios are enhanced or suppressed with respect to the SM
or CP asymmetries modified.
In the case of $S^s_{\mu^+\mu^-}$, that vanishes within the SM, we just look
at its sign. In the last two columns of Table~\ref{s23oases} we consider
\be
\Delta\mathcal{B}^{\mu^+\mu^-}_s\equiv\Delta\mathcal{B}(B_s\to\mu^+\mu^-), \qquad
\Delta\mathcal{B}^{\nu\bar\nu}_s\equiv
\Delta\mathcal{B}(B\to X_s \nu\bar\nu)\,.
\ee
\item
The oases with $i=2,4$ are very small and imply very concrete predictions
for various observables. In fact as we will soon see they are already
ruled out by the present data on $\mathcal{B}(B_s\to\mu^+\mu^-)$.  They correspond roughly to NP contribution to
$M^s_{12}$ twice as large as the SM one but carrying opposite sign.
\item
The increase of $M_{Z^\prime}$ by a given factor allows to increase
 $\tilde s_{23}$ by the same factor. This structure is evident from the formulae
for $\Delta S(B_s)$. However, the inspection of the formulae for
$\Delta F=1$ transitions shows that this
change will have impact on rare decays, making the NP
effects in them with increased  $M_{Z^\prime}$ smaller. This is evident
from the correlations derived in Section~\ref{sec:3a} and has been
emphasized at the beginning of our paper.
\end{itemize}

We will next confine our
numerical analysis to these oases, investigating whether some of them can be
excluded by other constraints and studying correlations between various
observables. To this end we set the lepton couplings as given in
(\ref{leptonicset}).

 As a final comment, we observe that the oases reported in Table ~\ref{s23oases} and all other tables for other scenarios below actually describe squares in
the spaces $(\tilde s_{23},\delta_{23})$, $(\tilde s_{13},\delta_{13})$ and
 $(\tilde s_{12},\delta_{12})$,  while the corresponding regions in Fig.~\ref{fig:oasesBsLHS1} and analogous figures for other scenarios
have more complicated shapes. Indeed, in our numerical analysis of the various observables we have varied
the parameters in the {\it true} oases,  requiring that constraints
(\ref{C1})-(\ref{C3}) are satisfied.

\begin{table}[!tb]
\centering
\begin{tabular}{|c||c|c|c|c|c|c|}
\hline
 & $\tilde s_{23}$ & $\delta_{23}$   & $S^s_{\mu^+\mu^-}$ & $\Delta S_{\psi\phi}$ &
$\Delta\mathcal{B}^{\mu^+\mu^-}_s$ &   $\Delta\mathcal{B}^{\nu\bar\nu}_s$  \\
\hline
\hline
  \parbox[0pt][1.6em][c]{0cm}{} $A_1(S1)$ & $0.0016-0.0061$
& $49^\circ-129^\circ$ &  $+$ & $\pm$ & $\mp$ & $\mp$ \\
 \parbox[0pt][1.6em][c]{0cm}{}$A_2(S1)$&  $0.0176-0.0181$&$87^\circ-92^\circ$ & &&&\\
\parbox[0pt][1.6em][c]{0cm}{} $A_3(S1)$ & $0.0016-0.0061$
& $229^\circ-309^\circ$ &  $-$   & $\pm$&$\pm$ & $\pm$ \\
 \parbox[0pt][1.6em][c]{0cm}{}$A_4(S1)$&  $0.0176-0.0181$&$267^\circ-272^\circ$ & & && \\
\hline
\end{tabular}
\caption{\it Oases in the space $(\tilde s_{23},\delta_{23})$ for $M_{Z^\prime}=1\tev$ in LHS1.
The sign of $S^s_{\mu^+\mu^-}$ chooses the oasis
uniquely. The same applies to the pair $S_{\psi\phi}$ and
$\Delta\mathcal{B}(B_s\to\mu^+\mu^-)$ as discussed in the text. Basically the
same results are obtained in LHS2.
}\label{s23oases}~\\[-2mm]\hrule
\end{table}

% LHS1
% Big        s23->[0.00164,0.00606] delta23->[0.859,2.249]
% Small    s23->[0.01764,0.01809] delta23->[1.519,1.606]
%
% LHS2
% Big        s23->[0.0018,0.0062] delta23->[0.864,2.229]
% Small    s23->[0.01765,0.01812] delta23->[1.517,1.604]

\begin{figure}[!tb]
\centering
\includegraphics[width = 0.45\textwidth]{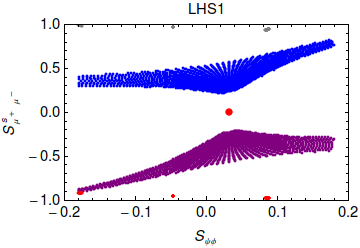}
\includegraphics[width = 0.45\textwidth]{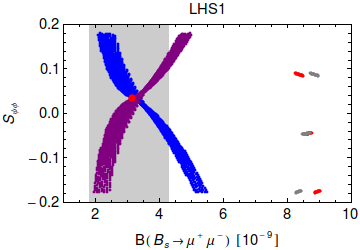}
\caption{\it$ S^s_{\mu^+\mu^-}$  versus $S_{\psi\phi}$ (left) and $S_{\psi\phi}$ versus $\mathcal{B}(B_s\to\mu^+\mu^-)$ (right) for
$M_{Z^\prime} = 1~$TeV in LHS1. $A_1$: blue, $A_3$: purple, $A_2$: red, $A_4$: gray. Gray region: exp 1$\sigma$ range  
$\mathcal{B}(B_s\to\mu^+\mu^-) = (2.9^{+1.4}_{-1.1})\cdot 10^{-9}$.  Red point: SM central value.}
 \label{fig:SmusvsSphiLHS1}~\\[-2mm]\hrule
\end{figure}

In Fig.~\ref{fig:SmusvsSphiLHS1} (left) we show $ S^s_{\mu^+\mu^-}$  vs $S_{\psi\phi}$. The
requirement of suppression of $\Delta M_s$ requires  $ S^s_{\mu^+\mu^-}$  to
be non-zero. A {\it positive} value of  $ S^s_{\mu^+\mu^-}$  chooses scenario $A_1$,
while a {\it negative} one scenario $A_3$. Note that in both scenarios the sign
of $S_{\psi\phi}$ is not fixed yet but it will be fixed by invoking
 $\mathcal{B}(B_s\to\mu^+\mu^-)$ below. We note
that in big oases for $M_{Z'}=1\tev$, $|S^s_{\mu^+\mu^-}|$ can reach values as high as $0.9$ when  $|S_{\psi\phi}|\approx 0.2$. Smaller values are found for larger  $M_{Z'}$. 
We also observe that the small oases represented by gray and red areas are indeed very small and imply $|S^s_{\mu^+\mu^-}|\approx 1$.

The fact that  $ S^s_{\mu^+\mu^-}$  is very
powerful in identifying the optimal oasis can be understood as follows.
 $S^s_{\mu^+\mu^-}$  is governed by the phase of the function $Y(B_s)$ that
originates in the $Z'$ contribution. It can distinguish between $A_1$ and
$A_3$ oasis because the new phase $\delta_{23}$  in these two oases differs by $180^\circ$
and consequently $\sin\delta_{23}$ relevant for this asymmetry differs by sign
in these two oases. Calculating the imaginary part of $Y(B_s)$ in
(\ref{YAB}) and taking into account that it is $\Delta_L^{sb}$ and not
$\Delta_L^{bs}$ that enters  $Y(B_s)$ one can convince oneself about the
definite sign of  $S^s_{\mu^+\mu^-}$ in $A_1$ and $A_3$ oases as stated
above.

The reason why  $\mathcal{B}(B_s\to\mu^+\mu^-)$ cannot be presently
powerful in the search for
oases is the significant experimental error  on $S_{\psi\phi}$ with which
this branching ratio is correlated. However, inspecting this correlation
in a given oasis
 constitutes
an important test of the model. We show this in Fig.~\ref{fig:SmusvsSphiLHS1} (right)\footnote{ The central values for
$\mathcal{B}(B_{d}\to\mu^+\mu^-)^{\rm SM}=1.0\times 10^{-10}$ and
$\mathcal{B}(B_{s}\to\mu^+\mu^-)^{\rm SM}=3.1\times 10^{-9}$ shown in the plots
correspond to fixed CKM parameters chosen by us and differ from the ones
listed in~(\ref{LHCb2}) and (\ref{LHCb3}) but are fully consistent
with them.}.
While in the
oasis $A_1$
$S_{\psi\phi}$ increases (decreases) uniquely with decreasing (increasing)
$\mathcal{B}(B_s\to\mu^+\mu^-)$, in the oasis $A_3$, the increase of
$S_{\psi\phi}$ implies uniquely an increase of $\mathcal{B}(B_s\to\mu^+\mu^-)$.
Therefore, while $\mathcal{B}(B_s\to\mu^+\mu^-)$ alone cannot uniquely
determine the optimal oasis, it can do in collaboration with $S_{\psi\phi}$.
Finding both these observables above or below their SM expectations,
would select the oasis $A_3$, while finding one of them enhanced and the
other suppressed (opposite sign in the case of $S_{\psi\phi}$) would
select $A_1$ as the optimal oasis. We indicate this pattern in
Table~\ref{s23oases}. In fact in the coming years it will be
$S_{\psi\phi}$ and $\mathcal{B}(B_s\to\mu^+\mu^-)$ which will be leading
this search as  $S^s_{\mu^+\mu^-}$ is much harder to measure.

If the favoured oasis will be found to  differ from the one found by means
of $S^s_{\mu^+\mu^-}$ one day
the
 model in question will be in trouble. Indeed, let us
assume that  $\mathcal{B}(B_s\to\mu^+\mu^-)$ will be found below its SM
value. Then the measurement of  $S_{\psi\phi}$ will uniquely tell us whether
$A_1$ or $A_3$ is the optimal scenario and consequently as seen in Fig.~\ref{fig:SmusvsSphiLHS1} (left) and
Table~\ref{s23oases}
we will be able to predict the sign of $S^s_{\mu^+\mu^-}$. Moreover, in the
case of $S^s_{\psi\phi}$ sufficiently different from zero, we will be able
to determine not only the sign but also the magnitude of  $S^s_{\mu^+\mu^-}$.

Probably the most important message from Fig.~\ref{fig:SmusvsSphiLHS1} (right) is the following one.
If NP is dominated by $Z'$ in the LHS1 scenario, then departure of
$S_{\psi\phi}$ from the SM implies automatically the departure of
 $\mathcal{B}(B_s\to\mu^+\mu^-)$ from its SM value and vice versa. Moreover, for
 $M_{Z'}=1\tev$  and $|S_{\psi\phi}|\approx 0.2$,
$\mathcal{B}(B_s\to\mu^+\mu^-)$ can
deviate from the SM value by
$\pm 60\%$. We also note that the small oases
are inconsistent with the LHCb data on  $\mathcal{B}(B_s\to\mu^+\mu^-)$ and
are already ruled out. Consequently  $|S^s_{\mu^+\mu^-}|\approx 1$
is also ruled out. Therefore we will omitt the results for small oases in
the subsequent plots for $B_s$ meson system.

\begin{figure}[!tb]
\centering
\includegraphics[width = 0.45\textwidth]{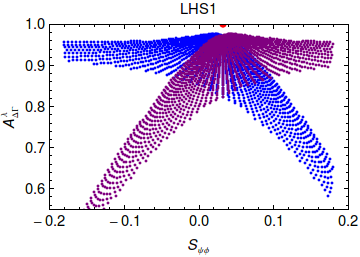}
\includegraphics[width = 0.45\textwidth]{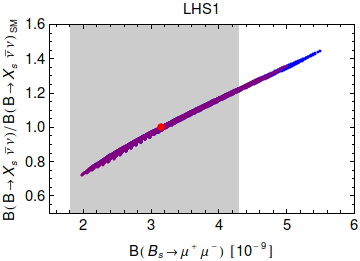}
\caption{\it $\mathcal{A}^\lambda_{\Delta\Gamma}$ versus $S_{\psi\phi}$ (left) and $\mathcal{B}(B\to X_s \nu\bar\nu)$ versus
$\mathcal{B}(B_s\to\mu^+\mu^-)$ (right) for
$M_{Z^\prime} = 1~$TeV in LHS1. $A_1$: blue, $A_3$: purple.  Gray region: exp 1$\sigma$ range  
$\mathcal{B}(B_s\to\mu^+\mu^-) = (2.9^{+1.4}_{-1.1})\cdot 10^{-9}$. Red point: SM central value.}
 \label{fig:AGammavsSphiLHS1}~\\[-2mm]\hrule
\end{figure}

In Fig.~\ref{fig:AGammavsSphiLHS1} (left) we plot $\mathcal{A}^\lambda_{\Delta\Gamma}$ vs  $S_{\psi\phi}$.
We observe that
for  $M_{Z'}=1\tev$  and $S_{\psi\phi}$ significantly different from
zero, $\mathcal{A}^\lambda_{\Delta\Gamma}$ can differ significantly from unity.
With $\mathcal{A}^\lambda_{\Delta\Gamma}$ as low as $0.6$ the
effect of $\Delta\Gamma_s$ on $\mathcal{B}(B_s\to\mu^+\mu^-)$  becomes
smaller.

In Fig.~\ref{fig:AGammavsSphiLHS1} (right) we show
$\mathcal{B}(B\to X_s \nu\bar\nu)$ vs
$\mathcal{B}(B_s\to\mu^+\mu^-)$. This correlation is valid in any oasis due
to the assumed equal sign of the leptonic couplings in (\ref{leptonicset}).
However, as seen in the plot the size of NP contribution may depend
on the oasis considered. We note that NP effects of $50\%$ are still
possible and suppression of   $\mathcal{B}(B_s\to\mu^+\mu^-)$ below the SM
value will also imply the suppression of $\mathcal{B}(B\to X_s \nu\bar\nu)$.
Yet, one should note that if the future data will disagree with this
pattern, the rescue could come from the flip of the signs in $\nu\bar\nu$
or $\mu^+\mu^-$ couplings provided this is allowed by leptonic decays of
$Z'$.

\begin{figure}[!tb]
\centering
\includegraphics[width = 0.45\textwidth]{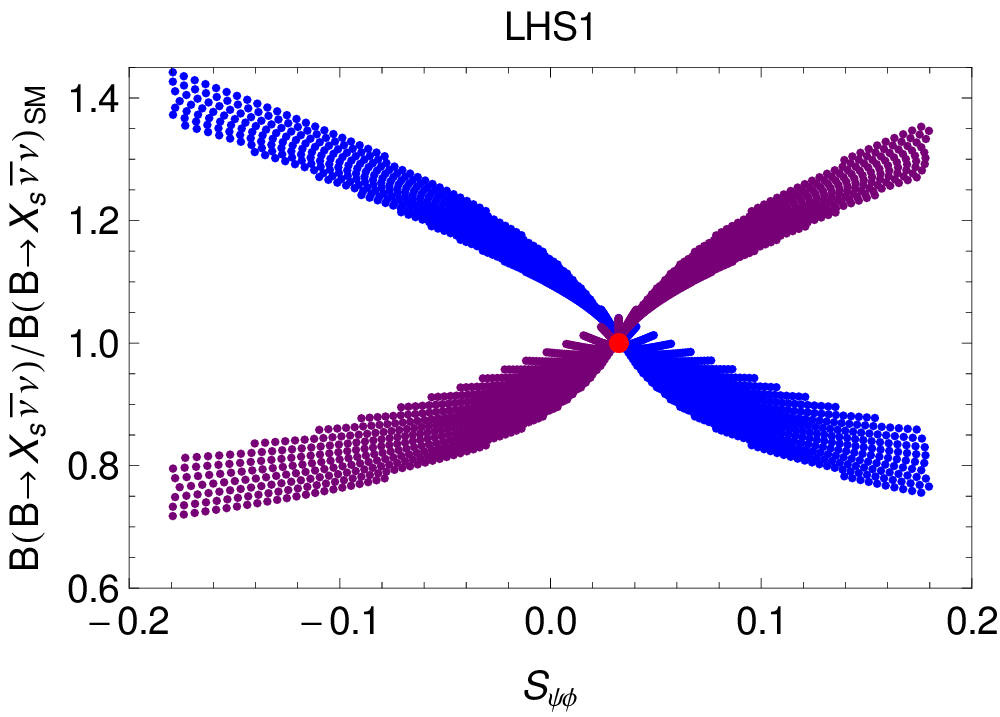}
\caption{\it $\mathcal{B}(B\to X_s \nu\bar\nu)$ versus
$\mathcal{B}(B_s\to\mu^+\mu^-)$ for
$M_{Z^\prime} = 1~$TeV in LHS1. $A_1$: blue, $A_3$: purple, $A_2$: red, $A_4$: gray. Red point: SM central value.}
 \label{fig:BXsnuvsSphiLHS1}~\\[-2mm]\hrule
\end{figure}

In Fig.~\ref{fig:BXsnuvsSphiLHS1} we show
$\mathcal{B}(B\to X_s \nu\bar\nu)$ vs $S_{\psi\phi}$ which could turn out
to be informative when $S_{\psi\phi}$ will be measured precisely one day.

\boldmath
\subsubsection{The $B_d$ Meson System}
\unboldmath

\begin{figure}[!tb]
\begin{center}
\includegraphics[width=0.45\textwidth] {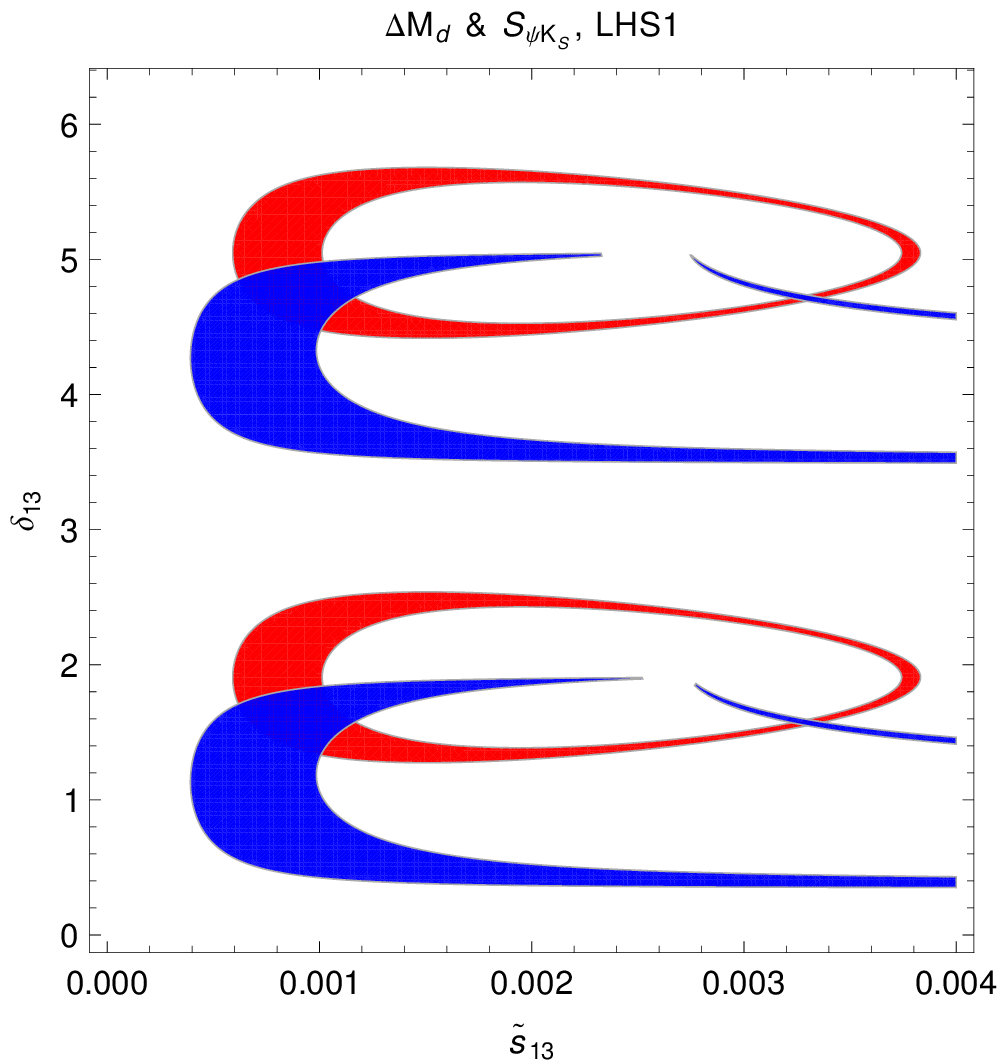}
\includegraphics[width=0.45\textwidth] {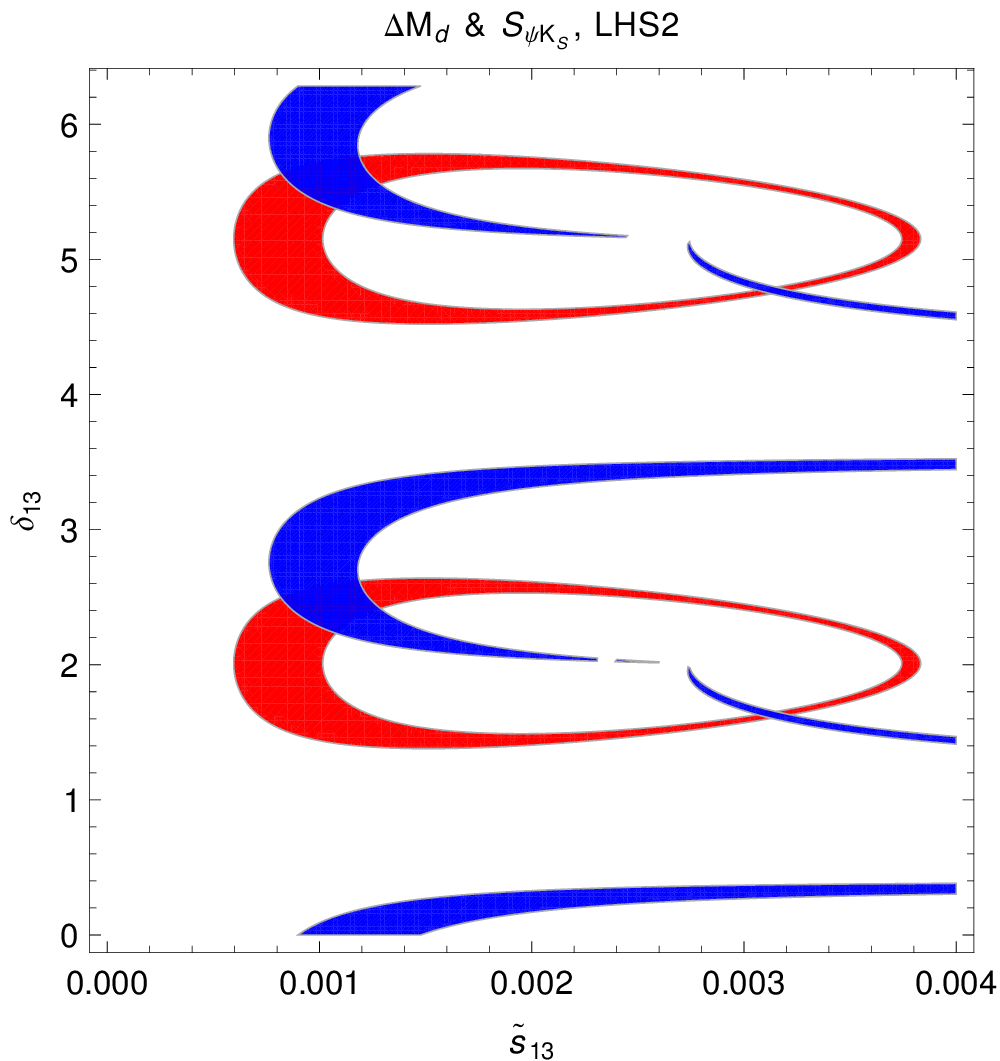}
\caption{\it  Ranges for $\Delta M_d$ (red region) and $S_{\psi K_S}$ (blue region) for $M_{Z^\prime}=1$ TeV in LHS1 (left) and
LHS2 (right) satisfying the bounds in Eq.~(\ref{C2}).
}\label{fig:oasesBdLHS}~\\[-2mm]\hrule
\end{center}
\end{figure}

We begin by searching for the allowed oases in this case. The result is shown
in Fig.~\ref{fig:oasesBdLHS} and Table~\ref{s13oases}. The general structure of the discrete
ambiguities is as in  Table~\ref{s23oases} but now as expected the selected
oases in S1 and S2 differ significantly from each other.

\begin{figure}[!tb]
\begin{center}
\includegraphics[width=0.45\textwidth] {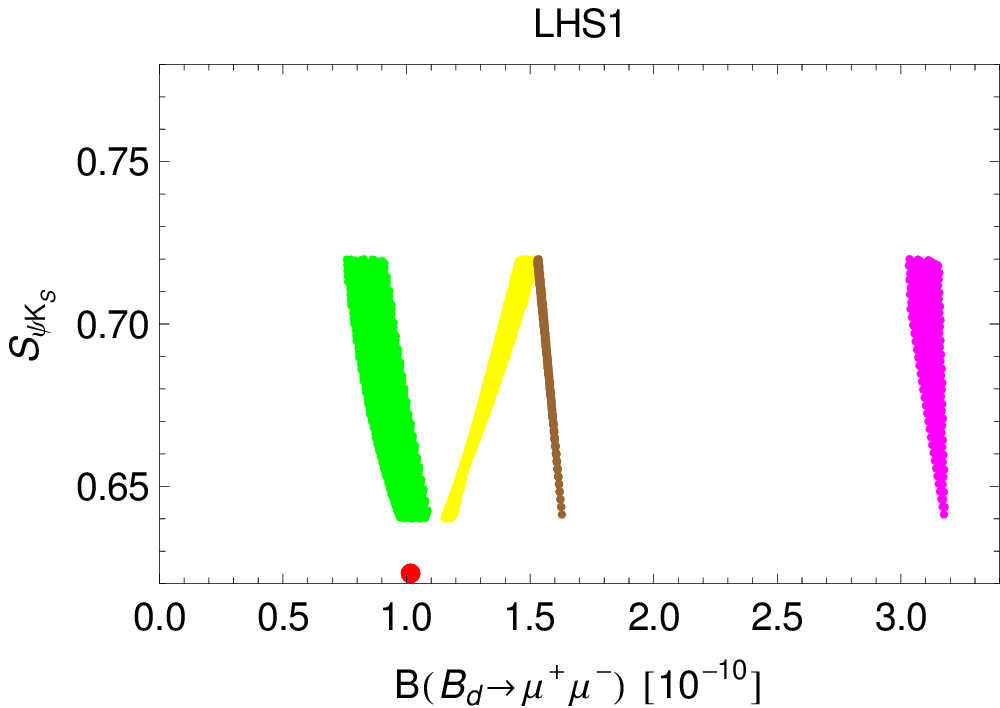}
\includegraphics[width=0.45\textwidth] {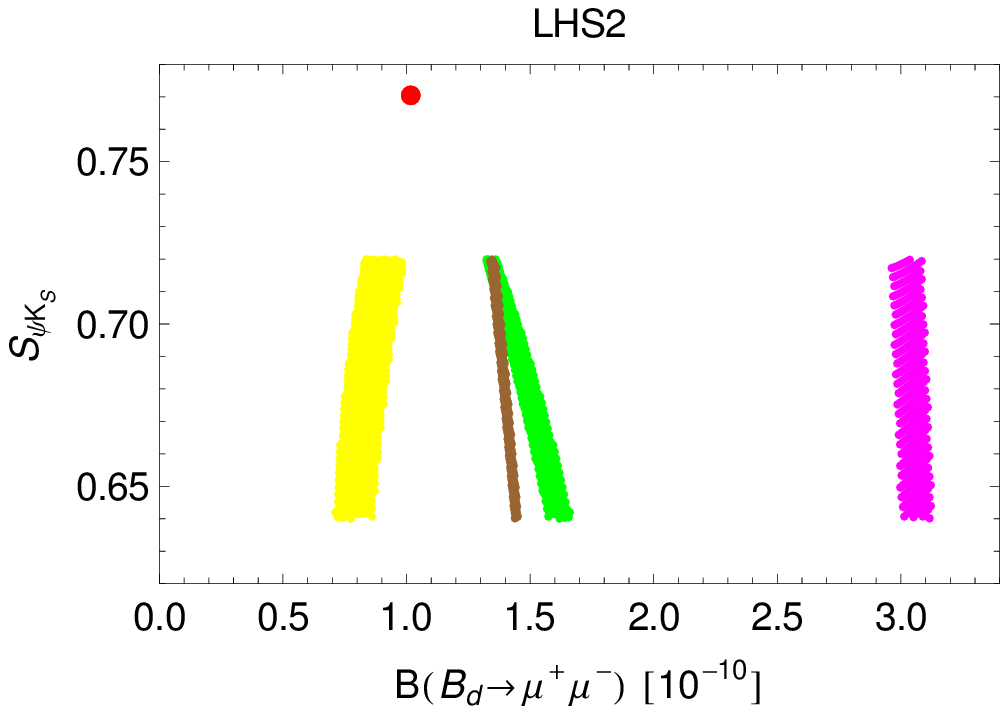}
\caption{\it  $S_{\psi K_S}$ versus   $\mathcal{B}(B_d\to\mu^+\mu^-)$  for $M_{Z^\prime}=1$ TeV in LHS1 (left) and
LHS2 (right). $B_1$: yellow, $B_3$: green, $B_2$: magenta, $B_4$: brown. Red point: SM central
value.}\label{fig:BdmuvsSKSLHS}~\\[-2mm]\hrule
\end{center}
\end{figure}

Let us first concentrate on S2 scenario that corresponds to the one already
analyzed in \cite{Buras:2012xx}.
In the right panel of Fig.~\ref{fig:BdmuvsSKSLHS} we show  $S_{\psi K_S}$ vs   $\mathcal{B}(B_d\to\mu^+\mu^-)$. The
requirement on $S_{\psi K_S}$ and $\Delta M_d$ forces
$\mathcal{B}(B_d\to\mu^+\mu^-)$ to differ from the SM value
but the sign of this
departure depends on the oasis considered. Here distinction is made
between $B_1$ and $B_3$ for which $\mathcal{B}(B_d\to\mu^+\mu^-)$ is
suppressed and enhanced with respect to the SM, respectively.
These enhancements and suppressions amount up to  $\pm 50\%$
for $M_{Z'}=1\tev$. They increase with decreasing  $S_{\psi K_S}$.

Note that because of the correlation between
$\mathcal{B}(B_d\to\mu^+\mu^-)$
and $S_{\psi K_S}$ and the fact that the latter is already
well determined, the range of $\delta_{13}$ cannot be large.
 $\mathcal{B}(B_d\to\mu^+\mu^-)$  can then distinguish between $B_1$ and
$B_3$ oases because $\cos\delta_{13}$ differs by sign in these two oases.
We find then destructive interference of $Z'$ contribution with the
SM contribution in oasis $B_1$ and constructive one in oasis $B_3$ implying
the results summarized in Table~\ref{s13oases} for this $\vub$ scenario.

\begin{table}[!tb]
\centering
\begin{tabular}{|c||c|c|c|c|}
\hline
 & $\tilde s_{13}$ & $\delta_{13}$   & $\Delta\mathcal{B}(B_d\to\mu^+\mu^-)$ &
$S^d_{\mu^+\mu^-}$\\
\hline
\hline
  \parbox[0pt][1.6em][c]{0cm}{} $B_1(S1)$ & $0.00062-0.00117$
& $76^\circ-105^\circ$ & $+(0)$ & $+$ \\
 \parbox[0pt][1.6em][c]{0cm}{}$B_2(S1)$&  $0.00322-0.00337$ &$89^\circ-91^\circ$ & & $+$ \\
\parbox[0pt][1.6em][c]{0cm}{} $B_3(S1)$ & $0.00062-0.00117$
& $256^\circ-285^\circ$ & $-(0)$ & $-$ \\
\parbox[0pt][1.6em][c]{0cm}{}$B_4(S1)$&  $0.00322-0.00337  $&$269^\circ-271^\circ$ & & $-$\\
\hline
\hline
  \parbox[0pt][1.6em][c]{0cm}{} $B_1(S2)$ & $0.00081-0.00128$
& $128^\circ-150^\circ$ & $-$ & $+$\\
 \parbox[0pt][1.6em][c]{0cm}{}$B_2(S2)$&  $0.00306-0.00322$&$92^\circ-95^\circ$ & & $+$\\
\parbox[0pt][1.6em][c]{0cm}{} $B_3(S2)$ & $0.00081-0.00128$
& $308-330^\circ$ & $+$ & $-$\\
\parbox[0pt][1.6em][c]{0cm}{}$B_4(S2)$&  $0.00306-0.00322 $&$272^\circ-275^\circ$ & & $-$\\
\hline
\end{tabular}
\caption{\it Oases in the space $(\tilde s_{13},\delta_{13})$ for $M_{Z^\prime}=1\tev$ in LHS
and two scenarios for $\vub$. The enhancement or suppression of $\mathcal{B}(B_d\to\mu^+\mu^-)$ with respect to the SM value chooses the oasis uniquely in LHS2 .
The sign of $S^d_{\mu^+\mu^-}$ chooses the oasis for both LHS1 and LHS2.
}\label{s13oases}~\\[-2mm]\hrule
\end{table}

% LHS1
% Big        s13->[0.00062,0.00117] delta13->[1.328,1.84] ok
% Small    s13->[0.00322,0.00337] delta13->[1.55,1.575] ok
%
% LHS2
% Big        s13->[0.00081,0.00128] delta13->[2.23,2.62] ok
% Small    s13->[0.00306,0.00322] delta13->[1.603,1.653] ok

\begin{figure}[!tb]
\begin{center}
\includegraphics[width=0.45\textwidth] {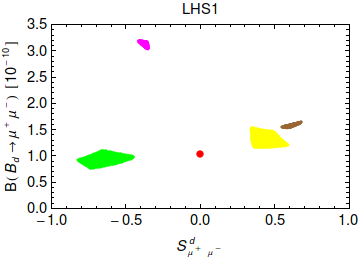}
\includegraphics[width=0.45\textwidth] {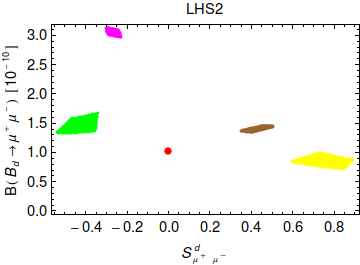}
\caption{\it $\mathcal{B}(B_d\to\mu\bar\mu)$ versus $S_{\mu^+\mu^-}^d$    for $M_{Z^\prime}=1$ TeV in LHS1 (left) and
LHS2 (right). $B_1$: yellow, $B_3$: green, $B_2$: magenta, $B_4$: brown. Red point: SM central
value.}\label{fig:BdmuvsSmudLHS}~\\[-2mm]\hrule
\end{center}
\end{figure}

We also observe  in Table~\ref{s13oases} that $S_{\mu^+\mu^-}^d$
can also help
by means of its  sign to distinguish between different oases. Fig.~\ref{fig:BdmuvsSmudLHS} (right panel)
also shows that in LHS2
the sign of  $S_{\mu^+\mu^-}^d$ is opposite to the  sign
of the shift in the corresponding branching ratio (except for the small oasis $B_2(S2)$), which can easily
be understood by inspecting the ranges of $\delta_{13}$. Moreover,
the predictions for $S_{\mu^+\mu^-}^d$ are rather precise.
This is in particular the case for small oases, which in the $B_d$-system
cannot be ruled out. In fact in the $B_2$ oasis $\mathcal{B}(B_d\to\mu^+\mu^-)$ can still be by a factor of three enhanced with respect to its SM value.
Finally, we observe that  $S_{\mu^+\mu^-}^d$ can be large, although not 
as large as  $S_{\mu^+\mu^-}^s$.

We next turn to LHS1 scenario for $\vub$ which is novel with respect to the
analysis in \cite{Buras:2012xx}. We observe that the phase $\delta_{13}$
is lower for big oases than in  the case of scenario S2 but $\tilde s_{13}$ is basically
the same. We observe that while the sign of $S^d_{\mu^+\mu^-}$ can still
distinguish between the big oases,
$B_d\to\mu^+\mu^-$ cannot do it as well. This is related to the fact
that with $\vub$ as low as 0.0031 we have to enhance slightly $S_{\psi K_S}$
in certain range of parameters involved. These features are seen
in the left panels in Figs.~\ref{fig:BdmuvsSKSLHS} and
\ref{fig:BdmuvsSmudLHS}.

What distinguishes
LHS1 from LHS2 is the sign of the correlation between $S^d_{\mu^+\mu^-}$ and
$\mathcal{B}(B_d\to\mu^+\mu^-)$. A positive $S^d_{\mu^+\mu^-}$ implies
enhancement of $\mathcal{B}(B_d\to\mu^+\mu^-)$ in LHS1 but suppression in
LHS2. Note that this pattern is independent of the sign of $Z'\mu^+\mu^-$
coupling as this coupling enters both observables. On the other hand the
flip of this sign would also flip signs in the last two columns in
Table~\ref{s13oases} and thereby interchange colours 
in Figs.~\ref{fig:BdmuvsSKSLHS} and
\ref{fig:BdmuvsSmudLHS}.

\boldmath
\subsubsection{The $K$ Meson System}
\unboldmath

In the $\model$ model, which was governed by S2 scenario, NP effects in rare $K$ decays were very small due to suppression of both $\bar s d Z'$ and
$\bar\nu\nu Z'$   couplings in this model.

However in a general $Z'$ model the possibility for S1 scenario for $\vub$ and
enhanced values of leptonic couplings
with respect to the ones found in the $\model$ model allow to find large NP
effects in rare $K$ decays. This allows to find
 interesting correlations between relevant branching ratios that we would
like to exhibit here.

\begin{figure}[!tb]
\begin{center}
\includegraphics[width=0.45\textwidth] {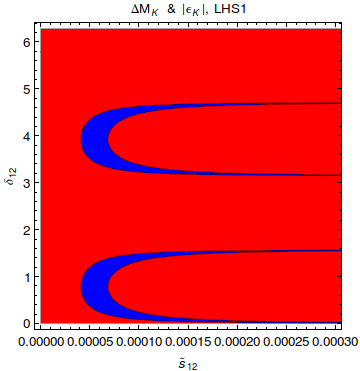}
\includegraphics[width=0.45\textwidth] {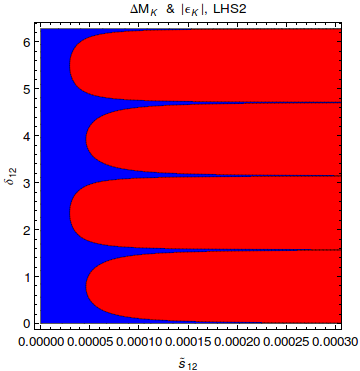}
\caption{\it  Ranges for $\Delta M_K$ (red region) and $\varepsilon_K$ (blue region) (LHS1: left, LHS2: right) for
$M_{Z^\prime}=1$ TeV  satisfying the bounds in Eq.~(\ref{C3}).
}\label{fig:oasesKLHS}~\\[-2mm]\hrule
\end{center}
\end{figure}

As seen in (\ref{C3}) the constraints from $\Delta F=2$ observables are
weaker than in previous cases. Yet as seen in Fig.~\ref{fig:oasesKLHS}
it is possible to identify the allowed oases. These plots have the same structure
as the plot in Fig.~2 of \cite{Blanke:2009pq} with the S1 and
S2 scenario for $\vub$ on the left and on the right, respectively.
We observe that the small oases are
absent now as $\varepsilon_K$ and $\Delta M_K$ are governed respectively
by imaginary and
real parts of $M_{12}^K$ and not by their absolute values like in the case
of $\Delta M_{s,d}$.
Therefore the solutions with very large NP contributions but opposite
signs to the SM contributions corresponding to small oases in the latter case
are not allowed here.

Due to weaker constraints in the $K$ system the oases are rather large.
We have two oases in S1:
\be\label{KoasesS1}
C_1(S1):~~0^\circ \le \delta_{12} \le 90^\circ,\qquad C_2(S1):~~180^\circ \le \delta_{12} \le 270^\circ
\ee
 and only one oasis in S2:
\be\label{KoasesS2}
C_1(S2):~~0^\circ \le \delta_{12} \le 360^\circ.
\ee

As emphasized in \cite{Blanke:2009pq} of particular interest are
the values
\be\label{delta12}
\delta_{12}= n\frac{\pi}{2}, \qquad n=0,1,2,3
\ee
for which NP contributions to $\varepsilon_K$ vanish. As seen in Fig.~\ref{fig:oasesKLHS}
this is only allowed for scenario S2 for which SM agrees well with the data
and NP contributions are not required. In this scenario  $\tilde s_{12}$
can even vanish. In scenario S1, in which NP contributions are required
to reproduce the data,  $\tilde s_{12}$ is bounded from below and
$\delta_{12}$ cannot satisfy (\ref{delta12}) but for sufficiently large
 $\tilde s_{12}$ can satisfy it approximately. As at these values of
$\delta_{12}$, the mass difference $\Delta M_K$ is non-zero,  $\tilde s_{12}$
is bounded from above but due to the weak  $\Delta M_K$-constraint this is not
seen in the plot.

In  \cite{Blanke:2009pq} an extensive analysis of the interplay between
 $\varepsilon_K$ and in $K\to\pi\nu\bar\nu$ in different NP scenarios
has been performed but the case of tree-level $Z'$ contributions has
not be discussed. As the latter contributions are much more specific
and simpler than the NP models discussed in  \cite{Blanke:2009pq},
it will be interesting to see how correlations between
$\varepsilon_K$ and $K\to\pi\nu\bar\nu$ in the eight scenarios
in (\ref{generalS}) compare with the findings of \cite{Blanke:2009pq}.

To this end for the LHS scenarios we find for the quantities
defined in \cite{Blanke:2009pq} \footnote{In \cite{Blanke:2009pq} $\rho$ was denoted by $\epsilon$.}
\be\label{Blanke1}
\phi_{K\to\pi\nu\bar\nu}=\phi_{\Delta S=2}=-\delta_{12}
\ee
\be\label{Blanke2}
R_{K\to\pi\nu\bar\nu}=-\frac{\Delta_L^{\nu\bar\nu}(Z')\tilde s_{12}}{g^2_{\rm SM}M^2_{Z'}},\qquad
R_{\Delta S=2}=-\frac{2\sqrt{\tilde r}\tilde s_{12}}{g_{\rm SM}M_{Z'}},
\ee
implying
\be\label{Blanke3}
\rho\equiv\frac{R_{K\to\pi\nu\bar\nu}}{R_{\Delta S=2}}=\frac{1}{2\sqrt{\tilde r}}
\frac{\Delta_L^{\nu\bar\nu}(Z')}{g_{\rm SM}M_{Z'}}=
1.2\Delta_L^{\nu\bar\nu}(Z') \frac{1\tev}{M_{Z'}}.
\ee
For our choice of $\Delta_L^{\nu\bar\nu}(Z')$ we find $\rho\approx 0.6$ for
$M_{Z'}=1\tev$. On
the basis of  \cite{Blanke:2009pq} we expect for this value of $\rho$ strict
correlation  between $\mathcal{B}(\kpn)$ and
$\mathcal{B}(\klpn)$  familiar from the LHT model \cite{Blanke:2009am}. It is
interesting that $\rho$ depends only on the size of $\Delta_L^{\nu\bar\nu}(Z')$
 and $M_{Z'}$. This will have important implications for the study of
flavour-violating $Z$ couplings considered in Section~\ref{sec:ZSM}.

\begin{figure}[!tb]
\begin{center}
\includegraphics[width=0.45\textwidth] {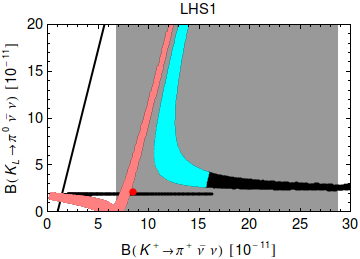}
\includegraphics[width=0.45\textwidth] {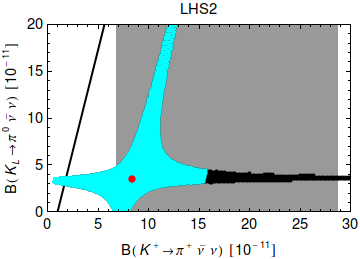}\\
\vspace{0.3cm}
\includegraphics[width=0.45\textwidth] {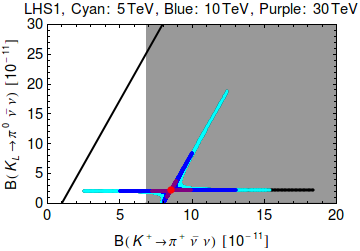}
\includegraphics[width=0.45\textwidth] {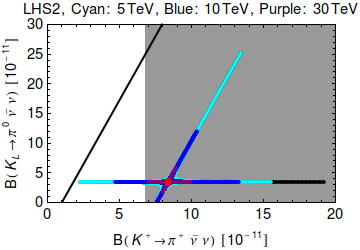}
\caption{\it  $\mathcal{B}(\klpn)$ versus
$\mathcal{B}(\kpn)$ for $M_{Z^\prime} = 1~$TeV (upper panels,  $C_1$: cyan, $C_2$: pink.) and $M_{Z^\prime} = 5~$TeV (cyan),
10~TeV (blue) and 30~TeV (purple) (lower panels) in LHS1 (left) and LHS2
(right).   Black regions are excluded by the upper bound $\mathcal{B}(K_L\to \mu^+\mu^-)\leq 2.5\cdot
10^{-9}$. Red point: SM central
value. Gray region:
experimental range of $\mathcal{B}(\kpn)$. }\label{fig:KLvsKpLHS}~\\[-2mm]\hrule
\end{center}
\end{figure}

In the upper panels of  Fig.~\ref{fig:KLvsKpLHS} we show this correlation in LHS1 and LHS2
 for $M_{Z'}=1\tev$. We observe the following pattern of deviations from
the SM expectations:
\begin{itemize}
\item
There are two branches in both scenarios. The difference between LHS1
and LHS2 originates from required NP contributions in LHS1 in order to
agree with the data on $\varepsilon_K$ and the fact that in LHS1 there
are two oases and only one in LHS2.
\item
The horizontal branch in both plots corresponds to $n=0,2$ in (\ref{delta12}), for which
NP contribution to $K\to\pi\nu\bar\nu$ is real and vanishes in the
case of $\klpn$.
\item
The second branch corresponds to $n=1,3$ in (\ref{delta12}), for which NP
contribution is purely imaginary. It is parallel to the Grossman-Nir (GN) bound
\cite{Grossman:1997sk}
that is represented by the solid line.
\end{itemize}

This pattern agrees with
general results of \cite{Blanke:2009pq}. In fact the structure of plots in
the Fig.~3 and 4 of the latter paper agrees for $\rho\approx 1$
very well with our findings for LHS2 and LHS1 scenarios, respectively.
What is striking is the fact that still large deviations from the SM
predictions are allowed, significantly larger than in the case of
rare $B$ decays. This is a consequence of the weaker constraint from
$\Delta S=2$ processes than $\Delta B=2$ and the fact that rare $K$ decays
are stronger suppressed than rare $B$ decays within the SM. 
Yet as we will soon see some of these large values
will be ruled out through the correlation with $K_L\to\mu^+\mu^-$.

\begin{figure}[!tb]
\begin{center}
\includegraphics[width=0.45\textwidth] {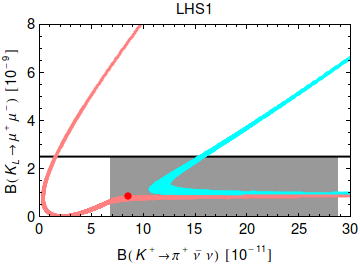}
\includegraphics[width=0.45\textwidth] {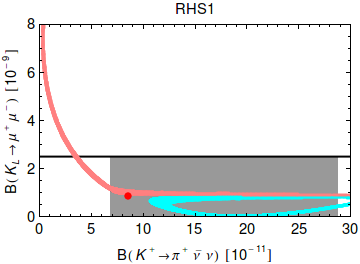}
\caption{\it  $\mathcal{B}(K_L\to\mu^+\mu^-)$ versus  $\mathcal{B}(\kpn)$ for $M_{Z^\prime} = 1~$TeV in LHS1 (left) and RHS1
(right).   $C_1$: cyan, $C_2$: pink. Red
point: SM central
value. Gray region: experimental range of
$\mathcal{B}(\kpn)$ and horizontal black line: upper bound of
$\mathcal{B}(K_L\to\mu^+\mu^-)$.}\label{fig:KLmuvsKpLHS}~\\[-2mm]\hrule
\end{center}
\end{figure}

In the left panel of Fig.~\ref{fig:KLmuvsKpLHS} we show the correlation between  $\mathcal{B}(\kpn)$ and  $\mathcal{B}(K_L\to\mu^+\mu^-)$
for LHS1. We note a correlation analogous to the one found in
the LHT model \cite{Blanke:2009am} but due to fewer free parameters
in $Z'$ model this correlation depends whether oasis $C_1$ or $C_2$ is considered.
Very similar correlation is
found in LHS2 scenario but as here only one very big oases is present only
{\it cyan} regions appear. We will return to the right panel in this
figure in the context of RHS1 scenario below.

While at first sight the correlation in  Fig.~\ref{fig:KLmuvsKpLHS} is 
similar in shape to the one
in Fig.~\ref{fig:KLvsKpLHS}, one should note that $K_L\to\mu^+\mu^-$ is
governed by the real part of the involved master function and not
imaginary part as was the case of $\klpn$. Therefore the horizonal line
in Fig.~\ref{fig:KLmuvsKpLHS} corresponds this time to $n=1,3$ in (\ref{delta12}), for which NP
contribution is purely imaginary, while the other branches correspond
 to $n=0,2$ in (\ref{delta12}), for which
NP contribution to $K\to\pi\nu\bar\nu$ is real and vanishes in the
case of $\klpn$.

We observe again that NP effects in both decays can be large and the
upper bound on  $\mathcal{B}(K_L\to\mu^+\mu^-)$ in (\ref{eq:KLmm-bound}) represented by the horizontal black line can easily be violated. The impact
of this bound on the results in Fig.~\ref{fig:KLvsKpLHS} is represented
by the black areas that violate this bound.

Combining the information from Figs.~\ref{fig:KLvsKpLHS} and ~\ref{fig:KLmuvsKpLHS} we obtain the following result:
\begin{itemize}
\item
In the case of the dominance of real NP contributions we find in $C_1(S1)$ 
for $M_{Z'}=1\tev$
\be\label{UPERBOUND}
\mathcal{B}(\kpn)\le  16\cdot 10^{-11}.
\ee
In this case $\klpn$ is SM-like and $\mathcal{B}(K_L\to\mu^+\mu^-)$ reaches
the upper bound in (\ref{eq:KLmm-bound}). On the other hand
$C_2(S1)$ oasis in this case is excluded through the simultaneous consideration
of both decays.
\item
In the case of the dominance of  imaginary  NP contributions the bound
on $\mathcal{B}(K_L\to\mu^+\mu^-)$ is ineffective and both
$\mathcal{B}(\kpn)$ and $\mathcal{B}(\klpn)$ can be significantly larger
than the SM predictions and $\mathcal{B}(\kpn)$ can also be larger than its
present experimental central value. We also find that for such large values
the branching ratios are strongly correlated. Inspecting in the LHS2 
scenario when the branch parallel to the GN bound leaves the grey region 
corresponding to the $1\sigma$ region in (\ref{EXP1})
we find a rough upper bound
\be\label{const}
\mathcal{B}(\klpn)\le 85\cdot 10^{-11},
\ee
\end{itemize}
which is much stronger than the present experimental upper bound in 
(\ref{EXP2}).

We conclude therefore that $K\to\pi\nu\bar\nu$ decays provide an important
portal to flavour-violating $Z'$ with masses outside the reach of the LHC 
before its upgrade and even in its second phase.
In the
lower part of
Fig.~\ref{fig:KLvsKpLHS}
 we show therefore 
how the plots in the upper part of this figure  would 
look like  for $M_{Z'}=5\tev$, $10\tev$ and $30\tev$. We observe that even 
at  $M_{Z'}=10\tev$
both branching ratios can still differ by much from SM predictions and for 
 $M_{Z'}\le 20\tev$ NP effects in these decays, in particular 
$\klpn$  should be detectable in the flavour precision era. For  $M_{Z'}=30\tev$ and higher scales it will be very difficult.

%We also note that the upper bound in (\ref{UPERBOUND}) did not change
%as the $M_{Z'}$ dependence  of NP contributions to
%$\mathcal{B}(\kpn)$ and $\mathcal{B}(K_L\to\mu^+\mu^-)$
%on the horizontal line is the same.

% \begin{figure}[!tb]
% \begin{center}
% \includegraphics[width=0.45\textwidth] {pKLvsKpLHS30TeV2.png}
% \includegraphics[width=0.45\textwidth] {pKLvsKpLHS50TeV2.png}
% \caption{\it  $\mathcal{B}(\klpn)$ versus
% $\mathcal{B}(\kpn)$ for $M_{Z^\prime} = 30~$TeV (left) and $M_{Z^\prime} = 50~$TeV (right) in LHS2.   $C_1$: cyan. Red point: SM
% central
% value. Gray region:
% experimental range of $\mathcal{B}(\kpn)$. }\label{fig:KLvsKpLHS100TeV}~\\[-2mm]\hrule
% \end{center}
% \end{figure}

\begin{figure}
\includegraphics[width=0.45\textwidth] {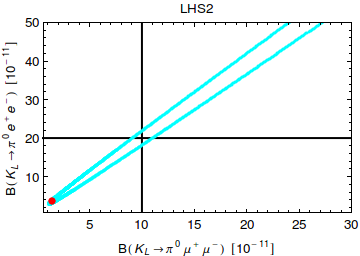}
\includegraphics[width=0.45\textwidth] {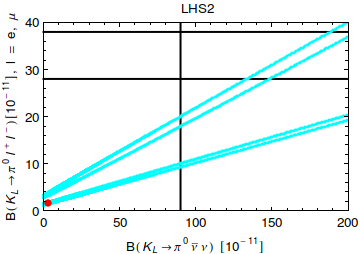}

\caption{\it $\mathcal{B}(K_L \to \pi^0 e^+e^-)$ as a
  function of $\mathcal{B}(K_L\to \pi^0 \mu^+\mu^-)$ (left panel) and 
 $\mathcal{B}(K_L\to \pi^0 e^+e^-)$ (upper curve) and  
$\mathcal{B}(K_L \to \pi^0\mu^+\mu^-)$ (lower curve) as functions of 
$\mathcal{B}(\klpn)$ (right panel)
in LHS2 for $M_{Z'}=1\tev$. The red points
represent SM
predictions.}
\label{fig:KmuKe}~\\[-2mm]\hrule
\end{figure}

In the left panel of Fig.\ \ref{fig:KmuKe} we show the correlation between 
$\mathcal{B}(K_L\to \pi^0 e^+e^-)$ and $\mathcal{B}(K_L \to \pi^0 \mu^+\mu^-)$
that has {first been} investigated in 
\cite{Isidori:2004rb,Friot:2004yr,Mescia:2006jd}. We have shown only the 
results in LHS2 as our main goal here is to find out whether large enhancements 
of the branching ratios for $\klpn$ and $\kpn$ can be affected 
by these decays. To this end we have also assumed constructive interference between SM and NP 
contributions.

We observe a strong correlation between $\mathcal{B}(K_L\to \pi^0 e^+e^-)$ and 
$\mathcal{B}(K_L \to \pi^0 \mu^+\mu^-)$, similar to the case of LHT \cite{Blanke:2006eb} and RSc  \cite{Blanke:2008yr} models. 
Indeed such a correlation is common to all models with no scalar operators contributing to the decays in question
\cite{Isidori:2004rb,Friot:2004yr,Mescia:2006jd}.
We also observe that both branching ratios can be in principle enhanced by an 
order of magnitude
over
the SM values in (\ref{eq:KLpmm}). However, the correlation with 
$\kpn$ and $\klpn$ does not allow for such large values. Indeed the experimental upper bound on $\mathcal{B}(\kpn)$ implies in a given scenario an upper bound 
on  $\mathcal{B}(\klpn)$ as given in (\ref{const}) and this in turn implies  upper bounds 
on  $\mathcal{B}(K_L\to \pi^0 e^+e^-)$ and 
$\mathcal{B}(K_L \to \pi^0 \mu^+\mu^-)$ that are strongly correlated with 
$\mathcal{B}(\klpn)$. This correlation is evident in $Z'$ scenarios if one 
compares the expressions for the relevant amplitudes.

In the right panel of  Fig.\ \ref{fig:KmuKe}  we show this correlation
 that has already been found in the LHT model 
\cite{Blanke:2006eb} and in the RSc model \cite{Blanke:2008yr}.
 We note that
a large enhancement of $\mathcal{B}(\klpn)$ automatically implies significant 
enhancements of $\mathcal{B}(K_L\to \pi^0\ell^+\ell^-)$, although the NP effects in
$\klpn$ are stronger. This is related to the fact that small or moderate 
NP effects in 
$K_L\to \pi^0\ell^+\ell^-$ are shadowed by the dominant indirectly
 CP-violating contribution. However, for large NP contributions when 
the directly  CP-violating contribution becomes more important NP effects 
in $\mathcal{B}(\klpn)$ and $\mathcal{B}(K_L\to \pi^0\ell^+\ell^-)$ are 
comparable in size.
The correlations in  Fig.~~\ref{fig:KmuKe} constitute a powerful test of the model considered.

The vertical solid line in the right panel of  Fig.~~\ref{fig:KmuKe} corresponds to the rough bound in (\ref{const}). It implies upper bounds 
on  $\mathcal{B}(K_L\to \pi^0\ell^+\ell^-)$ that are stronger than the present 
experimental bounds. We indicate these bounds by horizontal and vertical 
solid lines in the left panel of this figure.

This analysis shows that in LHS scenarios  the present 
upper bounds on $\mathcal{B}(K_L\to \pi^0\ell^+\ell^-)$ do not preclude 
large NP effects found in $\mathcal{B}(\kpn)$ and 
 $\mathcal{B}(\klpn)$ but the bounds on the latter branching ratios 
have an impact on $\mathcal{B}(K_L\to \pi^0\ell^+\ell^-)$. This property remains for higher values of $M_{Z'}$.

\subsection{The RHS1 and RHS2 Scenarios}

\subsubsection{First Observations}

We will now investigate $Z'$ scenario with exclusively RH couplings to quarks.
We should emphasize that such scenario is not artificial as in certain
extensions of the SM, the corrections to left-handed neutral gauge boson
couplings are suppressed due to some custodial symmetries. This is for instance
the case of the ordinary $Z$ gauge boson in
Randall-Sundrum scenarios with custodial symmetry in the bulk (RSc).
In such a case the phenomenology of flavour violation is dominated by
right-handed couplings \cite{Blanke:2008yr}.

Now in the RHS1 and RHS2 scenarios only RH couplings to quarks
are present in $Z'$ contributions. As QCD
is parity conserving, the hadronic matrix elements for operators with
RH currents as well as QCD corrections remain unchanged. The expressions
for $\Delta F=2$ observables in RHS1 and RHS2 scenarios as well as the
corresponding constraints have precisely the same structure as in the
LHS1 and LHS2 cases just discussed. Therefore the oases in the space of
parameters related to RH currents are precisely the same as those given in
Tables~\ref{s23oases} and \ref{s13oases}, except that the parameters $\tilde s_{ij}$
and $\delta_{ij}$ parametrize now RH and not LH currents. Anticipating this
result we have not introduced separate description of LH and RH oases.
Yet, in the case of $\Delta F=1$ observables several changes are present which
allow in principle to distinguish the RHS1 and RHS2 scenarios from the corresponding
LHS1 and LHS2, which we just analyzed in detail.

In what follows we will list all changes in the three meson systems one by one.
We use for the oases the same notation as in the LHS cases. The basic
rule for modifications of correlations between various observables
is as follows:
\begin{itemize}
\item
In $\Delta F=2$ observables nothing changes as stated above. Therefore we
do not show any plots for allowed oases.
\item
In $\Delta F=1$ observables governed by the functions $Y$, that is processes
with  muons in the final state, there is {\it a change
of  sign} of NP contributions in a given oasis. See (\ref{YAK}) and (\ref{YAB}).
\item
In $\Delta F=1$ observables governed by the functions $X$, that is processes
with  neutrinos in the final state, there is {\it no change
of sign} of NP contributions in a given oasis. The consequences
of it are straightforward in the case of $\kpn$ and $\klpn$ but
in the case of $b\to s\nu\bar\nu$ transitions the implications are
richer as we have four observables to our disposal that are
sensitive to the RH currents in a different manner. See (\ref{eq:BKnn})-(\ref{eq:Xsnn})  and  (\ref{eq:epseta-FL}). The same comments apply to
$B\to K^*\mu^+\mu^-$ and $B\to K \mu^+\mu^-$ which now receive contributions
from primed operators $Q_9^\prime$ and $Q_{10}^\prime$.
\end{itemize}

\boldmath
\subsubsection{The $B_s$ Meson System}
\unboldmath

In Fig.~\ref{fig:SmusvsSphiLHS1} we have shown  $S^s_{\mu^+\mu^-}$  vs $S_{\psi\phi}$ in the LHS1
scenario. This plot is also valid for RHS1 scenario except that now
a  {\it negative} value of  $ S^s_{\mu^+\mu^-}$  chooses scenario $A_1$,
while a {\it positive} one scenario $A_3$. The size of NP effects is the
same in LHS1 and RHS1, only the oases are interchanged. The same comments apply to large $\vub$ scenario.

We conclude therefore that on the basis of  $S^s_{\mu^+\mu^-}$  and  $S_{\psi\phi}$
alone it is not
possible to distinguish between LHS1 and RHS1 scenarios because in the
RHS1 scenario one can simply interchange the two big oases or two small oases
 to obtain the same
physical results as in LHS1 scenario. Therefore let us look at other observables.

In Fig.~\ref{fig:SmusvsSphiLHS1} (right) we have shown $S_{\psi\phi}$ vs $\mathcal{B}(B_s\to\mu^+\mu^-)$ in
the LHS1 scenario. This plot is also valid for the RHS1 scenario but
again the oases $A_1$ and $A_3$ are interchanged.
 While in the
oasis $A_1$
$S_{\psi\phi}$ increases (decreases) uniquely with increasing (decreasing)
$\mathcal{B}(B_s\to\mu^+\mu^-)$, in the oasis $A_3$, the increase of
$S_{\psi\phi}$ implies uniquely a decrease of $\mathcal{B}(B_s\to\mu^+\mu^-)$.

Clearly as in the LHS1 scenario this result represents a test of the
RHS1 scenario but if one day we will have precise measurements of
 $S^s_{\mu^+\mu^-}$,  $S_{\psi\phi}$ and $\mathcal{B}(B_s\to\mu^+\mu^-)$
we will still not be able to distinguish for instance whether we deal
with LHS1 scenario in oasis $A_1$ or RHS1 scenario in oasis $A_3$.

Fortunately, as we will see in subsection~\ref{bsllc}, we will be able to make a clear distinction between LHS and RHS scenarios by considering model independent bounds
from $B\to K^*\mu^+\mu^-$ and  $B\to K \mu^+\mu^-$ on the Wilson coefficients
of primed operator $Q_{10}^\prime$. Also, as seen in  (\ref{eq:BKnn})-(\ref{eq:Xsnn})  and  (\ref{eq:epseta-FL}),
$b\to s\nu\bar\nu$ transitions being sensitive to
RH currents will be very helpful in this respect.
In LHS scenarios the first  three
observables where affected in the same manner and $F_L$ was unaffected.
Therefore the comparison of these
different structures in RHS and LHS scenarios will give us a powerful
insight in the $Z'$ couplings.  As these issues will also be relevant
for LR and ALR scenarios, we will discuss $b\to s \nu\bar\nu$ observables
in the four scenarios in a separate subsection at the end of this section.

 \boldmath
\subsubsection{The $B_d$ Meson System}
\unboldmath

Similarly to the $B_s$ case the structure of oases is as
in Fig.~\ref{fig:oasesBdLHS} and Table~\ref{s13oases}.
In Fig.~\ref{fig:BdmuvsSKSLHS} we have shown  $S_{\psi K_S}$ vs  $\mathcal{B}(B_d\to\mu^+\mu^-)$
for the LHS2 scenario. This plot is also valid for RHS2 scenario and
$\mathcal{B}(B_d\to\mu^+\mu^-)$ can distinguish between  $B_1$ and $B_3$
oases. But now the behaviour of  $\mathcal{B}(B_d\to\mu^+\mu^-)$ in
these two oases is interchanged. It is enhanced and suppressed with respect
to the SM  in $B_1$ and $B_3$, respectively. Thus we cannot distinguish
between LHS2 and RHS2 on the basis of these observables. Clearly
the study of $b\to d \mu^+\bar\mu^-$ and  $b\to d \nu\bar\nu$ transitions could help in this context
but they are more challenging both theoretically and experimentally.

Analogous comments apply to RHS1 which cannot be distinguished from LHS1
on the basis of observables considered.

\boldmath
\subsubsection{The $K$ Meson System}
\unboldmath

In Fig.~\ref{fig:KLvsKpLHS} we have shown the correlation between $\mathcal{B}(\kpn)$ and
$\mathcal{B}(\klpn)$ in the LHS1 and LHS2  scenario. It is evident from~(\ref{Xeff}) that this plot applies identically to RHS1 and RHS2
scenarios
as well. Thus $\kpn$ and $\klpn$ decays are not useful for the search of RH currents as they are sensitive only the vector parts of $Z'$ couplings to quarks. 
 As expected, we also find that the results for $K_L\to\pi^0\ell^+\ell^-$ decays are the same as in LHS scenarios.

However, as known already from different studies,
in particular in RSc scenario \cite{Blanke:2008yr}, the correlation  between $\mathcal{B}(\kpn)$ and
 $\mathcal{B}(K_L\to\mu^+\mu^-)$ brings the rescue to this problematic as the
latter
decay is sensitive to the axial-vector couplings. In  the right
panel of Fig.~\ref{fig:KLmuvsKpLHS}
we show this
correlation for the RHS1 scenario.  Indeed the correlations in both oases
differ from the ones in LHS1.

We also note that in the case of the dominance of imaginary NP contributions
corresponding to the horizontal line,  $\mathcal{B}(\kpn)$ and
$\mathcal{B}(\klpn)$ can be large. But otherwise  $\mathcal{B}(\kpn)$ is
suppressed with respect to its SM value and  $\mathcal{B}(\klpn)$ is SM-like.

We should emphasize at this point that the impact of $K_L\to\mu^+\mu^-$
on $\kpn$ and $\klpn$ and its correlation with them depends on the sign
of leptonic couplings. For a negative $\Delta_A^{\mu\bar\mu}$ the results in
LHS and RHS
scenarios would be interchanged.

\subsection{The LRS1 and LRS2 Scenarios}

\subsubsection{First Observations}

If both LH and RH currents are present in NP contributions, the pattern
of flavour violation can differ from the scenarios considered until now
in a profound manner. If the LH and RH couplings differ from each other,
the number of parameters increases and it is harder to get clear cut
conclusions without some underlying fundamental theory. On the other hand
some of the ``symmetries'' between LHS and RHS scenarios identified above
are broken and
the effect of RH currents in certain cases could in principle be better
visible.

Here in order to keep
the same number of parameters as in previous scenarios we will assume
a left-right symmetry in the $Z'$-couplings to quarks. That is the
LH couplings $\Delta_L$ are equal in magnitudes and phases to the corresponding
RH couplings $\Delta_R$. In this manner we can also keep the same
parametrization of couplings as in previous scenarios.

Before entering the details let us emphasize two new features relative
to the cases in which either LH or RH couplings in NP contributions
were present:
\begin{itemize}
\item
NP contributions to $\Delta F=2$ observables are dominated now by new LR
operators, whose contributions are enhanced through renormalization group
effects relative to LL and RR operators and in the case of $\varepsilon_K$
also through chirally enhanced hadronic matrix elements. Consequently
the oases will differ from the previous ones.
\item
NP contributions to $B_{d,s}\to\mu^+\mu^-$ and $K_L\to\mu^+\mu^-$ vanish
eliminating in this manner $S^{s,d}_{\mu^+\mu^-}$ and $\mathcal{B}(B_{s,d}\to\mu^+\mu^-)$ as basic observables in the identification of acceptable oases.
On the other hand $B\to K^*\mu^+\mu^-$ and $B\to K\mu^+\mu^-$ receive
still NP contributions and can help in this context.
\item
Also NP contributions to decays with neutrinos
in the final state, that is $\kpn$, $\klpn$ and the $b\to s \nu\bar\nu$
transitions are important for testing the LRS1 and LRS2 scenarios.
\end{itemize}

While  $S^{s,d}_{\mu^+\mu^-}$ cannot help in the identification of the 
optimal 
oasis in the LR scenarios they are non-vanishing as seen in (\ref{Smumus}) and  (\ref{Smumud}):

\be\label{SmumuLR}
S_{\mu^+\mu^-}^q=-\sin(2\varphi_{B_q}).
\ee
While rather small they offer a clean test of LR scenarios.

\boldmath
\subsubsection{The $B_s$ Meson System}
\begin{figure}[!tb]
\begin{center}
\includegraphics[width=0.45\textwidth] {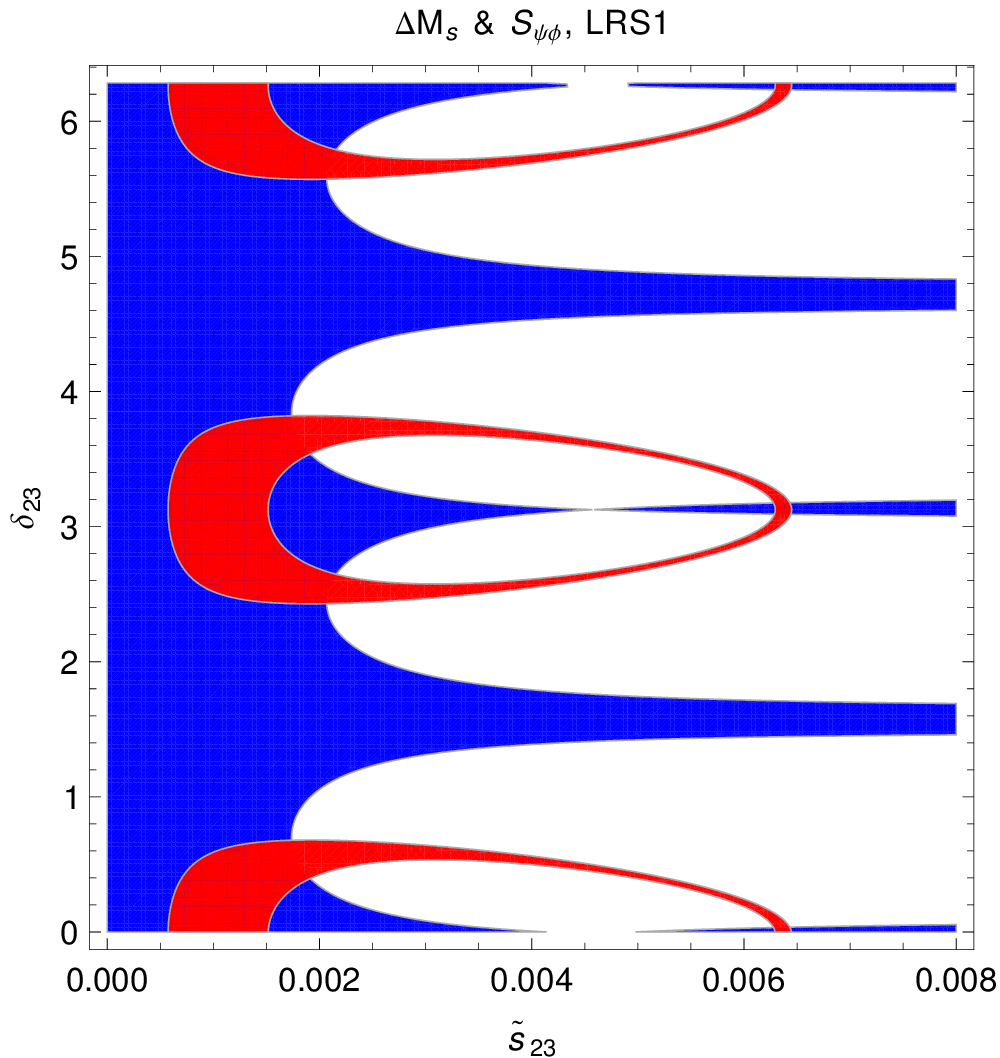}
\caption{\it  Ranges for $\Delta M_s$ (red region) and $S_{\psi \phi}$ (blue region) for $M_{Z^\prime}=1$~TeV in LRS1 satisfying the bounds
in
Eq.~(\ref{C1}).
}\label{fig:oasesBsLRS1}~\\[-2mm]\hrule
\end{center}
\end{figure}

\unboldmath
We begin the search for the oases with the $B_s$ system proceeding
with input parameters as in the previous scenarios.
The result of this search for $M_{Z'}=1\tev$ is shown in Fig.~\ref{fig:oasesBsLRS1},
where we show the allowed ranges
for $(\tilde s_{23},\delta_{23})$.
The {\it red} regions correspond to the allowed ranges for $\Delta M_{s}$,
while the {\it blue} ones to the corresponding ranges for  $S_{\psi\phi}$. The overlap between red and blue regions identifies the
oases we were looking for.

From these plots we extract several oases that are collected in
Table~\ref{s23oasesLR}. The notations are as in previous cases but it
should be kept in mind that the parameters  $(\tilde s_{23},\delta_{23})$
describe both LH and RH couplings. Note that the entries related
to $B_{s,d}\to \mu^+\mu^⁻$ are absent now as in this scenario there are
no NP contributions to these decays. Consequently the $\Delta\Gamma_s$
effects in $B_{s}\to \mu^+\mu^⁻$ are as in the SM:
$\mathcal{A}^\lambda_{\Delta\Gamma}=1.$

In order to understand the structure of oases in Table~\ref{s23oasesLR}, that differs from the ones
found so far, we note that the matrix element of the dominant $Q_1^{\rm LR}$
operator has the sign opposite to SM operators. Therefore, this operator
naturally suppresses $\Delta M_s$ with the phase $\delta_{23}$ centered in the
ballpark of $0^\circ$ and $180^\circ$, that is shifted down by roughly
$90^\circ$ relatively to the LHS scenarios. As the matrix element of
$Q_1^{\rm LR}$ is larger than that of the SM operator in LHS and RHS scenarios,
$\tilde s_{23}$ has to be sufficiently small to agree with data.

\begin{table}[!tb]
\centering
\begin{tabular}{|c||c|c|c|c|}
\hline
 & $\tilde s_{23}$ & $\delta_{23}$   &  $\Delta S_{\psi\phi}$
&   $\Delta\mathcal{B}^{\nu\bar\nu}_s$  \\
\hline
\hline
  \parbox[0pt][1.6em][c]{0cm}{} $A_1(S1)$ & $0.00059-0.00216$
& $139^\circ-219^\circ$ & $\pm$ & $\mp$ \\
 \parbox[0pt][1.6em][c]{0cm}{}$A_2(S1)$&  $0.00628-0.00644 $&$177^\circ-182^\circ$ & &\\
\parbox[0pt][1.6em][c]{0cm}{} $A_3(S1)$ & $0.00059-0.00216$
& $-41^\circ-39^\circ$ & $\pm$ & $\pm$ \\
 \parbox[0pt][1.6em][c]{0cm}{}$A_4(S1)$&  $0.00628-0.00644$&$-3^\circ-2^\circ$ & &  \\
\hline
\end{tabular}
\caption{\it Oases in the space $(\tilde s_{23},\delta_{23})$ for $M_{Z^\prime}=1\tev$ in LRS1. Basically the same
results are obtained for LRS2 scenario.
}\label{s23oasesLR}~\\[-2mm]\hrule
\end{table}

% LRS1
% Big        s23->[0.000587,0.002158] delta23->[2.429,3.820]
% Small    s23->[0.006276,0.00644] delta23->[3.0905,3.177]
%
% LRS2
% Big        s23->[0.0006576,0.00220] delta23->[2.434,3.799]
% Small    s23->[0.006447,0.00628] delta23->[3.088,3.175]

The crucial role in the $B_s$ meson system in this scenario,
in the absence of NP contributions to $B_{s,d}\to\mu^+\mu^-$ decays, is
now played by  $B\to K^*\mu^+\mu^-$, $B\to K\mu^+\mu^-$  and
$b\to s\nu\bar\nu$ transitions. We
will discuss the latter decays at the end of this section.

 \boldmath
\subsubsection{The $B_d$ Meson System}
\unboldmath

\begin{figure}[!tb]
\begin{center}
\includegraphics[width=0.45\textwidth] {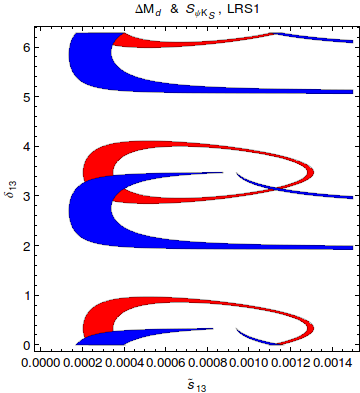}
\includegraphics[width=0.45\textwidth] {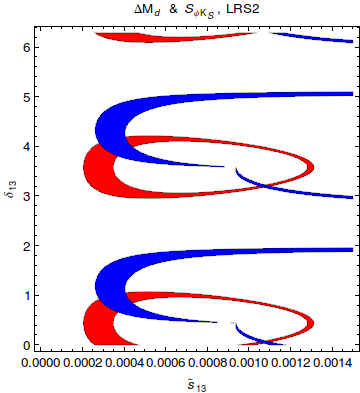}

\caption{\it  Ranges for $\Delta M_d$ (red region) and $S_{\psi K_S}$ (blue region) for $M_{Z^\prime}=1$~TeV in LRS1 (left) and
LRS2 (right) satisfying the bounds
in
Eq.~(\ref{C2}).
}\label{fig:oasesBdLRS1}~\\[-2mm]\hrule
\end{center}
\end{figure}

The structure of oases in this case is given
in Fig.~\ref{fig:oasesBdLRS1} and Table~\ref{s13oasesLR}. As we do not have
$\mathcal{B}(B_d\to\mu^+\mu^-)$ to our disposal and $b\to d\nu\bar\nu$
decays are challenging this system is not very useful to provide
tests of LRS scenarios without some fundamental theory.

Due to the sign
of the matrix element of the dominant $Q_1^{\rm LR}$ operator in both
LRS1 and LRS2 the mass difference $\Delta M_d$ is naturally suppressed.
The requested size of this suppression together with significant
suppression of $S_{\psi K_S}$ in LRS2 and slight enhancement of it in LRS1
governs the structure of the phases.

\begin{table}[!tb]
\centering
\begin{tabular}{|c||c|c|}
\hline
 & $\tilde s_{13}$ & $\delta_{13}$   \\
\hline
\hline
  \parbox[0pt][1.6em][c]{0cm}{} $B_1(S1)$ & $0.00021-0.00040$
& $166^\circ-195^\circ$ \\
 \parbox[0pt][1.6em][c]{0cm}{}$B_2(S1)$&  $0.00111-0.00117$ &$179^\circ-180^\circ$ \\
\parbox[0pt][1.6em][c]{0cm}{} $B_3(S1)$ & $0.00021-0.00040$
& $-14^\circ-15^\circ$ \\
\parbox[0pt][1.6em][c]{0cm}{}$B_4(S1)$&  $0.00111-0.00117$&$-1^\circ-0^\circ$ \\
\hline
\hline
  \parbox[0pt][1.6em][c]{0cm}{} $B_1(S2)$ & $0.00028-0.00044$
& $38^\circ-60^\circ$ \\
 \parbox[0pt][1.6em][c]{0cm}{}$B_2(S2)$&  $0.00106-0.00111$&$2^\circ-5^\circ$ \\
\parbox[0pt][1.6em][c]{0cm}{} $B_3(S2)$ & $0.00028-0.00044$
& $218^\circ-240^\circ$ \\
\parbox[0pt][1.6em][c]{0cm}{}$B_4(S2)$&  $0.00106-0.00111$&$182^\circ-185^\circ$ \\
\hline
\end{tabular}
\caption{\it Oases in the space $(\tilde s_{13},\delta_{13})$ for $M_{Z^\prime}=1\tev$
and two scenarios for $\vub$ in LR scenarios.
}\label{s13oasesLR}~\\[-2mm]\hrule
\end{table}

% LRS1
% Big        s13->[0.00021,0.00040] delta13->[2.9,3.4] ok
% Small    s13->[0.00111,0.00117] delta13->[3.12,3.15] ok
%
% LRS2
% Big        s13->[0.00028,0.00044] delta13->[0.66,1.05] ok
% Small    s13->[0.00106,0.00111] delta13->[0.032,0.083] ok

\boldmath
\subsubsection{The $K$ Meson System}
\unboldmath

\begin{figure}[!tb]
\begin{center}
\includegraphics[width=0.45\textwidth] {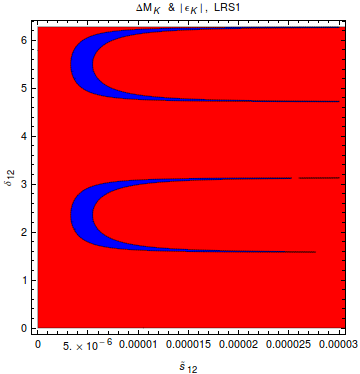}
\includegraphics[width=0.45\textwidth] {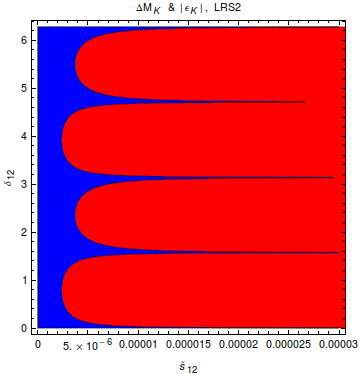}

\caption{\it  Ranges for $\Delta M_K$ (red region) and $\varepsilon_K$ (LRS1: left, LRS2: right) for
$M_{Z^\prime}=1$ TeV  satisfying the bounds in Eq.~(\ref{C3}).
}\label{fig:oasesKLRS}~\\[-2mm]\hrule
\end{center}
\end{figure}

In Fig.~\ref{fig:oasesKLRS} we show the oases in this system that due to the presence of
LR operators have a different structure than in previous scenarios. While
the shape of the single oasis in the LRS2 case is similar to the LHS2, for
LRS1 the oases are shifted by $90^\circ$:
\be\label{LRKoasesS1}
C_1(S1):~~90^\circ \le \delta_{12} \le 180^\circ,\qquad C_2(S1):~~270^\circ \le \delta_{12} \le 360^\circ.
\ee

\begin{figure}[!tb]
\begin{center}
\includegraphics[width=0.45\textwidth] {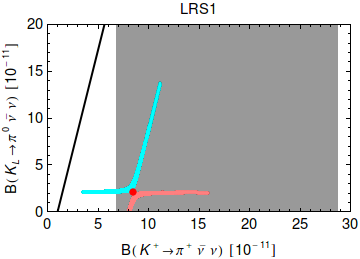}
\includegraphics[width=0.45\textwidth] {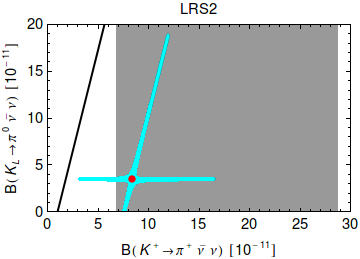}\\
\vspace{0.3cm}
\includegraphics[width=0.45\textwidth] {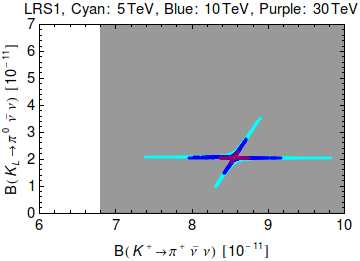}
\includegraphics[width=0.45\textwidth] {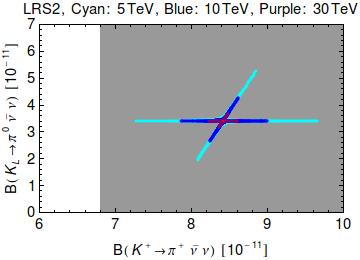}
\caption{\it $\mathcal{B}(\klpn)$ versus
$\mathcal{B}(\kpn)$ for $M_{Z^\prime} = 1~$TeV in LRS1 and LRS2 (upper panels,  $C_1$: cyan, $C_2$: pink.) and  $M_{Z^\prime}
=5~$TeV (cyan), 10~TeV (blue) and 30~TeV (purple)  in LRS1 and LRS2.   Red point: SM central
value. Gray region:
experimental range of $\mathcal{B}(\kpn)$.}\label{fig:KLvsKpLRS}~\\[-2mm]\hrule
\end{center}
\end{figure}

In Fig.~\ref{fig:KLvsKpLRS} we show
the correlation between $\mathcal{B}(\kpn)$ and
$\mathcal{B}(\klpn)$ in the LRS1 and LRS2 scenarios. Even if this correlation
has similar structure to the one found in previous scenarios, there are
differences:
 \begin{itemize}
\item
 Concentrating first on the $M_{Z'}=1\tev$ case we observe that
 the branches in both scenarios are much thiner than in the LHS1 and LHS2 cases
 which originates in the fact that NP contributions to $\varepsilon_K$ have
 to be kept under control in the presence of LR operators.
 \item
 In the LRS1 case the new structure of phases in (\ref{LRKoasesS1}) implies
 definite predictions  for both branching ratios in $C_1(S1)$ and $C_2(S1)$
 respectively { ($M_{Z'}=1\tev$)}:
 \be\label{LRK1}
 3\cdot 10^{-11}\le\mathcal{B}(\kpn)\le 12 \cdot 10^{-11}, \qquad
 2\cdot 10^{-11}\le\mathcal{B}(\klpn)\le 14 \cdot 10^{-11},
 \ee
 \be\label{LRK2}
 8\cdot 10^{-11}\le\mathcal{B}(\kpn)\le 16 \cdot 10^{-11}, \qquad
 0\le\mathcal{B}(\klpn)\le 3 \cdot 10^{-11}.
 \ee
 \item
 In LRS2 the effects are slightly larger than in LRS1 and in fact on the 
 horizontal line $\mathcal{B}(\kpn)$ can be larger than in LHS2 
 in Fig.~\ref{fig:KLvsKpLHS} as in LRS2 the $K_L\to\mu^+\mu^-$ bound is ineffective. 
 \item
 However, otherwise the effects in LRS1 and LRS2 scenarios are much smaller 
 than in LHS1 and LHS2 scenarios in accordance with the correlations 
 between $\Delta F=1$ and $\Delta F=2$ transitions derived in Subsection~\ref{CORR}. This is clearly seen  in lower panels in
Fig.~\ref{fig:KLvsKpLRS}, 
 where we show the results for higher values of $M_{Z'}$. While for 
 $M_{Z'}=5\tev$ NP effects in both branching ratio can still be distinguished 
 in the future from SM values, for higher masses of $M_{Z'}$ this will 
 be very difficult. Note that the SM values of $\mathcal{B}(\klpn)$ in LRS1 
 are visibly smaller than in LRS2 which is a clear consequence of the change 
 of $\vub$.
 \end{itemize}

This discussion shows that if  $M_{Z'}\ge 5\tev$ and both branching 
ratios will be found significantly larger than SM values the LHS1 and 
LHS2 scenarios will be favoured over LRS1 and LRS2.

These results have been obtained under the assumptions of the exact
LR symmetry. However, as already shown in
\cite{Buras:2010pz} and also discussed in \cite{Blanke:2009pq} if
the phase structure of LL and RR and LR contributions in $\Delta S=2$ 
transitions is not related to
each other the correlation between $\mathcal{B}(\kpn)$ and
$\mathcal{B}(\klpn)$ can change profoundly. This is for instance seen in 
Fig.~3 of \cite{Buras:2010pz}, where for such a general scenario one 
can find
a decrease of $\mathcal{B}(\klpn)$  with increasing $\mathcal{B}(\kpn)$,
the property which is absent in the results presented sofar. Indeed in 
this case the leading NP contribution in $\Delta S=2$ amplitudes 
is not proportional to the square of NP contribution to $\Delta S=1$ 
amplitudes implying a different correlation. As such a study can be more 
efficiently performed in a concrete model, we leave it for the future.

\begin{figure}[!tb]
\centering
\includegraphics[width=0.45\textwidth] {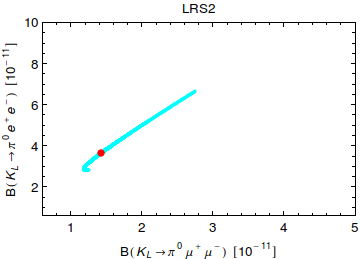}
\includegraphics[width=0.45\textwidth] {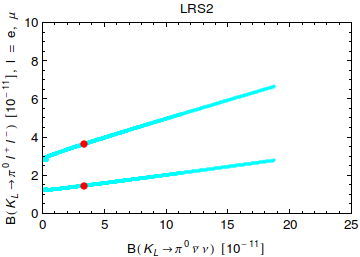}
\caption{\it $\mathcal{B}(K_L \to \pi^0 e^+ e^-)$ as a
  function of $\mathcal{B}(K_L\to \pi^0 \mu^+\mu^-)$ (left panel) and 
 $\mathcal{B}(K_L\to \pi^0 e^+e^-)$ (upper curve) and  
$\mathcal{B}(K_L \to \pi^0\mu^+\mu^-)$ (lower curve) as functions of 
$\mathcal{B}(\klpn)$ (right panel) in LRS2
at $M_{Z'}=1\tev$. The red points represent SM
predictions.}
\label{fig:LRKmuKe}~\\[-2mm]\hrule
\end{figure}

Finally we discuss $K_L\to \pi^0\ell^+\ell^-$ decays in this scenario.
In Fig.~\ref{fig:LRKmuKe} we show the results 
corresponding to those in Figs.~\ref{fig:KmuKe}  obtained in 
LHS2. We observe that NP effects in  $K_L\to \pi^0\ell^+\ell^-$ in 
 this scenario can be large but much smaller than in LHS2 in accordance with 
the correlations derived in Subsection~\ref{CORR}.
We have also 
checked that the present bounds on these decays do not remove sizable  NP 
effects found in LRS2 scenario for $\kpn$ and $\klpn$ decays. Similar 
results are found for LRS1.

\subsection{The ALRS1 and ALRS2 Scenarios}
We include this case as well because it has not been discussed in
the literature but it is an interesting NP scenario for the following
reasons:
\begin{itemize}
\item
NP contributions to $\Delta F=2$ observables are dominated as in
LRS scenarios by new LR
operators but as the sign of LR interference is flipped some differences
arise.
\item
NP contributions to $B_{d,s}\to\mu^+\mu^-$ enter again with full power.
Therefore these decays together with
$S^q_{\mu^+\mu^-}$ offer as in the LHS and RHS scenarios some help
in the identification of acceptable oases.
\item
The phase structure of the oases is as in LHS scenario but due to enhanced
hadronic matrix elements of LR operators the mixing parameters $\tilde s_{ij}$
are decreased typically by a factor of $3.5$.
\item
NP contributions to $\kpn$, $\klpn$ and $K_L\to \pi^0\ell^+\ell^-$ 
vanish in this scenario but
the $b\to s \nu\bar\nu$ transitions can still offer important tests.
\end{itemize}

In view of this simple structure of modifications with respect to LHS
scenario, all plots with the exception of decays with neutrinos in the
final state have the same structure as LH scenarios but NP effects are
smaller. {This is also seen by inspecting  (\ref{REL1}), (\ref{REL2}),
 (\ref{REL7}) and (\ref{REL8}).} 
Therefore we will  not show these plots. However when we
will move to consider flavour-violating $Z$ couplings this suppression
of NP effects in ALRS relative to LHS will turn out to be fortunate and
ALRS will be doing better than LHS.

Concerning  NP contributions to $K\to\pi\nu\bar\nu$ decays
this scenario could turn out one day to be interesting if
the data on observables in $B_s$ and $B_d$ systems will show the presence
of RH currents but negligible NP effects in $K\to\pi\nu\bar\nu$. On the
other hand as we will discuss soon
$b\to s \nu\bar\nu$ transitions can
help to distinguish this scenario from the previous ones.

\subsection{Implications of $b\to s \ell^+\ell^-$ Constraints}\label{bsllc}
Presently the NP effects found by us are consistent with the experimental
data on $B_{s,d}\to\mu^+\mu^-$, although a range of values above the
SM estimate of $\mathcal{B}(B_{s}\to\mu^+\mu^-)$ already slightly violates the existing upper bound. However,
also the data on $B\to X_s \ell^+\ell^-$, $B\to   K^*\ell^+\ell^-$  and
 $B\to   K\ell^+\ell^-$ improved
recently
 by much and it is of interest to see whether they have
an impact on our results. A very extensive model independent
analysis of the impact of the recent LHCb data on the Wilson coefficients
$C_9^{(\prime)}$ and  $C_{10}^{(\prime)}$ has been performed in
\cite{Altmannshofer:2012ir} and we can use these results in our case.

As seen in Subsection~\ref{sec:bqll} the Wilson coefficients $C_9^{(\prime)}$
depend on $\Delta_V^{\mu\bar\mu}(Z')$ couplings which did not enter our analysis.
They can be chosen to satify the constraints in question. Therefore we will
only check whether for the ranges of parameters considered
by us  the resulting coefficients $C_{10}^{(\prime)}$  satisfy the model
independent bounds in \cite{Altmannshofer:2012ir}. As these coefficients
are scale independent we can use the formulae in
Subsection~\ref{sec:bqll} and compare the resulting coefficients
with those in the latter paper. The allowed $2\sigma$ ranges of $C_{10}^{(\prime)}$ are shown in Figs.~1 and 2 of \cite{Altmannshofer:2012ir}. They are given
approximately as follows:
\begin{subequations}\label{equ:ASconstraint}
\begin{align}
 &-2\leq \Re(C_{10}^\prime)\leq 0\,, \quad-2.5\leq \Im(C_{10}^\prime)\leq 2.5\,,\\
&-0.8\leq \Re(C_{10}^\text{NP})\leq 1.8\,,\quad -3\leq \Im(C_{10})\leq 3\,.
\end{align}
\end{subequations}
Especially, 
the new data on
$B\to K^*\mu^+\mu^-$ allow only for  {\it negative} values of the real part
of $C^\prime_{10}$
\be \label{C10C}
\Re( C^\prime_{10}) \le 0
\ee
and this has an impact on  our results in RH and LR scenarios.

 We find that the NP effects in $C_{10}$ are within the $2\sigma$ bounds
 presented in \cite{Altmannshofer:2012ir} even if with improved data one will be able to remove
 certain range of parameters.  Some footprints of this can already been seen 
in Figs.~\ref{fig:SmusvsSphiLHS1} and \ref{fig:AGammavsSphiLHS1}.

\begin{figure}[!tb]
\centering
\includegraphics[width = 0.45\textwidth]{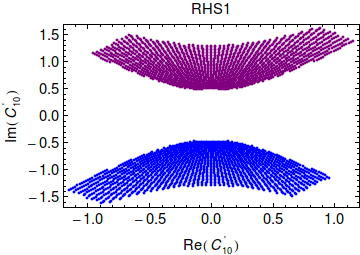}
\includegraphics[width = 0.45\textwidth]{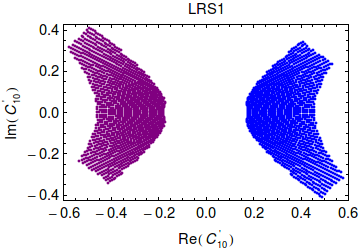}
\caption{\it $C_{10}^\prime$ for $M_{Z^\prime} = 1~$TeV in RHS1 (left) and LRS1 (right). $A_1$: blue, $A_3$: purple. }
 \label{fig:C10}~\\[-2mm]\hrule
\end{figure}

In Fig.~\ref{fig:C10} we show the results on $C^\prime_{10}$ in RH and LR scenarios
without the constraint in Eq.~(\ref{C10C}). We observe that imposing it removes
roughly half of each of oases
$A_1$ and $A_3$  in the RH case and the oasis
$A_1$ (blue)  in the case of LR. We will now investigate the implications
of this finding on the comparison between LHS and RHS scenarios.

\begin{figure}[!tb]
\centering
\includegraphics[width = 0.45\textwidth]{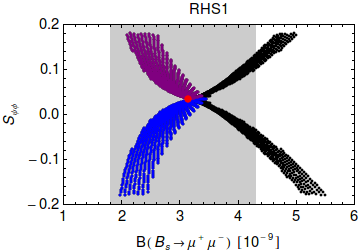}
\caption{\it $S_{\psi\phi}$ and $\mathcal{B}(B_s\to \mu^+\mu^-)$ for $M_{Z^\prime} = 1~$TeV in RHS1. $A_1$: blue, $A_3$: purple,
$A_2$: red, $A_4$: gray. Black: excluded due to $\Re(C_{10}^\prime)\geq 0$. Gray region: exp 1$\sigma$ range  
$\mathcal{B}(B_s\to\mu^+\mu^-) = (2.9^{+1.4}_{-1.1})\cdot 10^{-9}$. Red point: SM central value. }
 \label{fig:BsmuvsSphiRHS}~\\[-2mm]\hrule
\end{figure}

The most interesting impact of this constraint at present is
on the correlation of $S_{\psi\phi}$ and $\mathcal{B}(B_s\to \mu^+\mu^-)$
that for the LHS is given in Fig.~\ref{fig:SmusvsSphiLHS1}. As we discussed
previously the
same result is obtained in the RHS with the two oases $A_1$ (blue) and
$A_3$ (purple) interchanged. However taking the constraint
(\ref{C10C}) into account results in a modified
correlation within RH scenario that we show in Fig.~\ref{fig:BsmuvsSphiRHS}. The black
areas are excluded and in the RH scenario an enhancement of
 $\mathcal{B}(B_s\to \mu^+\mu^-)$ relative to the SM is excluded.
This feature distinguishes LH and RH scenarios. For
$\mathcal{B}(B_s\to \mu^+\mu^-)$ below its  SM value, the measurement
of these two observables cannot distinguish between LH and RH scenarios
as one can always move to the other oasis to obtain the same result.

\boldmath
\subsection{$b\to s\nu\bar\nu$ Observables in Different Scenarios.}
\begin{figure}[!tb]
\centering
\includegraphics[width = 0.7\textwidth]{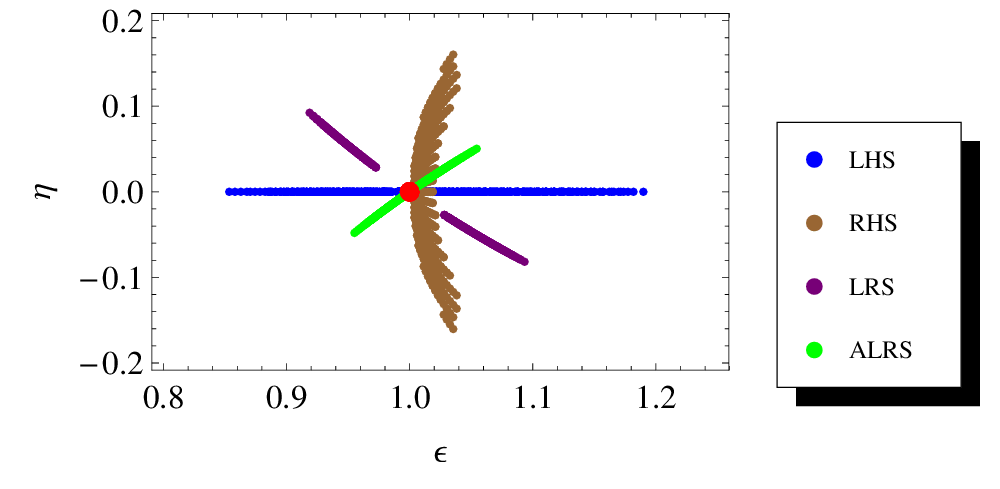}

\caption{\it {  $\eta$ versus $\epsilon$ for scenario LHS1, RHS1, LRS1 and ALRS1.}
}
 \label{fig:ep2vseta}~\\[-2mm]\hrule
\end{figure}

\unboldmath
In view of the important role of these transitions in the search for
RH currents we devote to them a separate subsection.
We begin with the $\epsilon-\eta$ plane proposed in
\cite{Altmannshofer:2009ma}. In Fig.~\ref{fig:ep2vseta} we show the
results for all four scenarios considered by us. Indeed they can be clearly distinguished in this plane. { Indeed a future
determination of $\epsilon$ and 
$\eta$ will tell us whether the nature chooses one of the scenario considered 
by us or a linear combination of them.}

With four observables and four scenarios for $Z'$-couplings there is a multitude of results for specific observables one could present at this stage. Here
we present only some of them that we consider most interesting.

\begin{figure}[!tb]
\centering
\includegraphics[width = 0.45\textwidth]{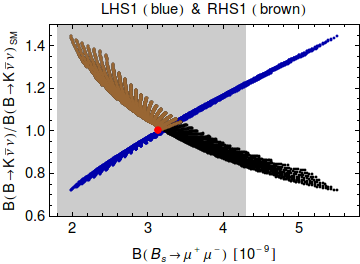}
\includegraphics[width = 0.45\textwidth]{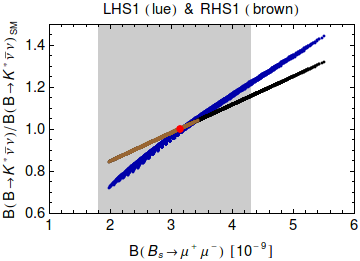}
\caption{\it  $\mathcal{B}(B\to K\nu\bar\nu)$ versus
$\mathcal{B}(B_s\to\mu^+\mu^-)$  (left) and  $\mathcal{B}(B\to K^\star\nu\bar\nu)$ versus
$\mathcal{B}(B_s\to\mu^+\mu^-)$  (right) for
$M_{Z^\prime} = 1~$TeV in LHS1 (blue for both oases $A_{1,3}$) and RHS1 (brown for both oases $A_{1,3}$)),  black points excluded by $b\to
s\ell^+\ell^-$.  Gray region: exp 1$\sigma$ range  
$\mathcal{B}(B_s\to\mu^+\mu^-) = (2.9^{+1.4}_{-1.1})\cdot 10^{-9}$. Red point: SM central
value.}
 \label{fig:BKnuvsBsmu}~\\[-2mm]\hrule
\end{figure}

On the left in Fig.~\ref{fig:BKnuvsBsmu} we show $\mathcal{B}(B\to K\nu\bar\nu)$ versus $\mathcal{B}(B_s\to\mu^+\mu^-)$ in LHS1 and RHS1 scenarios.
Even without the (\ref{C10C}) constraint which eliminates the black region
in the RH scenario, there is a clear distinction between these two scenarios
so that the measurement of these two observables can uniquely tell us whether
we deal with LHS or RHS case. Imposing (\ref{C10C}) we find that in the RH scenario
$\mathcal{B}(B\to K\nu\bar\nu)$ can only be enhanced and
 $\mathcal{B}(B_s\to\mu^+\mu^-)$ suppressed. In the case of
 $\mathcal{B}(B\to K^\star\nu\bar\nu)$ as seen  on the right in the same
figure  its branching ratio can only be suppressed relative to
the SM in RHS but otherwise the distinction between LHS and RHS is not as
pronounced as for $\mathcal{B}(B\to K\nu\bar\nu)$.

\begin{figure}[!tb]
\centering
\includegraphics[width = 0.45\textwidth]{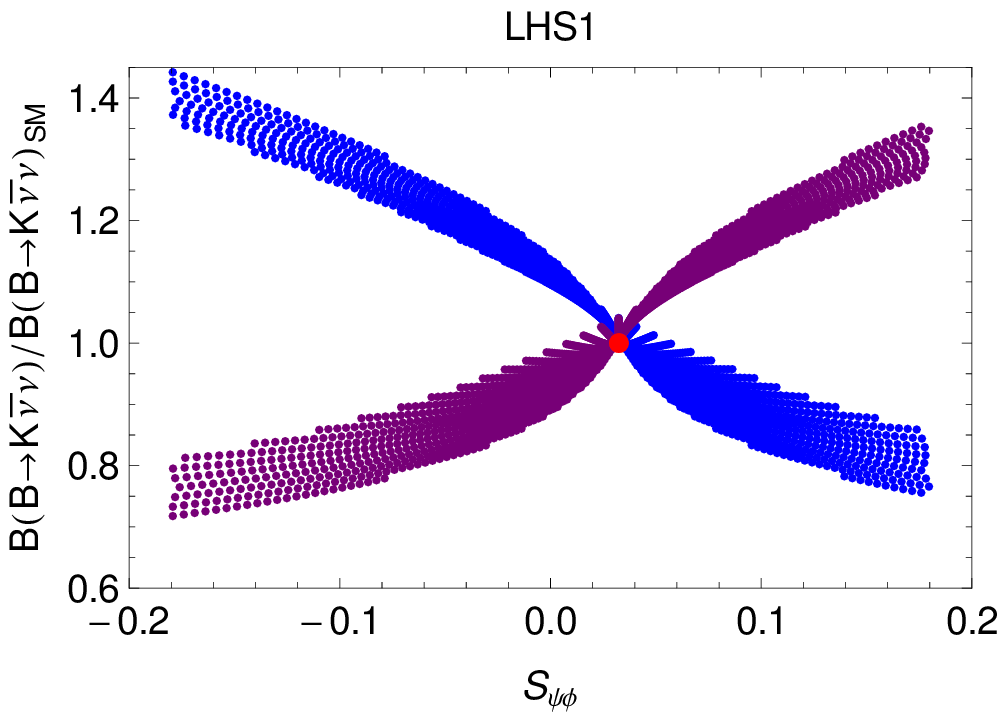}
\includegraphics[width = 0.45\textwidth]{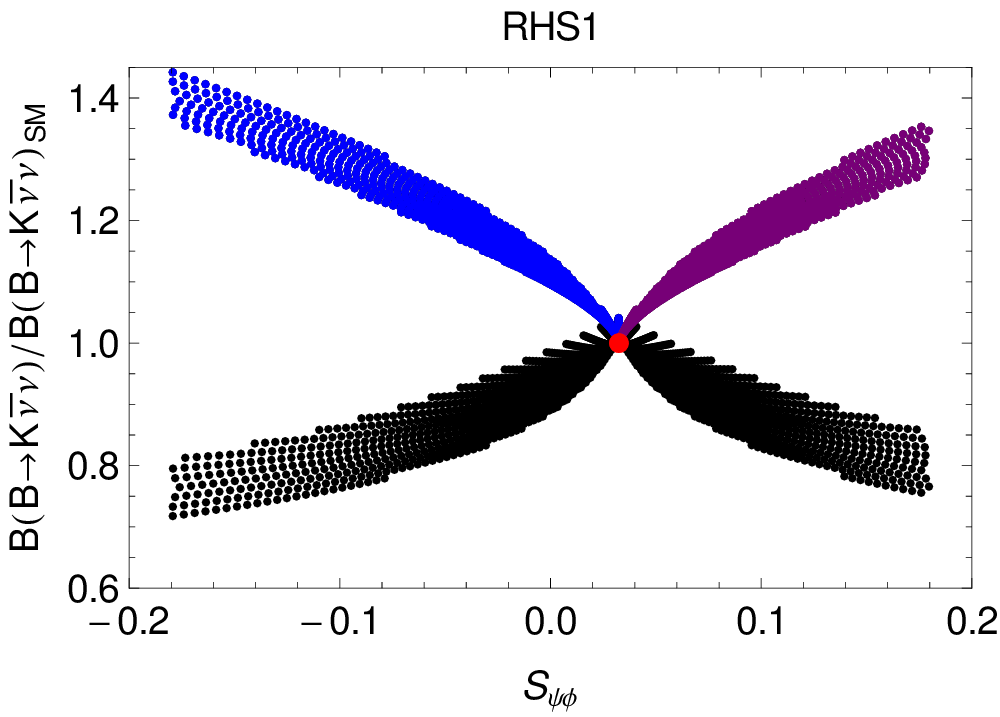}

\includegraphics[width = 0.45\textwidth]{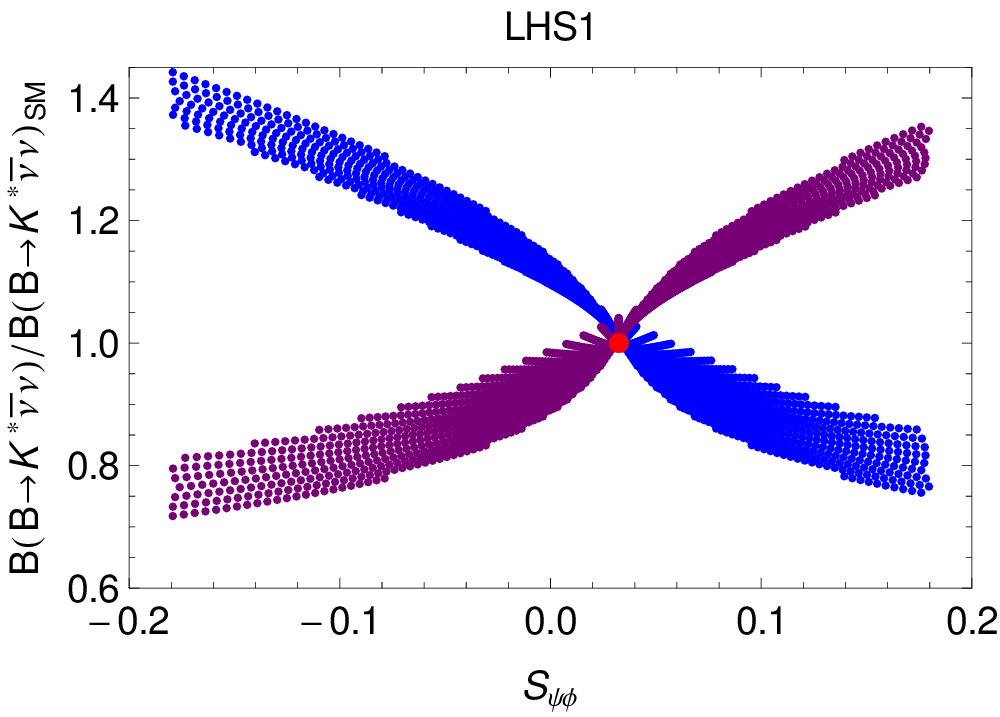}
\includegraphics[width = 0.45\textwidth]{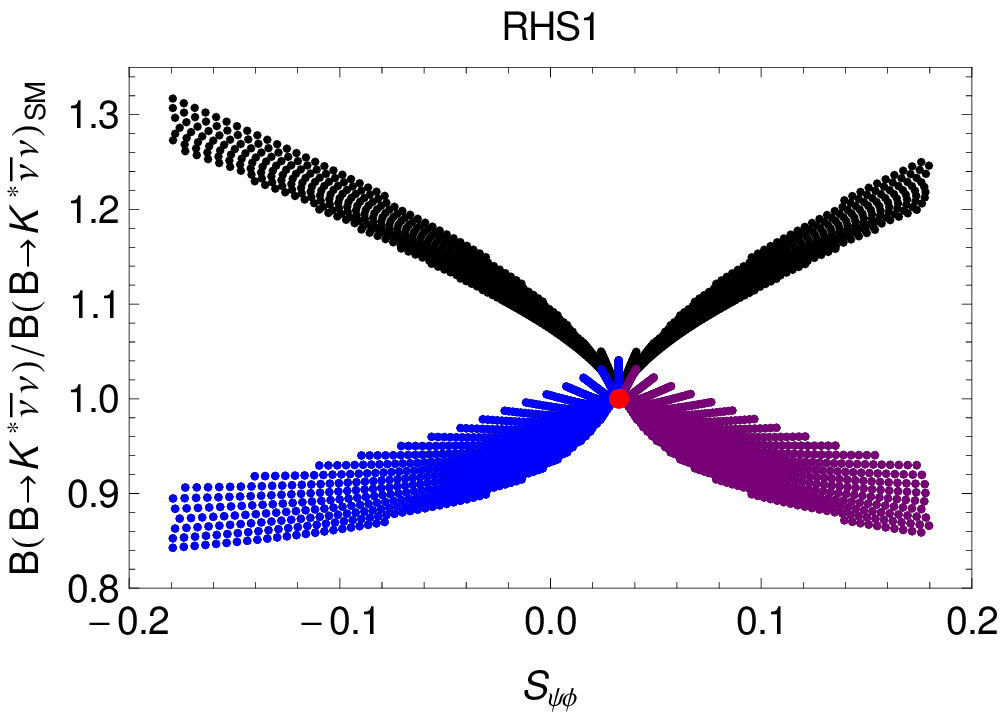}
\caption{\it  $\mathcal{B}(B\to K\nu\bar\nu)$ versus
$S_{\psi\phi}$  (top) and  $\mathcal{B}(B\to K^\star\nu\bar\nu)$ versus
$S_{\psi\phi}$  (down) for
$M_{Z^\prime} = 1~$TeV in LHS1 (left) and RHS1 (right). $A_1$: blue, $A_3$: purple, $A_2$: red, $A_4$: gray. Black points excluded by
$b\to
s\ell^+\ell^-$.  Red point: SM central value.}
 \label{fig:BKnuvsSphi}~\\[-2mm]\hrule
\end{figure}

In Fig.~\ref{fig:BKnuvsSphi}  we show the correlations between
$\mathcal{B}(B\to K\nu\bar\nu)$ and
$S_{\psi\phi}$  (top) and  $\mathcal{B}(B\to K^\star\nu\bar\nu)$ and
$S_{\psi\phi}$ (down) in LHS and RH scenarios with the black regions
excluded by the constraint in  (\ref{C10C}).
We note in particular that
in the RHS the measurement of $S_{\psi\phi}$, if different from the SM value,
will uniquely determine the allowed oasis.

\begin{figure}[!tb]
\centering
\includegraphics[width = 0.45\textwidth]{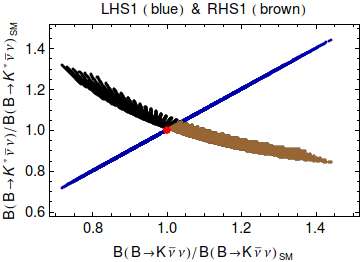}
\includegraphics[width = 0.45\textwidth]{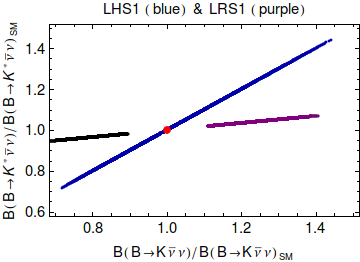}
\caption{\it  $\mathcal{B}(B\to K^\star\nu\bar\nu)$ versus
$\mathcal{B}(B\to K\nu\bar\nu)$  for
$M_{Z^\prime} = 1~$TeV in LHS1 (blue for both oases $A_{1,3}$),  RHS1 (brown for both oases $A_{1,3}$) and LRS1 (purple for both oases
$A_{1,3}$)), black points excluded by $b\to s\ell^+\ell^-$. Red point:
SM central value.}
 \label{fig:BKstarnuvsBKnu}~\\[-2mm]\hrule
\end{figure}

Finally, in Fig.~\ref{fig:BKstarnuvsBKnu} we show $\mathcal{B}(B\to K^\star\nu\bar\nu)$ versus
$\mathcal{B}(B\to K\nu\bar\nu)$ in RH and LR scenarios compared to LH scenario
with black regions excluded by the constraint in (\ref{C10C}). Also these
plots will test the presence of RH currents as emphasized in
\cite{Buras:2010pz}.

The observable $F_L$ being dependent only on $\eta$ can in principle
serve to identify the presence of RH currents. However in our case, as seen
in Fig.~\ref{fig:ep2vseta},
$|\eta|\le 0.16$, and NP effects in $F_L$ turn out to be small as can be deduced
from Fig.~4 in \cite{Altmannshofer:2009ma}.

\boldmath
\section{The $U(2)^3$ Limit}\label{sec:U(2)}
\unboldmath
We have investigated how the parameter space is further constrained when
the flavour $U(2)^3$ symmetry \cite{Barbieri:2011ci,Barbieri:2011fc,Barbieri:2012uh,Barbieri:2012bh,Crivellin:2011fb,Crivellin:2011sj,Crivellin:2008mq}
is imposed on the $Z'$ couplings.
 As pointed out in \cite{Buras:2012sd} in this case $\varphi_{B_d}=\varphi_{B_s}$
which in turn implies
not only  the correlation between CP asymmetries
$S_{\psi K_S}$ and $S_{\psi\phi}$
but also  a  triple $S_{\psi K_S}-S_{\psi\phi}-|V_{ub}|$ correlation.

As in \cite{Buras:2012sd} we will only consider the case of $U(2)^3$
broken by the minimal set of spurions: the $MU(2)^3$ case. Then only the
LHS1 and LHS2 are involved. We find then that $\tilde s_{ij}$ and
$\delta_{ij}$ are constrained at the fundamental level as follows.

In the $K$ system we have
\be
\tilde s_{12}=a \vtd\vts, \qquad \delta_{12}=\beta-\beta_s,
\ee
where $a\ge 0$ and real. Thus NP effects in $\varepsilon_K$, $\kpn$,
$\klpn$ and $K_L\to \mu^+\mu^-$ are described for fixed leptonic couplings
by a single real and positive definite parameter. Once this parameter is
fixed through one of these observables, the others are uniquely predicted.
Note that this is even more predictive than CMFV in which except for common
CKM couplings there is no relation between $\Delta F=2$ and $\Delta F=1$
transitions unless some ratios are constructed \cite{Buras:2003td}.

The observables in $B_d$ and $B_s$ systems are correlated with each other
due to the relations:
\be\label{equ:U23relation}
\frac{\tilde s_{13}}{\vtd}=\frac{\tilde s_{23}}{\vts}, \qquad
\delta_{13}-\delta_{23}=\beta-\beta_s.
\ee
Thus, once the allowed oases in the $B_d$ system are fixed, the oases in $B_s$
system are determined. Moreover, all observables in both systems are described
by only one real positive parameter and one phase, e.g. $({\tilde s}_{23},\delta_{23})$.

\begin{figure}[!tb]
\centering
\includegraphics[width = 0.45\textwidth]{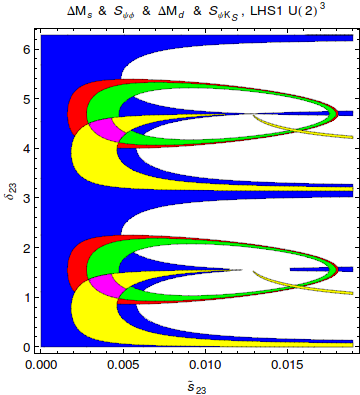}
\includegraphics[width = 0.45\textwidth]{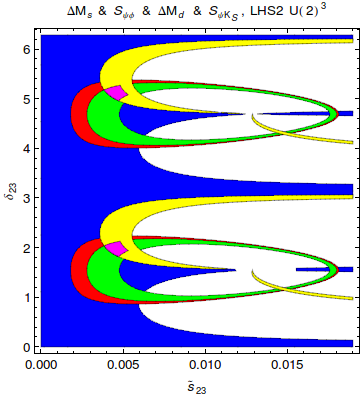}
\caption{\it Ranges for $\Delta M_s$ (red region), $S_{\psi \phi}$ (blue region), $\Delta M_d$ (green region) and $S_{\psi K_S}$
(yellow region)for $M_{Z^\prime}=1$~TeV in LHS1 (left) and LHS2 (right) in the $U(2)^3$ limit satisfying the bounds
in Eq.~(\ref{C1}) and ~(\ref{C2}). The overlap region of all four regions is shown in magenta. }
 \label{fig:oasesU2}~\\[-2mm]\hrule
\end{figure}

In Fig.~\ref{fig:oasesU2} we combine Figs.~\ref{fig:oasesBsLHS1} and~\ref{fig:oasesBdLHS} using the $U(2)^3$ symmetry
relations in (\ref{equ:U23relation}). In the $U(2)^3$ limit the small oases are eliminated and the big oases get smaller. This decrease turns out to be
not very pronounced in the case of $({\tilde s}_{13},\delta_{13})$ oases
as they were already small as seen in Fig.~\ref{fig:oasesBdLHS} but has
a significant impact on $({\tilde s}_{23},\delta_{23})$ oases which where
much larger as seen in  Fig.~\ref{fig:oasesBsLHS1}. Moreover the fact that
the results in the $B_d$ system depend on whether LHS1 or LHS2 is considered
is now transfered through the relations in (\ref{equ:U23relation}) into
the $B_s$ system. This is clearly seen in Fig.~\ref{fig:oasesU2}, in particular
the final oases (magenta) in LHS2 are  smaller than in LHS1 due to the required
shift of $S_{\psi K_S}$.

\begin{figure}[!tb]
\centering
\includegraphics[width = 0.45\textwidth]{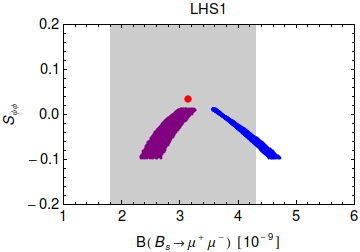}
\includegraphics[width = 0.45\textwidth]{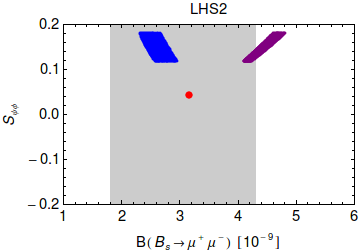}
\caption{\it $S_{\psi\phi}$ versus $\mathcal{B}(B_s\to\mu^+\mu^-)$  for
$M_{Z^\prime} = 1~$TeV in LHS1 (left) and LHS2 (right) in the $U(2)^3$ limit. Blue region corresponds to the lower magenta oases
in Fig.~\ref{fig:oasesU2} (former $A_1$) and the purple region corresponds to the upper magenta oases (former $A_3$).  Gray region: exp
1$\sigma$ range  
$\mathcal{B}(B_s\to\mu^+\mu^-) = (2.9^{+1.4}_{-1.1})\cdot 10^{-9}$. }
 \label{fig:BsmuvsSphiU2LHS}~\\[-2mm]\hrule
\end{figure}

This change of allowed oases in the $B_s$ system has a profound impact
on the correlation between $S_{\psi\phi}$ and
$\mathcal{B}(B_s\to\mu^+\mu^-)$. We show this in
Fig.~\ref{fig:BsmuvsSphiU2LHS} that should be compared with the corresponding
correlation in  Fig.~\ref{fig:SmusvsSphiLHS1}.
We observe that already the sign of $S_{\psi\phi}$ will decide whether LHS1 or
LHS2 is favoured. Moreover if $\mathcal{B}(B_s\to\mu^+\mu^-)$ will
turned out to be suppressed relatively to the SM then only one oasis
will survive in each scenario. Comparison with future precise values 
of $\vub$ will confirm or rule out this scenario of NP.
These correlations are particular
examples of the correlations in $MU(2)^3$ models pointed out in
\cite{Buras:2012sd}. What
is new here is that in a specific model considered by us the $\vub-S_{\psi\phi}$
correlation has now also implications for $\mathcal{B}(B_s\to\mu^+\mu^-)$.

We also note that in this case \cite{Buras:2012sd}
\be\label{Fleischer}
S^s_{\mu^+\mu^-}=S^d_{\mu^+\mu^-}=\sin (2\theta_Y-2\varphi_{\rm new}),
\ee
where
\be
\theta_Y=\theta_Y^{d}=\theta_Y^{s}, \quad \varphi_{\rm new}=\varphi_{B_d}=\varphi_{B_s}.
\ee
Moreover, as the CMFV relations for $\mathcal{B}(B_{s,d}\to\mu^+\mu^-)$, 
also apply in this case the result in (\ref{LHCb2corr}) allows 
to find \cite{Buras:2012sd}
\be\label{boundMFV1}
 \mathcal{B}(B_{d}\to\mu^+\mu^-)=(1.0^{+0.5}_{-0.3})\times 10^{-10}, \quad
 ({\rm CMFV,~MU(2)^3}).
 \ee

Finally we remark that within the $\model$ model, analyzed by us in
\cite{Buras:2012xx}, the imposition of $U(2)^3$ symmetry implies the
relations:
\be
\tilde s_{13}=b\vtd, \qquad \tilde s_{23}=b\vts, \qquad
\delta_1-\delta_2=\beta -\beta_s+\pi,
\ee
where $b\ge0$ and real.
Consequently NP in all three systems is described by only one real positive
definite parameter and one phase.

\section{Flavour Violating SM $Z$ Boson}\label{sec:ZSM}
\subsection{Preliminaries}
We will now turn our attention to flavour violating $Z$-couplings that
can be generated in the presence of other neutral gauge bosons and or
new heavy vectorial fermions with $+2/3$ and $-1/3$ electric charges.
RSc is an explicit model of this type \cite{Blanke:2008yr,Buras:2009ka}. 
See also \cite{delAguila:2011yd}.
Recently, an extensive analysis of flavour violation in the presence
of a vectorial $+2/3$ quark has been presented in
\cite{Botella:2012ju}, where
references to previous literature can be found.
 In the case considered by us, new quarks with  $-1/3$ charges are essential for generating
flavour violating couplings to SM down-quarks but the presence of  heavy
quarks with $+2/3$ charges could be relevant for charm physics. Moreover,
such heavy fermions could contribute to rare $K$ and $B$ decays through
loop diagrams. In what follows we will assume that these loop contributions
can be neglected in comparison with the tree-level effects discussed by
us. Of course in a concrete model one has to check whether this assumption is
justified.

The strategy and formalism developed in the previous sections can be used
in a straightforward manner  for the case of $Z$ flavour-violating couplings to
quarks.
In this case we have
\be
M_Z=91.2\gev, \quad \Delta_L^{\nu\bar\nu}(Z)=\Delta_A^{\mu\bar\mu}(Z)=0.372,
\quad  \Delta_V^{\mu\bar\mu}(Z)=-0.028
\ee
The implications of these changes are as follows:
\begin{itemize}
\item
The decrease of the neutral gauge boson mass by an order of magnitude relatively to the nominal value $M_{Z'}=1\tev$ used by us decreases the couplings
$\tilde s_{ij}$ by the same amount without any impact on the phases
$\delta_{ij}$ when the constraints from
$\Delta F=2$ processes are imposed.
\item
As already noticed in \cite{Buras:2012xx} and discussed at the beginning
of our paper once the parameters $\tilde s_{ij}$
are constrained through $\Delta F=2$ observables
 the decrease of neutral gauge boson mass enhances NP effects in rare $K$
and $B$ decays. This follows from the structure of tree-level contributions
to FCNC processes and is not generally the case when NP contributions are
governed by penguin and box diagrams. The formulae in Section~\ref{sec:3a}
exhibit this feature transparently.
\item
The latter fact implies that already the present experimental
bounds on $\mathcal{B}(\kpn)$ and $\mathcal{B}(B_{s,d}\to\mu^+\mu^-)$
as well as the data on
$B\to X_s \ell^+\ell^-$, $B\to   K^*\ell^+\ell^-$  and
 $B\to   K\ell^+\ell^-$  decays become more powerful than the $\Delta F=2$
transitions in constraining flavour violating couplings of $Z$ so that
effects in $\Delta F=2$ processes cannot be as large as in $Z'$ case.
\end{itemize}

We will now investigate how the flavour-violating $Z$ couplings perform
in the three meson systems.

 \boldmath
\subsection{The $B_s$ Meson System}
\begin{figure}[!tb]
\begin{center}
\centering
\includegraphics[width=0.45\textwidth] {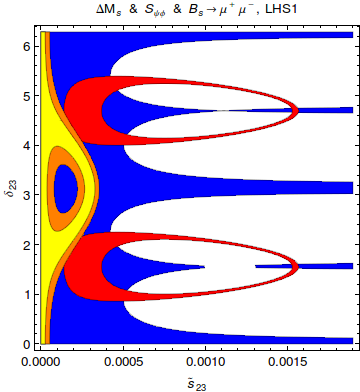}
\caption{\it Ranges for $\Delta M_s$ (red region), $S_{\psi \phi}$ (blue region) and $B_s\to \mu^+\mu^-$ (yellow)
in LHS1 satisfying the bounds
in Eq.~(\ref{C1}) and $\mathcal{B}(B_s\to\mu^+\mu^-)$ in 1$\sigma$  range $[1.8,4.3]\cdot 10^{-9}$ (yellow) and 
$[0.7,5.7]\cdot 10^{-9}$ (orange).  
}\label{fig:ZoasesBsLHS1}~\\[-2mm]\hrule
\end{center}
\end{figure}

\unboldmath
We used first only the  $\Delta M_s$  and $S_{\psi\phi}$ constraints finding
that small oases are excluded by the data on $B_s\to\mu^+\mu^-$.  
However, 
also in big oases $\mathcal{B}(B_s\to\mu^+\mu^-)$ is always larger than
its SM value and mostly above the data except in LRS case where NP contributions vanish. 
In Fig.~\ref{fig:ZoasesBsLHS1} (in the case of the LHS1 scenario) we show the
region allowed by $\Delta M_s$ and $S_{\psi\phi}$ contraint  together
with the yellow region allowed by $\mathcal{B}(B_s\to\mu^+\mu^-) = (2.9^{+1.4}_{-1.1})\cdot 10^{-9}$.
We observe no overlap between these regions.  A small orange region is still 
left when $2\sigma$ range for $\mathcal{B}(B_s\to\mu^+\mu^-)$ is considered.
Thus at first sight one could conclude that
in contrast to the $Z'$ scenarios 
 $\mathcal{B}(B_s\to\mu^+\mu^-)$ could reach values $2 \sigma$ 
away from its experimental value.

However, when the values of $C_{10}^{\rm NP}$ and  $C_{10}^{\prime}$ are
considered the situation gets worse. The $2\sigma$ bounds on
these coefficients from \cite{Altmannshofer:2012ir} (see Eq.~(\ref{equ:ASconstraint}))  are  violated in
LHS and RHS cases and for the LRS scenario only small room is left. Indeed we find
\be\label{CC1}
|\Im (C_{10})|\ge 3.5~~{\rm (LHS1)}, \qquad  |\Im (C_{10})^\prime|\ge 3.5~~{\rm (RHS1)}
\ee
and
\be\label{CC2}
|\Re( C_{10}^\text{NP})|\ge 1.5~~{\rm (LRS1)}, \qquad  |\Re (C_{10})^\prime|\ge 1.5~~{\rm (LRS1)}.
\ee

Therefore, the main message from this exercise is that when the above
constraints are taken into account it is very difficult to suppress
$\Delta M_s$ sufficiently in LHS, LRS and RHS scenarios
without violating the constraints from $b\to s \mu^+\mu^-$
transitions. We conclude therefore that 
this NP scenario appears to be strongly 
disfavoured  even if not fully ruled out because of assumed small 
hadronic uncertainties.

In ALR scenario we find
\be\label{CC3}
|\Re( C_{10}^\text{NP})|\ge 1.5~~{\rm (ALRS1)}, \qquad  |\Im( C_{10})^\prime|\ge 1.0~~
{\rm (ALRS1)}\,.
\ee
While this scenario is in a slightly better shape most of the allowed space
is ruled out as well.

We have also calculated $C_9$ and $C_{9}^\prime$ coefficients. Due to
the smallness of $\Delta^{\mu\bar\mu}_V$ in the SM, the present $2\sigma$
bounds are satisfied in all scenarios.  Yet, in view of the results for
$C_{10}$ it does not look that we should expect much from flavour-violating $Z$
couplings in $B_s$ system and consequently we will not consider
$b\to s\nu\bar\nu$ transitions. Similar conclusions have been reached in
\cite{Altmannshofer:2012ir,Beaujean:2012uj}.

 \boldmath
\subsection{The $B_d$ Meson System}
\begin{figure}[!tb]
\begin{center}
\includegraphics[width=0.45\textwidth] {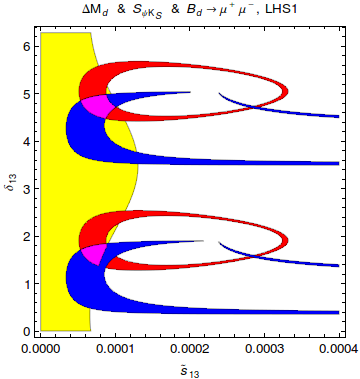}
\includegraphics[width=0.45\textwidth] {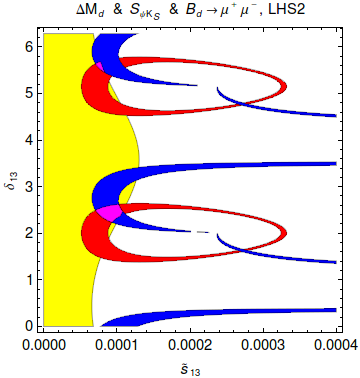}
\caption{\it  Ranges for $\Delta M_d$ (red region), $S_{\psi K_S}$ (blue region) and $B_d\to \mu^+\mu^-$ (yellow)
in LHS1 (left) and LHS2 (right) satisfying the bounds
in Eq.~(\ref{C2}) and $\mathcal{B}(B_d\to \mu^+\mu^-)\leq 9.4\cdot 10^{-10}$.
}\label{fig:ZoasesBdLHS1}~\\[-2mm]\hrule
\end{center}
\end{figure}

\begin{figure}[!tb]
\begin{center}
\includegraphics[width=0.45\textwidth] {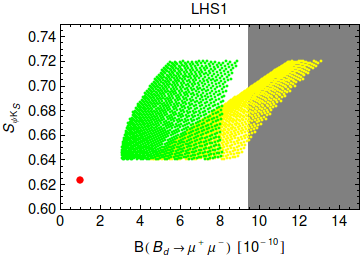}
\includegraphics[width=0.45\textwidth] {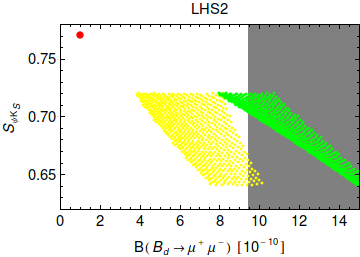}

\includegraphics[width=0.45\textwidth] {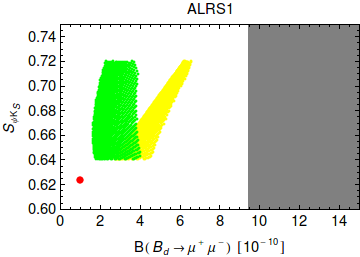}
\includegraphics[width=0.45\textwidth] {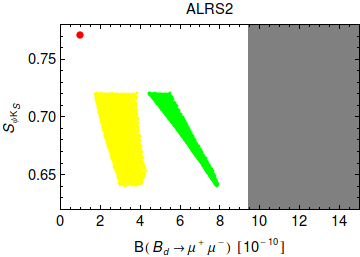}
\caption{\it  $S_{\psi K_S}$ versus   $\mathcal{B}(B_d\to\mu^+\mu^-)$  in LHS1, LHS2 (upper row) and ALRS1, ALRS2 (lower row).
$B_1$: yellow, $B_3$: green.  Red point:
SM central
value. Gray region: excluded by $\mathcal{B}(B_d\to\mu^+\mu^-)\leq 9.4\cdot 10^{-10}$.}\label{fig:ZBdmuvsSKSLHS}~\\[-2mm]\hrule
\end{center}
\end{figure}

\unboldmath
In the $B_d$ system using the same constraints as before we find
the allowed oases shown in Fig.~\ref{fig:ZoasesBdLHS1}. 
{ While the magenta regions
in this plot are also allowed by the upper bound on
$\mathcal{B}(B_d\to \mu^+\mu^-)$, 
Fig.~\ref{fig:ZBdmuvsSKSLHS} shows that the latter bound has already and 
impact on the LHS1 and LHS2 scenarios. This should  be compared with the 
$Z'$ case in Fig.~\ref{fig:BdmuvsSKSLHS}, where NP effects where much 
smaller.} This time we show also the
results in ALRS1 and ALRS2 scenarios in
which NP effects are smaller than in LHS1 and LHS2 scenarios. We only
show the results for big oases as small oases imply typicaly values
for $\mathcal{B}(B_d\to\mu^+\mu^-)$ of $8\cdot 10^{-9}$ and $4\cdot 10^{-9}$
for LHS and ALRS, respectively and are ruled out.

\begin{figure}[!tb]
\begin{center}
\includegraphics[width=0.45\textwidth] {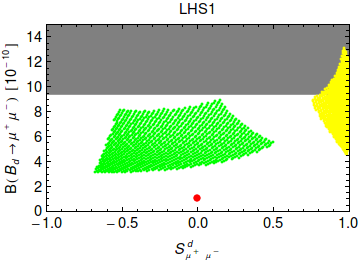}
\includegraphics[width=0.45\textwidth] {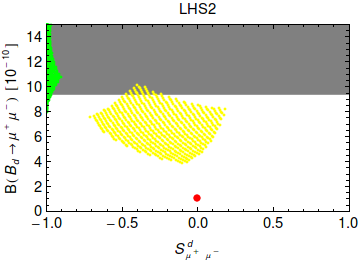}

\includegraphics[width=0.45\textwidth] {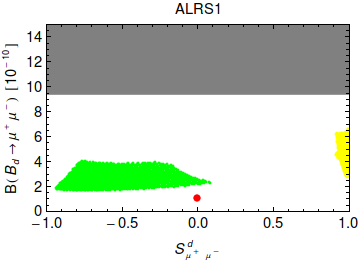}
\includegraphics[width=0.45\textwidth] {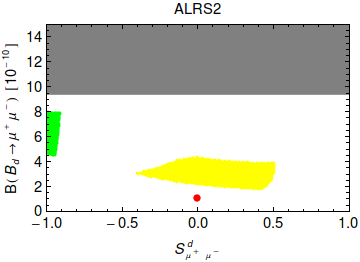}
\caption{\it    $\mathcal{B}(B_d\to\mu^+\mu^-)$ versus $S_{\mu^+\mu^-}^{d}$ in LHS1, LHS2 (upper row) and ALRS1, ALRS2 (lower
row).
$B_1$: yellow, $B_3$: green.  Red point:
SM central
value. Gray region: excluded by $\mathcal{B}(B_d\to\mu^+\mu^-)\leq 9.4\cdot 10^{-10}$.}\label{fig:ZBdmuvsSmudLHS}~\\[-2mm]\hrule
\end{center}
\end{figure}

We observe that $\mathcal{B}(B_d\to\mu^+\mu^-)$ can strongly be enhanced
in all shown scenarios so that with improved bound on this branching
ratio LHS1 and LHS2 scenarios could be put into difficulties, while in 
ALRS1 and ALRS2 one could easier satisfy these bounds. If such a situation
really took place and NP effects would be observed in this decay, this would
mean that both LH and RH $Z'$-couplings in the $B_d$ system would be required
but with opposite sign.

 Fig.~\ref{fig:ZBdmuvsSmudLHS} shows that not only $\mathcal{B}(B_d\to\mu^+\mu^-)$ but also the CP-asymmetry $S_{\mu^+\mu^-}^{d}$ can deviate significantly 
from SM expectation.

 \boldmath
\subsection{The $K$ Meson System}
\unboldmath
The effects of flavour violating $Z$ couplings
in $\kpn$ and $\klpn$ can be very large in LHS, RHS and LRS but
they can be bounded by the upper bound on
$K_L\to\mu^+\mu^-$ except for the LR scenarios and the case of purely
imaginary NP contributions in all these scenarios
 where this bound is ineffective.

\begin{figure}[!tb]
\begin{center}
\includegraphics[width=0.45\textwidth] {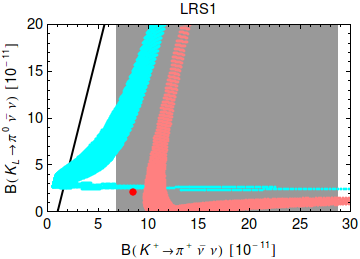}
\includegraphics[width=0.45\textwidth] {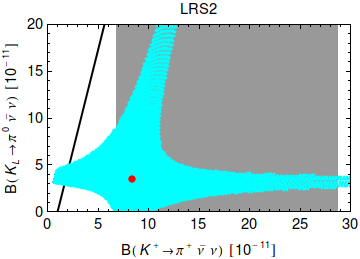}
\caption{\it $\mathcal{B}(\klpn)$ versus
$\mathcal{B}(\kpn)$in LRS1 (left) and LRS2 (right).   $C_1$: cyan, $C_2$: pink. Red point: SM central
value. Gray region:
experimental range of $\mathcal{B}(\kpn)$.}\label{fig:ZKLvsKpLRS}~\\[-2mm]\hrule
\end{center}
\end{figure}

We begin therefore with LRS1 and LRS2 scenarios and show in Fig.~\ref{fig:ZKLvsKpLRS}
$\mathcal{B}(\klpn)$ vs $\mathcal{B}(\kpn)$. Indeed the NP effects can be
much larger  than in the $Z'$ case shown in
Fig.~\ref{fig:KLvsKpLRS}. 

\begin{figure}[!tb]
\begin{center}
\includegraphics[width=0.45\textwidth] {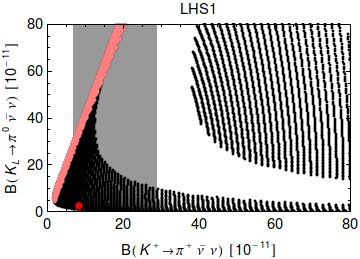}
\includegraphics[width=0.45\textwidth] {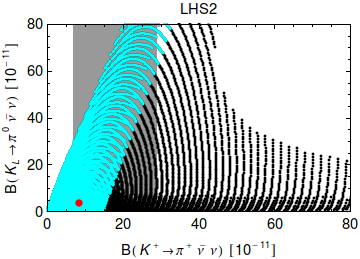}
\caption{\it $\mathcal{B}(\klpn)$ versus
$\mathcal{B}(\kpn)$in LHS1 (left) and LHS2 (right).   $C_1$: cyan, $C_2$: pink. Black points excluded due to $\mathcal{B}(K_L\to \mu^+\mu^-)\leq 2.5\cdot 10^{-9}$ constraint. Red point: SM central
value. Gray region:
experimental range of $\mathcal{B}(\kpn)$.}\label{fig:ZKLvsKpLHS}~\\[-2mm]\hrule
\end{center}
\end{figure}

\begin{figure}[!tb]
\begin{center}
\includegraphics[width=0.45\textwidth] {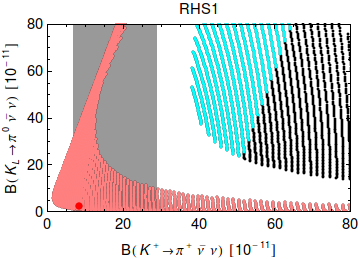}
\includegraphics[width=0.45\textwidth] {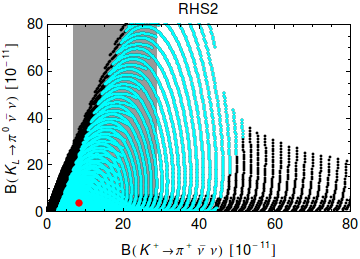}
\caption{\it $\mathcal{B}(\klpn)$ versus
$\mathcal{B}(\kpn)$in RHS1 (left) and RHS2 (right).   $C_1$: cyan, $C_2$: pink. Black points excluded due to $\mathcal{B}(K_L\to
\mu^+\mu^-)\leq 2.5\cdot 10^{-9}$ constraint. Red point: SM central
value. Gray region:
experimental range of $\mathcal{B}(\kpn)$.}\label{fig:ZKLvsKpRHS}~\\[-2mm]\hrule
\end{center}
\end{figure}

In Fig.~\ref{fig:ZKLvsKpLHS} we show analogous result for LHS1 and LHS2 case imposing
the $K_L\to\mu^+\mu^-$ constraint. We observe that in both scenarios only
the branch unaffected by the $K_L\to\mu^+\mu^-$ constraint survives. This is
$C_2(S1)$ and $C_1(S2)$. In  $C_1(S1)$
$\mathcal{B}(\kpn)\ge 3.5\cdot 10^{-10}$ and this case is ruled out. In Fig.~\ref{fig:ZKLvsKpRHS} we show the corresponding
results for RHS1 and RHS2 where the $K_L\to\mu^+\mu^-$ constraint has
a different impact than in LHS cases. Also here NP effects can be very large.

\begin{figure}[!tb]
\centering
\includegraphics[width = 0.45\textwidth]{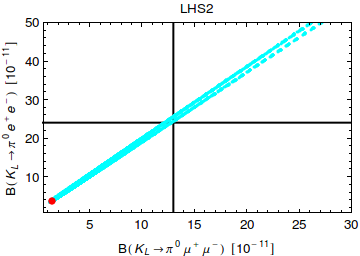}
\includegraphics[width = 0.45\textwidth]{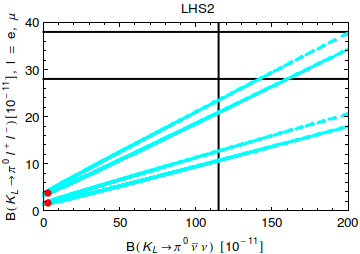}
\caption{\it $\mathcal{B}(K_L \to \pi^0 e^+ e^-)$ as a
  function of $\mathcal{B}(K_L\to \pi^0 \mu^+\mu^-)$ (left panel) and 
$\mathcal{B}(K_L\to \pi^0 e^+e^-)$ (upper curve) and  
$\mathcal{B}(K_L \to \pi^0\mu^+\mu^-)$ (lower curve) as functions of 
$\mathcal{B}Br(\klpn)$ (right panel) in LHS2. 
The red points represent SM
predictions.}
\label{fig:KmuKeZ}~\\[-2mm]\hrule
\end{figure}

\begin{figure}[!tb]
\centering
\includegraphics[width = 0.45\textwidth]{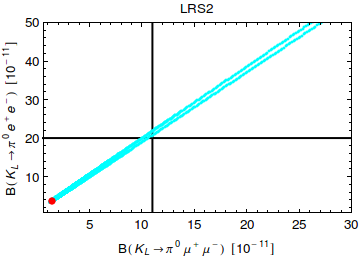}
\includegraphics[width = 0.45\textwidth]{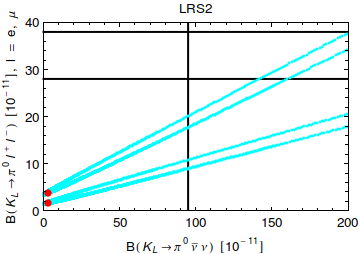}
\caption{\it $\mathcal{B}(K_L \to \pi^0 e^+ e^-)$ as a
  function of $\mathcal{B}(K_L\to \pi^0 \mu^+\mu^-)$ (left panel) and 
$\mathcal{B}(K_L\to \pi^0 e^+e^-)$ (upper curve) and  
$\mathcal{B}(K_L \to \pi^0\mu^+\mu^-)$ (lower curve) as functions of 
$\mathcal{B}Br(\klpn)$ (right panel) in LRS2. 
The red points represent SM
predictions.}
\label{fig:LRKmuKeZ}~\\[-2mm]\hrule
\end{figure}

Finally we discuss $K_L\to \pi^0\ell^+\ell^-$ decays.
In Figs.~\ref{fig:KmuKeZ} and \ref{fig:LRKmuKeZ}  we show the results corresponding to the ones found for $Z'$ LHS and LRS scenarios. As the results for 
S1 scenarios turn out to be very similar we only show the results for S2 scenarios.  The meaning of the curves is the same as in the case of $Z'$ results in 
Figs.~\ref{fig:KmuKe} and \ref{fig:LRKmuKe}. We observe that NP effects in 
 $K_L\to \pi^0\ell^+\ell^-$ decays can be very large but they are bounded 
by the upper bound on $\mathcal{B}(\klpn)$ which follows from the present bound 
on $\mathcal{B}(\kpn)$. We note that in LHS scenarios the upper bound on 
$\mathcal{B}(\klpn)$ practically coincides with the GN bound and 
amounts to
\be\label{constZ}
\mathcal{B}(\klpn)\le 115\cdot 10^{-11}.
\ee
It is slightly weaker in LRS scenarios. In any case the 
the present 
upper bounds on $\mathcal{B}(K_L\to \pi^0\ell^+\ell^-)$ do not preclude 
large NP effects found in $\mathcal{B}(\kpn)$ and 
 $\mathcal{B}(\klpn)$.

\boldmath
\subsection{Comments on $\epe$}
\unboldmath
The large NP effects in $\kpn$ and $\klpn$ both through $Z'$ and $Z$ 
tree level exchanges belong clearly to highlights of our paper. Yet we would like 
to emphasize at this point that in principle these large effects could  
be eliminated by the ratio $\epe$ if the relevant hadronic matrix elements 
were precisely known. As already pointed out in 
\cite{Buras:1998ed,Buras:1999da} there is a strong 
correlation between  $\kpn$ and $\klpn$ and $\epe$ because electroweak 
penguin contributions that are relevant for $\epe$ govern  $\kpn$ and $\klpn$ 
even if operators are different.
The strongest correlation is between $\mathcal{B}(\klpn)$ and $\epe$ because they 
are both CP-violating. However, if $\mathcal{B}(\klpn)$ 
is bounded by $\epe$ then 
automatically $\mathcal{B}(\kpn)$ is bounded on the branch parallel to the 
GN bound on which their
 ratio is 
approximately constant. 
On the second branch 
$\mathcal{B}(\kpn)$ is less affected but there the bound from $K_L\to\mu^+\mu^-$ plays a role unless we work in LRS scenario.

Unfortunately, in spite of the recent progress on the calculation of the 
hadronic matrix elements  relevant for $\epe$ 
\cite{Blum:2011pu,Blum:2011ng,Blum:2012uk},  the hadronic 
uncertainties in $\epe$ 
are still too large for reaching a clear cut conclusion on the impact of 
this ratio on rare $K$ decays. An analysis of $\epe$ in the LHT model demonstrates this problem
in explicit terms \cite{Blanke:2007wr}. If one uses hadronic matrix 
elements of QCD and electroweak penguin operators obtained 
in the large N approach, $(\epe)_{\rm SM}$ is in the ballpark of the 
experimental data and sizable departures of $\mathcal{B}(\klpn)$ from its SM 
value are not allowed. $\kpn$ being CP conserving and consequently 
not as strongly correlated with $\epe$ as $\mathcal{B}(\klpn)$ could still be 
enhanced by $50\%$ in LHT. On the other hand if hadronic matrix elements in question 
differ significantly from their large N values,  
$(\epe)_{\rm SM}$ disagrees with experiment and much more
room for enhancements of rare K decay branching ratios through
NP contributions is available.

\section{Summary and Conclusions}\label{sec:5}

\begin{table}[!ht]
{\renewcommand{\arraystretch}{1.1}
\begin{center}\small
\begin{tabular}{|c|c|c|c|}
\hline
$B_{d,s}$ systems: observables & scenarios & Figure & comments
\\
\hline
\hline
$\Delta M_s$ \& $S_{\psi\phi}$ oases & LHS/RHS $Z^\prime$ & \ref{fig:oasesBsLHS1} &\scriptsize{ oases ($A_1,A_2$,$A_3,A_4$) found}\\
 & LRS $Z^\prime$ & \ref{fig:oasesBsLRS1}&\scriptsize{ oases ($A_1,A_2$,$A_3,A_4$) found}\\
& LHS $U(2)^3$ $Z^\prime$ &\ref{fig:oasesU2} &\scriptsize{two common oases for $B_s$ and $B_d$}\\
 & LHS $Z$ & \ref{fig:ZoasesBsLHS1} & \scriptsize{ no oases found}\\
\hline
$S_{\psi\phi}$ vs. $B_s\to\mu^+\mu^-$ & LHS $Z^\prime$ & \ref{fig:SmusvsSphiLHS1} &\scriptsize{excludes ($A_2,\,A_4$)}\\
 &&&\scriptsize{(anti)correlation found in $A_3\,(A_1)$}\\
 & RHS $Z^\prime$ &\ref{fig:BsmuvsSphiRHS} &\scriptsize{(anti)correlation found in $A_1\,(A_3)$}\\
 & LHS $U(2)^3$ $Z^\prime$ &\ref{fig:BsmuvsSphiU2LHS} & \\
\hline
$S_{\mu^+\mu^-}^s$ vs. $S_{\psi\phi}$ & LHS $Z^\prime$ & \ref{fig:SmusvsSphiLHS1} &\scriptsize{$S_{\mu^+\mu^-}^s>0\,(<0)$  in $A_1 (A_3)$}\\
\hline
$B\to X_s\bar\nu\nu$ vs. $B_s\to\mu^+\mu^-$ &  LHS $Z^\prime$ &\ref{fig:AGammavsSphiLHS1} & \scriptsize{correlation found in
$A_3,\,A_1$}\\
\hline
$B\to X_s\bar\nu\nu$ vs. $S_{\psi\phi}$ & LHS $Z^\prime$ &\ref{fig:BXsnuvsSphiLHS1}  &\scriptsize{(anti)correlation found in
$A_3\,(A_1)$}\\
\hline
$B\to K^{(\star)}\bar\nu\nu$ vs. $B_s\to\mu^+\mu^-$ &  LHS/RHS $Z^\prime$ & \ref{fig:BKnuvsBsmu} & \scriptsize{$B\to K\bar\nu\nu$ :
(anti)correlation in
LHS(RHS)}\\
&&&\scriptsize{$B\to K^{\star}\bar\nu\nu$ : correlation in LHS and RHS} \\
\hline
$B\to K^{(\star)}\bar\nu\nu$ vs. $S_{\psi\phi}$ &  LHS/RHS $Z^\prime$ &\ref{fig:BKnuvsSphi}  &\scriptsize{LHS: (anti)correlation  in
$A_3\,(A_1)$} \\
&&& \scriptsize{RHS: $K$ case: above SM, } \\
&&& \scriptsize{(anti)correlation  in $A_3\,(A_1)$; }
\\
&&& \scriptsize{opposite in $ K^{\star}$ case: below SM}\\
\hline
$B\to K^{\star}\bar\nu\nu$ vs. $B\to K\bar\nu\nu$ &  LHS/RHS/LRS $Z^\prime$ &\ref{fig:BKstarnuvsBKnu} & \scriptsize{ (anti)correlation in
LHS and LRS (RHS)} \\
\hline
$Im(C^\prime_{10})$ vs. $Re(C^\prime_{10})$ &  RHS/LRS $Z^\prime$ &\ref{fig:C10}  &\scriptsize{  RHS: $Im(C^\prime_{10})>0\,(<0)$ in
$A_3(A_1)$} \\
&&&\scriptsize{  $Re(C^\prime_{10})>0\,(<0)$ in $A_1(A_3)$} \\
\hline
$\eta$ vs. $\epsilon$ ($b\to s\bar\nu\nu$) &  LHS/RHS/LRS/ALRS $Z^\prime$ &\ref{fig:ep2vseta}  &\scriptsize{ $\eta=0$ in LHS; no
dependence on $\epsilon$ in
RHS}\\
&&& \scriptsize{(anti)correlation in ALRS(LRS)}\\
\hline
\hline
$\Delta M_d$ \& $S_{\psi K_S}$ oases & LHS/RHS $Z^\prime$ & \ref{fig:oasesBdLHS} &\scriptsize{ oases ($B_1,B_2$,$B_3,B_4$) found}\\
 & LRS $Z^\prime$ & \ref{fig:oasesBdLRS1}&\scriptsize{ oases ($B_1,B_2$,$B_3,B_4$) found}\\
& LHS $U(2)^3$ $Z^\prime$ & \ref{fig:oasesU2} &\scriptsize{two common oases for $B_s$ and $B_d$}\\
 & LHS $Z$ &\ref{fig:ZoasesBdLHS1} &\scriptsize{only two oases found} \\
\hline
$S_{\psi K_S}$ vs. $B_d\to\mu^+\mu^-$ & LHS $Z^\prime$ & \ref{fig:BdmuvsSKSLHS} &\scriptsize{ $B(B_d\to\mu^+\mu^-)$ above SM in $
B_2,B_4$}\\
&&& \scriptsize{ and in $B_1$ (LHS1) or in $B_3$ (LHS2)} \\
 & LHS/ALRS $Z$ & \ref{fig:ZBdmuvsSKSLHS} & \scriptsize{$B_d\to\mu^+\mu^-$ always above SM;}\\
\hline
 $B_d\to\mu^+\mu^-$ vs. $S_{\mu^+\mu^-}^d$ & LHS $Z^\prime$ &\ref{fig:BdmuvsSmudLHS} &\scriptsize{$S_{\mu^+\mu^-}^d>0 \,(<0)$ in
$B_1,B_4(B_2,B_3)$}\\
 $B_d\to\mu^+\mu^-$ vs. $S_{\mu^+\mu^-}^d$ & LHS/ALRS $Z$ & \ref{fig:ZBdmuvsSmudLHS} &\scriptsize{}\\
\hline
\end{tabular}
\end{center}}
\caption{\it Overview of correlation plots in $B_d$ and $B_s$ sectors.
\label{tab:corrB}}
\end{table}

\begin{table}[!ht]
{\renewcommand{\arraystretch}{1.1}
\begin{center}\small
\begin{tabular}{|c|c|c|c|}
\hline
$K$ system: observables & scenarios & Figure & comments
\\
\hline
\hline
$\Delta M_K$ \& $\varepsilon_K$ oases & LHS/RHS $Z^\prime$ &\ref{fig:oasesKLHS}  &\scriptsize{oases ($C_1,C_2$) found in S1; only $C_1$ in
S2}\\
 & LRS $Z^\prime$ &\ref{fig:oasesKLRS}&\scriptsize{oases ($C_1,C_2$) found in S1; only $C_1$ in S2}\\
\hline
$ K_L\to \pi^0\bar\nu\nu$ vs. $K^+\to \pi^+\bar\nu\nu$ & LHS $Z^\prime$ & \ref{fig:KLvsKpLHS} &\scriptsize{two branch structure}\\
& LRS $Z^\prime$ &\ref{fig:KLvsKpLRS} &\scriptsize{two branch structure}\\
& LHS $Z$ &\ref{fig:ZKLvsKpLHS} &\scriptsize{only one branch allowed}\\
& LRS $Z$ &\ref{fig:ZKLvsKpLRS} &\scriptsize{two branch structure}\\
& RHS $Z$ &\ref{fig:ZKLvsKpRHS}&\scriptsize{two branch structure}\\
\hline
$K_L\to\mu^+\mu^-$ vs. $K^+\to \pi^+\bar\nu\nu$ & LHS/RHS $Z^\prime$ & \ref{fig:KLmuvsKpLHS} &\scriptsize{(anti)correlation in LHS (RHS)}\\
\hline
$K_L\to \pi^0e^+e^-$ vs. $K_L\to \pi^0\mu^+\mu^-$ & LHS $Z^\prime$ & \ref{fig:KmuKe} &\scriptsize{correlation}\\
 & LRS $Z^\prime$ & \ref{fig:LRKmuKe} &\scriptsize{correlation}\\
& LHS $Z$ & \ref{fig:KmuKeZ} &\scriptsize{correlation}\\
& LRS $Z$ & \ref{fig:LRKmuKeZ} &\scriptsize{correlation}\\
\hline
$K_L\to \pi^0\ell^+\ell^-$ vs. $K_L\to \pi^0\bar\nu\nu$ & LHS $Z^\prime$ & \ref{fig:KmuKe} &\scriptsize{correlation}\\
& LRS $Z^\prime$ & \ref{fig:LRKmuKe} &\scriptsize{correlation}\\
& LHS $Z$ &  \ref{fig:KmuKeZ} &\scriptsize{correlation}\\
& LRS $Z$ & \ref{fig:LRKmuKeZ} &\scriptsize{correlation}\\
\hline

\end{tabular}
\end{center}}
\caption{\it Overview of correlation plots in the $K$ sector.
\label{tab:corrK}}
\end{table}

In this paper we exhibited the pattern of flavour violation in models in which
NP effects are dominated by tree-level $Z^\prime$ exchanges under the assumption that the theoretical and experimental errors on
various input parameters will
decrease with time. In particular we have identified a number of correlations
between $\Delta F=2$ and $\Delta F=1$ processes that will enable in due time
to test this NP scenario.
Our detailed
analysis of these correlations in Section~\ref{sec:Excursion} shows that  in the three meson
systems considered significant foot prints of $Z^\prime$ will be seen
provided $M_{Z'}\le 5\tev$. But only in the $K$ system these effects can
be detected if $M_{Z'}$ is larger and outside the LHC reach.

In view of these findings
we have investigated whether larger effects could be found if the
SM Z boson couplings were flavour violating. Indeed, in this case
the stronger constraints come from $\Delta F=1$ and not $\Delta F=2$
processes. We find then that imposition of the present constraints
from $b\to s \mu^+\mu^-$ transitions makes NP effects in $\Delta F=2$
transitions in the $B_s$ system very small precluding also large NP effects
in $b\to s \nu\bar\nu$ transitions.

The situation is different 
in the $B_d$ system and in particular in the $K$ system where effects
from flavour-violating $Z$ couplings are still allowed
to be large.

Our results are summarized in a number of plots that have been obtained
in various scenarios for the $Z^\prime$ couplings and for inclusive and
exclusive values of $\vub$. Also a number of plots have been shown
for the case of flavour-violating $Z$- couplings.
The overview of all correlations found by us and of the related figures 
      is given in Tables~\ref{tab:corrB} and~\ref{tab:corrK}. There we collect a short description of all those plots, in
particular stressing
whether correlations or anticorrelations are found among the various observables.
 We list here only
few highlights:
\begin{itemize}
\item
For each scenario we have identified allowed oases in the parameter space
of the model. In each oasis particular structure of correlations between
various observables will in the future either favour  or exclude a given oasis.
\item
For the near future the correlations involving $S_{\psi\phi}$ and
$\mathcal{B}(B_{s,d}\to\mu^+\mu^-)$   will be most interesting as the
data on these three observables will be improved in the coming months
sharping the outcome of our analysis, possibly ruling out some oases
and scenarios of the couplings.
The plots in Figs.~\ref{fig:SmusvsSphiLHS1}, \ref{fig:BsmuvsSphiRHS} and
\ref{fig:BdmuvsSKSLHS}
will be helpful in monitoring these developments.
\item
Of particular interest will be the study of the effects of right-handed
currents. Here the recent constraints on the Wilson coefficients
of primed operators from $b\to s \mu^+\mu^-$ transitions had already
impact on our results. In the future an important role in testing
RH currents will be played by $b\to s\nu\bar\nu$ transitions. The
plots in Fig.~\ref{fig:BKnuvsBsmu}, \ref{fig:BKnuvsSphi}  and \ref{fig:BKstarnuvsBKnu} exhibit the power of these decays in this
context.
\item
While in $B_{s,d}$ decays for $M_{Z'}\ge 5\tev$ $Z'$ effects are predicted to be
small, in $\kpn$ and $\klpn$ decays they can be important in LHS and RHS 
scenarios even at
$M_{Z'}=10\tev$  and slightly larger scales. This is seen in
Fig.~\ref{fig:KLvsKpLHS}. On the other hand as seen in Fig.~\ref{fig:KLvsKpLRS}
in LRS scenarios for  $M_{Z'}\ge 5\tev$ it will be difficult to identify 
NP in these decays.
\item
We have demonstrated that the imposition of $U(2)^3$ symmetry on the $Z'$
couplings has a profound impact on the correlation of $S_{\psi\phi}$ with
$B_s\to\mu^+\mu^-$ sharping the predictions significantly. We show this
in Fig.~\ref{fig:BsmuvsSphiU2LHS}.
\item
Our analysis of flavour-violating $Z$-couplings shows that in the case of
$B_d$ and $K$ system they could constitute and important portal to NP with
only small effects still allowed in the $B_s$ system. The plots in
Figs.~\ref{fig:ZoasesBdLHS1},   \ref{fig:ZBdmuvsSKSLHS}, \ref{fig:ZBdmuvsSmudLHS}, \ref{fig:ZKLvsKpLRS}, \ref{fig:ZKLvsKpLHS}, \ref{fig:ZKLvsKpRHS},
\ref{fig:KmuKeZ} and \ref{fig:LRKmuKeZ}
 illustrate these
findings.
\item
We have reemphasized, following \cite{Buras:1998ed,Buras:1999da,Blanke:2007wr}, 
that large NP effects in rare $K$ decays could be 
softened by the correlations with $\epe$ if the hadronic matrix elements 
relevant for this ratio were better known.
\end{itemize}

 We are aware of the fact that some of the correlations presented
by us would be washed out if we included all existing uncertainties. Yet, our
simplified numerical analysis had as the main goal to illustrate how the
decrease of theoretical, parametric and experimental uncertainties in the
coming years might allow to exhibit certain features of NP, even if
deviations from the SM will be only moderate.
In this manner we have uncovered a world of correlations present
in NP scenario, where new effects are dominated by flavour-violating couplings
of a heavy neutral gauge bosons and/or SM $Z$ boson.
In fact within the
coming years the size of the assumed uncetainties in our analysis
 could likely become
reality.

We are looking forward to improved experimental data  and improved lattice
calculations. The correlations identified in this paper will allow to
monitor how  the simple NP scenarios discussed by us
face the future precision flavour data.

{\bf Acknowledgements}\\
We thank Pietro Colangelo, Robert Fleischer, Robert Knegjens, and Robert Ziegler for discussions.
This research was financially supported by the ERC Advanced Grant project ``FLAVOUR'' (267104).

\bibliographystyle{JHEP}
\bibliography{allrefs}
\end{document}